\RequirePackage{lineno}
\documentclass[aps,amsmath]{revtex4}
\usepackage{graphicx}  
\usepackage{dcolumn}   
\usepackage{amssymb}   
\usepackage{rotating}  
\usepackage{appendix}
\def\MET{{\mbox{$E\kern-0.57em\raise0.19ex\hbox{/}_{T}$}}}
\def\met{{\mbox{$E\kern-0.57em\raise0.19ex\hbox{/}_{T}$}}}
\def\DZ{D0 }
\def\DZero{D0 }
\def\Dzero{D0 }

\def\ifb{~fb$^{-1}$}
\def\pp{$p\bar{p}$}
\def\bb{$b\bar{b}$}
\def\cc{$c\bar{c}$}
\def\ttbar{$t\bar{t}$}
\def\WH{$WH\rightarrow \ell\nu b\bar{b}$}

\def\lmet{$WH\rightarrow \ell\kern-0.45em\raise0.19ex\hbox{/} \nu b\bar{b}$}
\def\ZH{$ZH\rightarrow \nu\bar{\nu} b\bar{b}$}
\def\ZHll{$ZH\rightarrow \ell^+ \ell^- b\bar{b}$}

\def\vww{$VH \rightarrow \ell^{\pm}\ell'^{\pm} + X$}

\def\hww{$H\rightarrow W^+ W^-$}
\def\hbb{$H\rightarrow b\bar{b}$}

\def\tevE{$\sqrt{s}=1.96$~TeV}

\begin{document}

\rightline{FERMILAB-CONF-11-354-E}
\rightline{CDF Note 10606}
\rightline{\DZ Note 6226}
\vskip0.5in

\title{Combined CDF and \DZ Upper Limits on Standard Model Higgs Boson Production with up to 8.6 fb$^{-1}$ of Data\\[2.5cm]}

\author{
The TEVNPH Working Group\footnote{The Tevatron
New-Phenomena and Higgs Working Group can be contacted at
TEVNPHWG@fnal.gov. More information can be found at http://tevnphwg.fnal.gov/.}
 }
\affiliation{\vskip0.3cm for the CDF and \DZ Collaborations\\ \vskip0.2cm
\today}
\begin{abstract}
\vskip0.3in
We combine results from CDF and D0 on direct searches for the standard model (SM)
Higgs boson ($H$) in \pp~collisions at the Fermilab Tevatron at $\sqrt{s}=1.96$~TeV.
Compared to the previous Tevatron Higgs boson search combination more data have been added,
additional channels have been incorporated, and some previously used channels
have been reanalyzed to gain sensitivity.  We use the MSTW08 parton distribution
functions and the latest theoretical cross sections when comparing our
limits to the SM predictions.  With up to 8.2\ifb\ of data analyzed at CDF and up
to 8.6\ifb\ at D0, the 95\% C.L. our upper limits on Higgs boson production are factors of 
1.17, 1.71, and 0.48 times the values of the SM cross section for 
Higgs bosons of mass $m_{H}=$115~GeV/$c^2$, 140~GeV/$c^2$,
and 165~GeV/$c^2$, respectively.  The corresponding median upper limits expected in the absence
of Higgs boson production are 1.16, 1.16, and 0.57.  
There is a small ($\approx 1\sigma$) excess of data events
with respect to the background estimation in searches for the Higgs boson in the mass range $125<m_{H}<155$~GeV/$c^{2}$.
We exclude, at the 95\% C.L., a new and larger region at high mass between
$156<m_{H}<177$~GeV/$c^{2}$, with an expected exclusion region of $148<m_{H}<180$~GeV/$c^{2}$.

\vskip 2cm
{\hspace*{5.5cm}\em Preliminary Results}
\end{abstract}

\maketitle

\newpage
\section{Introduction} 

The search for a mechanism for electroweak symmetry breaking, and in
particular for a standard model (SM) Higgs boson, has been a major
goal of particle physics for many years, and is a central part of the
Fermilab Tevatron physics program. Both the CDF and \Dzero collaborations
have performed new combinations~\cite{CDFHiggs,DZHiggs} of multiple
direct searches for the SM Higgs boson.  The new searches include more
data, additional channels, and improved analysis techniques
compared to previous analyses.  The sensitivities of these new combinations
significantly exceed those of previous combinations~\cite{prevhiggs,WWPRLhiggs}.

In this note, we combine the most recent results of all such
searches in \pp~collisions at~\tevE.  The analyses combined
here seek signals of Higgs bosons produced in association with
a vector boson ($q\bar{q}\rightarrow W/ZH$), through gluon-gluon
fusion ($gg\rightarrow H$), and through vector boson fusion (VBF)
($q\bar{q}\rightarrow q^{\prime}\bar{q}^{\prime}H$) corresponding
to integrated luminosities up to 8.2\ifb~at CDF and up to
8.6\ifb~at D0.  The Higgs
boson decay modes studied are $H\rightarrow b{\bar{b}}$, $H\rightarrow
W^+W^-$, $H\rightarrow ZZ$, $H\rightarrow\tau^+\tau^-$
and $H\rightarrow \gamma\gamma$.

To simplify the combination, the searches are separated into
165 mutually exclusive final states (71 for CDF and 94 for D0;
see Tables~\ref{tab:cdfacc} and~\ref{tab:dzacc}) referred to
as ``analysis sub-channels'' in this note.  The selection
procedures for each analysis are detailed in Refs.~\cite{cdfWH2J}
through~\cite{dzHgg}, and are briefly described below.

\section{Acceptance, Backgrounds, and Luminosity}  

Event selections are similar for the corresponding CDF and D0 analyses, consisting
typically of a preselection followed by the use of a multivariate analysis technique
with a final discriminating variable to separate signal and background.
For the case of \WH, an isolated lepton ($\ell=$ electron or muon)
and two or three jets required, with one or more $b$-tagged jets, i.e.,
identified as containing a weakly-decaying $b$ hadron.  Selected
events must also display a significant imbalance in transverse momentum
(referred to as missing transverse energy or \met).  Events with more
than one isolated lepton are rejected.

For the D0 \WH\ analyses, the
data are split by lepton type and jet multiplicity (two or three jet
sub-channels), and whether there are one or two $b$-tagged jets.
As with other D0 analyses targeting the \hbb\ decay,  the \WH\ analyses use a new
boosted decision tree based $b$-tagging algorithm for this combination. The new
algorithm is an upgraded version
of the neural network $b$-tagger used previously~\cite{Abazov:2010ab}, and includes
more information relating to the lifetime of the jet and results in a better
discrimination between $b$ and light jets. Unlike previous
versions of the analysis the same ``loose'' $b$-tagging criterion is applied to
both the single (LST) and double (LDT)
tag samples, with the output of the $b$-tagger now being used as an input to the final discriminant.
This loose $b$-tagging criterion corresponds to an identification efficiency of
$\approx 80\%$ for true $b$-jets for a mis-identification rate of $\approx 10\%$.
Each sub-channel is analyzed separately.  The outputs of boosted decision trees, trained
separately for each sample and for
each Higgs boson mass, are used as the final discriminating variables in the
limit setting procedure. In addition for this combination D0 now uses 8.5 fb$^{-1}$ of data.

For the CDF \WH\ analyses, events are analyzed in two and three jet
sub-channels separately, and in each of these samples the events
are grouped into various lepton and $b$-tag categories.  Events are
broken into separate analysis categories based on the quality of the
identified lepton.  Separate categories are used for events with a
high quality muon or central electron candidate, an isolated track
or identified loose muon in the extended muon coverage, a forward
electron candidate, and a loose central electron or isolated track
candidate.  The final two lepton categories, which
provide some acceptance for lower quality electrons and single prong
tau decays, are used only in the case of two-jet events. Within the
lepton categories there are four $b$-tagging categories considered
for two-jet events: two tight $b$-tags (TDT), one tight $b$-tag and
one loose $b$-tag (LDT), one tight $b$-tag and one looser $b$-tag
(LDTX), and a single, tight, $b$-tag (ST).  For three jet events
there is no LDTX tagging category and the corresponding events are
included within the ST  category.  In the case of the two jet
events, a Bayesian neural network discriminant is trained at each
Higgs boson mass within the test range for each of the specific categories
(defined by lepton type, $b$-tagging type, and number of jets),
while matrix element (ME) discriminants are used for each three jet
event category.

For the \ZH\ analyses, the selection is similar to the $WH$ selection,
except all events with isolated leptons are rejected and stronger multijet
background suppression techniques are applied.  Both the CDF and D0 analyses
use a track-based missing transverse momentum calculation as a discriminant
against false \met. In addition both CDF and D0 utilize multi-variate
techniques, a boosted decision tree at D0 and a neural network at CDF, to
further discriminate against the multijet background before $b$-tagging.
There is a sizable fraction of the \WH\ signal in which the lepton is
undetected that is selected in the \ZH\ samples,  so these analyses are
also referred to as $VH \rightarrow \met b \bar{b}$.  The CDF analysis
uses three non-overlapping categories of $b$-tagged events (TDT, LDT and ST)
D0 uses the same loose single (LST) and
double tag (LDT) criteria as the \WH\ analyses, with the output of the $b$-tagger
again being used as an input to the final discriminant. CDF uses neural-network outputs for the final
discriminating variables, while D0 uses boosted decision tree outputs.
For this combination D0 has analysed the 2-jet sample in an exclusive manner, updating both
the 1- and 2 $b$-tag samples to use 8.4 fb$^{-1}$ of data. The exclusive 3-jet sample is currently not
included.

The \ZHll\ analyses require two isolated leptons and at least two jets.
D0's \ZHll\ analyses separate events into non-overlapping samples of
events with either one tight $b$-tag (TST) or both one tight and one
loose $b$-tags (TLDT).
CDF separates events into single tag (ST), double tag (TDT) and loose
double tag (LDT) samples. To increase signal acceptance D0 loosens
the selection criteria for one of the leptons to include an isolated
track not reconstructed in the muon detector ($\mu\mu_{trk}$) or an
electron from the inter-cryostat region of the D0 detector ($ee_{ICR}$).
Combined with the dielectron ($ee$) and dimuon ($\mu\mu$) analyses,
these provide four orthogonal analyses, and each uses 8.6 fb$^{-1}$ of
data in this combination. CDF uses neural networks to select loose
dielectron and dimuon candidates.
D0 applies a kinematic fit to optimize reconstruction, while CDF corrects
jet energies for \met\ using a neural network approach.  D0 uses random
forests of decision trees to provide the final variables for setting
limits.  CDF utilizes a multi-layer discriminant based on neural networks
where two discriminant functions are used to define three separate
regions of the final discriminant function.

For the \hww~analyses, signal events are characterized by large \met~and
two opposite-signed, isolated leptons.  The presence of neutrinos in the
final state prevents the accurate reconstruction of the candidate Higgs
boson mass.
D0 selects events containing electrons and/or muons, dividing the data sample
into three final states: $e^+e^-$, $e^\pm \mu^\mp$, and $\mu^+\mu^-$. Each final state is further
subdivided according to the number of jets in the event: 0, 1, or 2 or more (``2+'') jets.
Each of the three final states each uses 8.1 fb$^{-1}$ of data. Decays involving tau
leptons are included in two orthogonal ways. A dedicated
analysis ($\mu\tau_{\rm{had}}$) using 7.3 fb$^{-1}$
of data studying the final state involving a muon and a hadronic tau decay is included in the
Tevatron combination. Final states involving
other tau decays and mis-identified hadronic tau decays are included in the $e^+e^-$, $e^\pm \mu^\mp$,
and $\mu^+\mu^-$ final state analyses.
CDF separates the \hww\ events in five non-overlapping samples, split
into ``high $s/b$'' and ``low $s/b$'' categories defined by lepton
types and the number of reconstructed jets: 0, 1, or 2+ jets.  The sample
with two or more jets is not split into low $s/b$ and high $s/b$ lepton
categories due to the smaller statistics in this channel.  A sixth CDF
channel is the low dilepton mass ($m_{\ell^+\ell^-}$) channel, which
accepts events with $m_{\ell^+\ell^-}<16$~GeV.  A new improvement of the
CDF analysis is the ability to recover events with lepton pairs
that lie within each other's isolation cones.  This improvement leads to
a significant increase in sensitivity from the low $m_{\ell^+\ell^-}$
channel, in particular.

The division of events into categories based on the number of reconstructed
jets allows the analysis discriminants
to separate differing contributions of signal and background processes more
effectively.  The signal production mechanisms considered are $gg\rightarrow
H\rightarrow W^+W^-$, $WH/ZH\rightarrow jjW^+W^-$, and vector-boson fusion.  The
relative fractions of the contributions from each of the three signal processes
and background processes, notably $W^+W^-$ production and $t{\bar{t}}$ production,
are very different in the different jet categories.  Dividing our data into these
categories provides more statistical discrimination, but introduces the need to
evaluate the systematic uncertainties carefully in each jet category.  A discussion
of these uncertainties is found in Section~\ref{sec:signal}.

The D0 $e^+e^-$, $e^\pm \mu^\mp$, and $\mu^+\mu^-$ final state channels use
boosted decision tree outputs as the final discriminants while
the $\mu\tau_{\rm{had}}$ channel uses neural networks.
CDF uses neural-network outputs, including likelihoods constructed from
calculated matrix-element probabilities as additional inputs for the 0-jet bin.

D0 includes \vww\ analyses in which the associated vector boson and the $W$ boson from the Higgs
boson decay are required to decay leptonically, with like-sign leptons present,
thereby defining three like-sign dilepton final states ($e^\pm e^\pm$, $e^\pm
\mu^\pm$, and $\mu^{\pm}\mu^{\pm}$).  The combined output of two decision
trees, trained against the instrumental and diboson backgrounds respectively,
is used as the final discriminant. CDF also includes a separate analysis
of events with same-sign leptons to incorporate additional potential signal
from associated production events in which the two leptons (one from the
associated vector boson and one from a $W$ boson produced in the Higgs boson decay)
have the same charge.  CDF additionally incorporates three tri-lepton channels
to include additional associated production contributions in which leptons result
from the associated $W$ boson and the two $W$ bosons produced in the Higgs boson decay
or where an associated $Z$ boson decays into a dilepton pair and a third lepton
is produced in the decay of either of the $W$ bosons resulting from the Higgs boson
decay.  In the latter case, CDF separates the sample into one jet and two or
more jet sub-channels to take advantage of the fact that the Higgs boson candidate
mass can be reconstructed from the invariant mass of the two jets, the lepton, and
the missing transverse energy.

For the first time CDF includes a search for $H \rightarrow ZZ$ using four
lepton events.  A simple four-lepton invariant mass discriminant is used to
separate potential Higgs boson signal events from the non-resonant $ZZ$ background.
CDF has also updated its opposite-sign channels in which one of the two
lepton candidates is a hadronic tau.  Events are separated into $e$-$\tau$
and $\mu$-$\tau$ channels.  The final discriminants are obtained from boosted
decision trees which incorporate both hadronic tau identification and kinematic
event variables as inputs.  D0 also includes channels in which one of the $W$
bosons in the $H \rightarrow W^+W^-$ process decays leptonically and the other
decays hadronically.  Electron and muon final states are studied separately,
each with 5.3 fb$^{-1}$ of data.  Random forests are used for the final
discriminants.

CDF includes an updated, generic analysis searching for Higgs bosons decaying
to tau lepton pairs incorporating contributions from direct $gg \rightarrow H$
production, associated $WH$ or $ZH$ production, and vector boson production.
CDF also includes for the first time an analysis of events that contain one
reconstructed lepton ($\ell$ = $e$ or $\mu$) in addition to a tau lepton
pair focusing on associated production where $H \rightarrow \tau \tau$ and
an additional lepton is produced in the decay of the $W$ or $Z$ boson.  For
the generic search, events with either one or two jets are separated into two
independent analysis channels.  The final discriminant for setting limits
is obtained using four boosted decision tree discriminants, each designed to
discriminate the signal against one of the major backgrounds (QCD multijets,
$W$ plus jets, $Z/\gamma^{*} \rightarrow \tau^+\tau^-$, and $Z/\gamma^{*}
\rightarrow \ell^+\ell^-$ where $\ell = e$ or $\mu$).  In the new analysis
events are separated into three tri-lepton categories ($e$-$\mu$-$\tau_{\rm{had}}$,
$\ell$-$\ell$-$\tau_{\rm{had}}$, and $\ell$-$\tau_{\rm{had}}$-$\tau_{\rm{had}}$).  The final
discriminants are likelihoods based on Support Vector Machine (SVM)~\cite{svm} outputs
obtained using separate trainings for the signal against each of the primary
backgrounds ($Z$ plus jets, $t\bar{t}$, and dibosons).
The D0 $\ell^\pm \tau^{\mp}_{\rm{had}}jj$ analysis likewise includes
direct $gg \rightarrow H$ production, associated $WH$ or $ZH$ production, and vector boson production.
Decay of the Higgs boson to both tau and $W$ boson pairs is considered. A final state consisting
of one leptonic tau decay, one hadronic hadronic tau decay and two jets is required.
Both muonic and electronic sub-channels use 4.3 fb$^{-1}$
The output of boosted decision trees is used as the final discriminant.

CDF incorporates an older all-hadronic analysis, which results in two
$b$-tagging sub-channels (TDT and LDT) for both $WH/ZH$ and VBF production
to the $jjb{\bar{b}}$ final state.  Events with either four or five
reconstructed jets are selected, and at least two must be $b$-tagged.  The
large QCD multijet backgrounds are modeled from the data by applying a
measured mistag probability to the non $b$-tagged jets in events containing
a single $b$-tag.  Neural network discriminants based on kinematic event
variables including those designed to separate quark and gluon jets are used
to obtain the final limits.

D0 and CDF both contribute analyses searching for Higgs bosons decaying into
diphoton pairs.  The CDF analysis looks for a signal peak in the diphoton
invariant mass spectrum above the smooth background originating from standard
QCD production.  Signal acceptance has been increased in the updated analysis
by including forward (plug) calorimeter candidates as well as central photon
conversion candidates.  Events are now separated into four independent
analysis channels based on the photon candidates contained within the event:
two central candidates (CC), one central and one plug candidate (CP), one
central and one central conversion candidate (CC-Conv), or one plug and
one central conversion candidate (PC-Conv).
In the D0 analysis the contribution of jets misidentified as photons is reduced
by combining information sensitive to differences in the energy deposition from
these particles in the tracker, calorimeter and central preshower in a neural
network.  The output of boosted decision trees, rather than the diphoton invariant mass,
is used as the final discriminating variable.
The transverse energies of the leading two photons along with the
azimuthal opening angle between them and the diphoton invariant mass and
transverse momentum are used as input variables. A sizeable improvement in
sensitivity ($\approx 30\%$) beyond that achieved with the invariant mass is obtained.

CDF for the first time includes three non-overlapping sets of analysis channels
searching for the process $t \bar{t} H \rightarrow t \bar{t} b \bar{b}$.
One set of channels selects events with a reconstructed lepton, large
missing transverse energy, and four or five reconstructed jets.  These
events are further sub-divided into five $b$-tagging categories (three
tight $b$-tags (TTT), two tight and one loose $b$-tags (TTL) , one tight and
two loose $b$-tags (TLL), two tight $b$-tags (TDT), and one tight and one
loose $b$-tags (LDT)).  Ensembles of neural network discriminants trained at
each mass point are used to set limits.  A second set of channels selects
events with no reconstructed lepton.  These events
are separated into two categories, one containing events with large missing
transverse energy and five to nine reconstructed jets and another containing
events with low missing transverse energy and seven to ten reconstructed jets.
Events in these two channels are required to have a minimum of two $b$-tagged
jets based on a neural network tagging algorithm.  Events with three or more
$b$-tags are analyzed in separate channels from those with exactly two tags.
Two stages of neural network discriminants are used (the first to help reject
large multijet backgrounds and the second to separate potential $t\bar{t}H$
signal events from $t\bar{t}$ background events).

For both CDF and D0, events from QCD multijet (instrumental) backgrounds are
typically measured in independent data samples using several different methods.
For CDF, backgrounds from SM processes with electroweak gauge bosons or top
quarks were generated using \textsc{PYTHIA}~\cite{pythia}, \textsc{ALPGEN}~\cite{alpgen},
\textsc{MC@NLO}~\cite{MC@NLO}, and \textsc{HERWIG}~\cite{herwig} programs.
For D0, these backgrounds were generated using \textsc{PYTHIA}, \textsc{ALPGEN},
and \textsc{COMPHEP}~\cite{comphep}, with \textsc{PYTHIA} providing parton-showering
and hadronization for all the generators.  These background processes were normalized
using either experimental data or next-to-leading order calculations (including
\textsc{MCFM}~\cite{mcfm} for the $W+$ heavy flavor process).

\section{Signal Predictions}
\label{sec:signal}

We normalize our Higgs boson signal predictions to the most recent high-order
calculations available.  The $gg\rightarrow H$ production cross section we use is
calculated at next-to-next-to leading order (NNLO) in QCD with a next-to-next-to leading log (NNLL)
resummation of soft gluons; the calculation also includes two-loop electroweak effects and
handling of the running $b$ quark mass~\cite{anastasiou,grazzinideflorian}.
The numerical values in Table~\ref{tab:higgsxsec} are updates~\cite{grazziniprivate}
of these predictions with $m_t$ set to 173.1~GeV/$c^2$~\cite{tevtop09},
and with a treatment of the massive top and bottom loop corrections up to
next-to-leading-order (NLO) + next-to-leading-log (NLL) accuracy. The
factorization and renormalization scale choice for this calculation is $\mu_F=\mu_R=m_H$.
These calculations are refinements of the earlier NNLO calculations of the $gg\rightarrow H$
production cross section~\cite{harlanderkilgore2002,anastasioumelnikov2002,ravindran2003}.
Electroweak corrections were computed in Refs.~\cite{actis2008,aglietti,maltoni,aglietti1}. Soft gluon
resummation was introduced in the prediction of the $gg\rightarrow H$ production cross
section in Ref.~\cite{catani2003}.  The $gg\rightarrow H$ production cross section depends strongly
on the gluon parton density function, and the accompanying value
of $\alpha_s(q^2)$.  The cross sections used here are calculated
with the MSTW~2008 NNLO PDF set~\cite{mstw2008}, as recommended by the PDF4LHC working group~\cite{pdf4lhc}.  The inclusive
Higgs boson production cross sections are
listed in Table~\ref{tab:higgsxsec}.

For analyses that consider inclusive $gg\rightarrow H$ production but do not split it into separate channels
based on the number of reconstructed jets, we use the inclusive uncertainties from the simultaneous variation
of the factorization and renormalization scale up and down by a factor of two.  We use the
prescription of the PDF4LHC working group for evaluating PDF uncertainties on the inclusive production
cross section.  QCD scale uncertainties that affect the cross section via their impacts on the PDFs are included
as a correlated part of the total scale uncertainty.  The remainder of the PDF uncertainty is treated as
uncorrelated with the QCD scale uncertainty.

For analyses seeking $gg\rightarrow H$ production that
divide events into categories based on the number of reconstructed jets, we employ a new
approach for evaluating the impacts of the scale uncertainties.  Following the recommendations of Ref.~\cite{bnlaccord},
we treat the QCD scale uncertainties obtained from the NNLL inclusive~\cite{grazzinideflorian,anastasiou}, NLO one or
more jets~\cite{anastasiouwebber}, and NLO two or more jets~\cite{campbellh2j}
cross section calculations as uncorrelated with one another.  We then obtain
QCD scale uncertainties for the exclusive $gg\rightarrow H+0$~jet, 1~jet, and 2~or more jet categories
by propagating the uncertainties on the inclusive cross section predictions through the subtractions
needed to predict the exclusive rates.  For example, the $H$+0~jet cross section is obtained by
subtracting the NLO $H+1$~or more jet cross section from the inclusive NNLL+NNLO cross section.
We now assign three separate, uncorrelated scale uncertainties
which lead to correlated and anticorrelated uncertainty contributions between exclusive jet categories.
The procedure in Ref.~\cite{anastasiouwebber} is used
to determine PDF model uncertainties.  These are obtained
separately for each jet bin and treated as 100\% correlated
between jet bins and between D0 and CDF.

The scale choice affects the $p_T$ spectrum of the Higgs boson when
produced in gluon-gluon fusion, and this effect changes the acceptance
of the selection requirements and also the shapes of the distributions
of the final discriminants.  The effect of the acceptance change is
included in the calculations of Ref.~\cite{anastasiouwebber} and
Ref.~\cite{campbellh2j}, as the experimental requirements are simulated
in these calculations. The effects on the final discriminant shapes
are obtained by reweighting the $p_T$ spectrum of the Higgs boson
production in our Monte Carlo simulation to higher-order calculations.
The Monte Carlo signal simulation used by CDF and D0 is provided by
the LO generator {\sc pythia}~\cite{pythia} which includes a parton
shower and fragmentation and hadronization models.  We reweight the
Higgs boson $p_T$ spectra in our {\sc pythia} Monte Carlo samples to
that predicted by {\sc hqt}~\cite{hqt} when making predictions of
differential distributions of $gg\rightarrow H$ signal events. To
evaluate the impact of the scale uncertainty on our differential
spectra, we use the {\sc resbos}~\cite{resbos} generator, and apply
the scale-dependent differences in the Higgs boson $p_T$ spectrum to
the {\sc hqt} prediction, and propagate these to our final
discriminants as a systematic uncertainty on the shape, which is
included in the calculation of the limits.

We include all significant Higgs boson production modes in the high-mass
search.   Besides gluon-gluon fusion through virtual quark loops
(ggH), we include Higgs boson production in association with a $W$
or $Z$ vector boson (VH), and vector boson  fusion (VBF). For the low-mass searches,
we target the $WH$, $ZH$, VBF, and  $t{\bar{t}}H$ production
modes with specific searches, including also those signal components
not specifically targeted but which fall in the acceptance nonetheless.
Our $WH$ and $ZH$ cross sections are from Ref.~\cite{djouadibaglio}.
This calculation starts with the NLO calculation of
{\sc v2hv}~\cite{v2hv} and includes NNLO QCD contributions~\cite{vhnnloqcd}, as well
as one-loop electroweak corrections~\cite{vhewcorr}.
We use the VBF cross section computed at NNLO in QCD in Ref.~\cite{vbfnnlo}.
Electroweak corrections to the VBF production cross section are computed
with the {\sc hawk} program~\cite{hawk}, and are small and negative (2-3\%)
in the Higgs boson mass range considered here.  We include these corrections in the VBF
cross sections used for this result.  The $t{\bar{t}}H$ production cross
sections we use are from Ref.~\cite{tth}.

In order to predict the kinematic distributions of Higgs boson signal events, CDF and D0
use the \textsc{PYTHIA}~\cite{pythia} Monte Carlo program, with
\textsc{CTEQ5L} and \textsc{CTEQ6L1}~\cite{cteq} leading-order (LO)
parton distribution functions.  

The Higgs boson decay branching ratio
predictions used for this result are those of Ref.~\cite{lhcxs}.  In this calculation,
the partial decay widths for all Higgs boson decays except to pairs of $W$ and $Z$ bosons
are computed with \textsc{HDECAY}~\cite{hdecay}, and the $W$ and $Z$ pair decay widths are
computed with {\sc Prophecy4f}~\cite{prophecy4f}.
The relevant decay branching ratios are listed in Table~\ref{tab:higgsxsec}.
The uncertainties on the predicted branching ratios from uncertainties in $m_b$, $m_c$, and
$\alpha_s$ are presented in Ref.~\cite{dblittlelhc}.


\begin{sidewaystable}
\begin{center}
\caption{
The production cross sections and decay branching fractions for the SM
Higgs boson assumed for the combination.}
\vspace{0.2cm}
\label{tab:higgsxsec}
\begin{tabular}{|c|c|c|c|c|c|c|c|c|c|c|c|}\hline
$m_H$ & $\sigma_{gg\rightarrow H}$ & $\sigma_{WH}$ & $\sigma_{ZH}$ & $\sigma_{VBF}$ & $\sigma_{t{\bar{t}}H}$  &
$B(H\rightarrow b{\bar{b}})$ & $B(H\rightarrow c{\bar{c}})$ & $B(H\rightarrow \tau^+{\tau^-})$ & $B(H\rightarrow W^+W^-)$ & $B(H\rightarrow ZZ)$ & $B(H\rightarrow\gamma\gamma)$ \\
(GeV/$c^2$) & (fb)  & (fb)    & (fb)    & (fb)   & (fb)     & (\%)   & (\%)    & (\%)  & (\%)   & (\%)     & (\%) \\ \hline
\hline
   100 &  1821.8    &  291.90 & 169.8   &  97.2  &  8.000   & 79.1   & 3.68     & 8.36    & 1.11   & 0.113  & 0.159   \\
   105 &  1584.7    &  248.40 & 145.9   &  89.7  &  7.062   & 77.3   & 3.59     & 8.25    & 2.43   & 0.215  & 0.178   \\
   110 &  1385.0    &  212.00 & 125.7   &  82.7  &  6.233   & 74.5   & 3.46     & 8.03    & 4.82   & 0.439  & 0.197   \\
   115 &  1215.9    &  174.50 & 103.9   &  76.4  &  5.502   & 70.5   & 3.27     & 7.65    & 8.67   & 0.873  & 0.213   \\
   120 &  1072.3    &  150.10 &  90.2   &  70.7  &  4.857   & 64.9   & 3.01     & 7.11    & 14.3   & 1.60   & 0.225   \\
   125 &   949.3    &  129.50 &  78.5   &  65.3  &  4.279   & 57.8   & 2.68     & 6.37    & 21.6   & 2.67   & 0.230   \\
   130 &   842.9    &  112.00 &  68.5   &  60.4  &  3.769   & 49.4   & 2.29     & 5.49    & 30.5   & 4.02   & 0.226   \\
   135 &   750.8    &   97.20 &  60.0   &  55.9  &  3.320   & 40.4   & 1.87     & 4.52    & 40.3   & 5.51   & 0.214   \\
   140 &   670.6    &   84.60 &  52.7   &  51.8  &  2.925   & 31.4   & 1.46     & 3.54    & 50.4   & 6.92   & 0.194   \\
   145 &   600.6    &   73.70 &  46.3   &  48.1  &  2.593   & 23.1   & 1.07     & 2.62    & 60.3   & 7.96   & 0.168   \\
   150 &   539.1    &   64.40 &  40.8   &  44.6  &  2.298   & 15.7   & 0.725    & 1.79    & 69.9   & 8.28   & 0.137   \\
   155 &   484.0    &   56.20 &  35.9   &  41.2  &  2.037   & 9.18   & 0.425    & 1.06    & 79.6   & 7.36   & 0.100   \\
   160 &   432.3    &   48.50 &  31.4   &  38.2  &  1.806   & 3.44   & 0.159    & 0.397   & 90.9   & 4.16   & 0.0533  \\
   165 &   383.7    &   43.60 &  28.4   &  36.0  &  1.607   & 1.19   & 0.0549   & 0.138   & 96.0   & 2.22   & 0.0230  \\
   170 &   344.0    &   38.50 &  25.3   &  33.4  &  1.430   & 0.787  & 0.0364   & 0.0920  & 96.5   & 2.36   & 0.0158  \\
   175 &   309.7    &   34.00 &  22.5   &  31.0  &  1.272   & 0.612  & 0.0283   & 0.0719  & 95.8   & 3.23   & 0.0123  \\
   180 &   279.2    &   30.10 &  20.0   &  28.8  &  1.132   & 0.497  & 0.0230   & 0.0587  & 93.2   & 6.02   & 0.0102  \\
   185 &   252.1    &   26.90 &  17.9   &  26.9  &  1.004   & 0.385  & 0.0178   & 0.0457  & 84.4   & 15.0   & 0.00809 \\
   190 &   228.0    &   24.00 &  16.1   &  25.0  &  0.890   & 0.315  & 0.0146   & 0.0376  & 78.6   & 20.9   & 0.00674 \\
   195 &   207.2    &   21.40 &  14.4   &  23.3  &  0.789   & 0.270  & 0.0125   & 0.0324  & 75.7   & 23.9   & 0.00589 \\
   200 &   189.1    &   19.10 &  13.0   &  21.6  &  0.700   & 0.238  & 0.0110   & 0.0287  & 74.1   & 25.6   & 0.00526 \\ \hline
\end{tabular}	
\end{center}	
\end{sidewaystable}

Tables~\ref{tab:cdfacc} and~\ref{tab:dzacc} summarize, for CDF and D0 respectively,
the integrated luminosities, the Higgs boson mass ranges over which the searches are performed,
and references to further details for each analysis.


\begin{table}[h]
\caption{\label{tab:cdfacc}Luminosity, explored mass range and references
for the different processes and final states ($\ell$ = $e$ or $\mu$) for
the CDF analyses.  The generic labels ``$2\times$'' and ``$4\times$'' refer
to separations based on lepton categories.}
\begin{ruledtabular}
\begin{tabular}{lccc} \\
Channel & Luminosity  & $m_H$ range & Reference \\
        & (fb$^{-1}$) & (GeV/$c^2$) &           \\ \hline
$WH\rightarrow \ell\nu b\bar{b}$ 2-jet channels\ \ \ 4$\times$(TDT,LDT,ST,LDTX)                                                    & 7.5  & 100-150 & \cite{cdfWH2J} \\
$WH\rightarrow \ell\nu b\bar{b}$ 3-jet channels\ \ \ 2$\times$(TDT,LDT,ST)                                                         & 5.6  & 100-150 & \cite{cdfWH3J} \\
$ZH\rightarrow \nu\bar{\nu} b\bar{b}$ \ \ \ (TDT,LDT,ST)                                                                           & 7.8  & 100-150 & \cite{cdfmetbb} \\
$ZH\rightarrow \ell^+\ell^- b\bar{b}$ \ \ \ 2$\times$(TDT,LDT,ST)                                                                  & 7.7  & 100-150 & \cite{cdfZHll1,cdfZHll2} \\
$H\rightarrow W^+ W^-$ \ \ \ 2$\times$(0 jets,1 jet)+(2 or more jets)+(low-$m_{\ell\ell}$)+($e$-$\tau_{\rm{had}}$)+($\mu$-$\tau_{\rm{had}}$) & 8.2  & 110-200 & \cite{cdfHWW} \\
$WH \rightarrow WW^+ W^-$ \ \ \ (same-sign leptons)+(tri-leptons)                                                                  & 8.2  & 110-200 & \cite{cdfHWW} \\
$ZH \rightarrow ZW^+ W^-$ \ \ \ (tri-leptons with 1 jet)+(tri-leptons with 2 or more jets)                                         & 8.2  & 110-200 & \cite{cdfHWW} \\
& 8.2  & 110-200 & \cite{cdfHZZ} \\
$H$ + $X\rightarrow \tau^+ \tau^-$ \ \ \ (1 jet)+(2 jets)                                                                          & 6.0  & 100-150 & \cite{cdfHtt} \\
$WH \rightarrow \ell \nu \tau^+ \tau^-$/$ZH \rightarrow \ell^+ \ell^- \tau^+ \tau^-$\ \ \ ($\ell$-$\ell$-$\tau_{\rm{had}}$)+($e$-$\mu$-$\tau_{\rm{had}}$)+($\ell$-$\tau_{\rm{had}}$-$\tau_{\rm{had}}$)
                                                                                                                                   & 6.2  & 110-150 & \cite{cdfVHtt} \\
$WH+ZH\rightarrow jjb{\bar{b}}$ \ \ \  (GF,VBF)$\times$(TDT,LDT)                                                                   & 4.0  & 100-150 & \cite{cdfjjbb} \\
$H \rightarrow \gamma \gamma$ \ \ \  (CC,CP,CC-Conv,PC-Conv)                                                                       & 7.0  & 100-150 & \cite{cdfHgg} \\
$t\bar{t}H \rightarrow W W b\bar{b} b\bar{b}$ (lepton) \ \ \  (4jet,5jet)$\times$(TTT,TTL,TLL,TDT,LDT)                             & 6.3  & 100-150 & \cite{cdfttHLep} \\
$t\bar{t}H \rightarrow W W b\bar{b} b\bar{b}$ (no lepton) \ \ \  (low met,high met)$\times$(2 tags,3 or more tags)                 & 5.7  & 100-150 & \cite{cdfttHnoLep} \\
\end{tabular}
\end{ruledtabular}
\end{table}

\vglue 0.5cm

\begin{table}[h]
\caption{\label{tab:dzacc}Luminosity, explored mass range and references
for the different processes
and final states ($\ell=e, \mu$) for the D0 analyses.
}
\begin{ruledtabular}
\begin{tabular}{lccc} \\
Channel & Luminosity  & $m_H$ range & Reference \\
        & (fb$^{-1}$) & (GeV/$c^2$) &           \\ \hline
$WH\rightarrow \ell\nu b\bar{b}$ \ \ \ (LST,LDT,2,3 jet)             & 8.5  & 100-150 & \cite{dzWHl} \\
$ZH\rightarrow \nu\bar{\nu} b\bar{b}$ \ \ \ (LST,LDT)   & 8.4  & 100-150 & \cite{dzZHv2} \\
$ZH\rightarrow \ell^+\ell^- b\bar{b}$ \ \ \ (TST,TLDT,$ee$,$\mu\mu$,$ee_{ICR}$,$\mu\mu_{trk}$) & 8.6  & 100-150 & \cite{dzZHll1} \\
%
$H$+$X$$\rightarrow$$ \ell^\pm \tau^{\mp}_{\rm{had}}jj$  \ \ \      & 4.3  & 105-200 & \cite{dzVHt2} \\
$VH \rightarrow \ell^\pm \ell^\pm\ + X $ \ \ \  & 5.3  & 115-200 & \cite{dzWWW} \\
$H\rightarrow W^+ W^- \rightarrow \ell^\pm\nu \ell^\mp\nu$ \ \ \ (0,1,2+ jet)     & 8.1  & 115-200 & \cite{dzHWW}\\
$H\rightarrow W^+ W^- \rightarrow \mu\nu \tau_{\rm{had}}\nu$ \ \ \      & 7.3  & 115-200 & \cite{dzHWWtau}\\
$H\rightarrow W^+ W^- \rightarrow \ell\bar{\nu} jj$      & 5.4  & 130-200 & \cite{dzHWWjj}\\
$H \rightarrow \gamma \gamma$                                 & 8.2  & 100-150 & \cite{dzHgg} \\
\end{tabular}
\end{ruledtabular}
\end{table}

\section{Distributions of Candidates} 

All analyses provide binned histograms of the final discriminant variables
for the signal and background predictions, itemized separately for each
source, and the observed data.
The number of channels combined is large, and the number of bins
in each channel is large.  Therefore, the task of assembling
histograms and checking whether the expected and observed limits are
consistent with the input predictions and observed data is difficult.
We therefore provide histograms that aggregate all channels' signal,
background, and data together.  In order to preserve most of the
sensitivity gain that is achieved by the analyses by binning the data
instead of collecting them all together and counting, we aggregate the
data and predictions in narrow bins of signal-to-background ratio,
$s/b$.  Data with similar $s/b$ may be added together with no loss in
sensitivity, assuming similar systematic errors on the predictions.
The aggregate histograms do not show the effects of systematic
uncertainties, but instead compare the data with the central
predictions supplied by each analysis.

The range of $s/b$ is quite large in each analysis, and so
$\log_{10}(s/b)$ is chosen as the plotting variable.  Plots of the
distributions of $\log_{10}(s/b)$ are shown for Higgs boson masses
of 115, 140, and 165~GeV/$c^2$ in Figure~\ref{fig:lnsb}.  These
distributions can be integrated from the high-$s/b$ side downwards,
showing the sums of signal, background, and data for the most pure
portions of the selection of all channels added together.  These
integrals can be seen in Figure~\ref{fig:integ}.  The most significant
candidates are found in the bins with the highest $s/b$; an excess
in these bins relative to the background prediction drives the Higgs
boson cross section limit upwards, while a deficit drives it downwards.
The lower-$s/b$ bins show that the modeling of the rates and kinematic
distributions of the backgrounds is very good.  The integrated plots
show a slight excess of events in the highest-$s/b$ bins for the
analyses seeking a Higgs boson mass of 115~GeV/$c^2$ and 140~GeV/$c^2$, and a slight
deficit of events in the highest-$s/b$ bins for the analyses seeking
a Higgs boson of mass 165~GeV/$c^2$.

We also show the distributions of the data after subtracting the
expected background, and compare that with the expected signal yield
for a Standard Model Higgs boson, after collecting all bins in all
channels sorted by $s/b$.  These background-subtracted distributions
are shown in Figure~\ref{fig:bgsub} for Higgs boson masses of 115, 140, and
165~GeV/$c^2$.  These graphs also show the
remaining uncertainty on the background prediction after fitting the
background model to the data within the systematic uncertainties on
the rates and shapes in each contributing channel.

 \begin{figure}[t]
 \begin{centering}
 \includegraphics[width=0.4\textwidth]{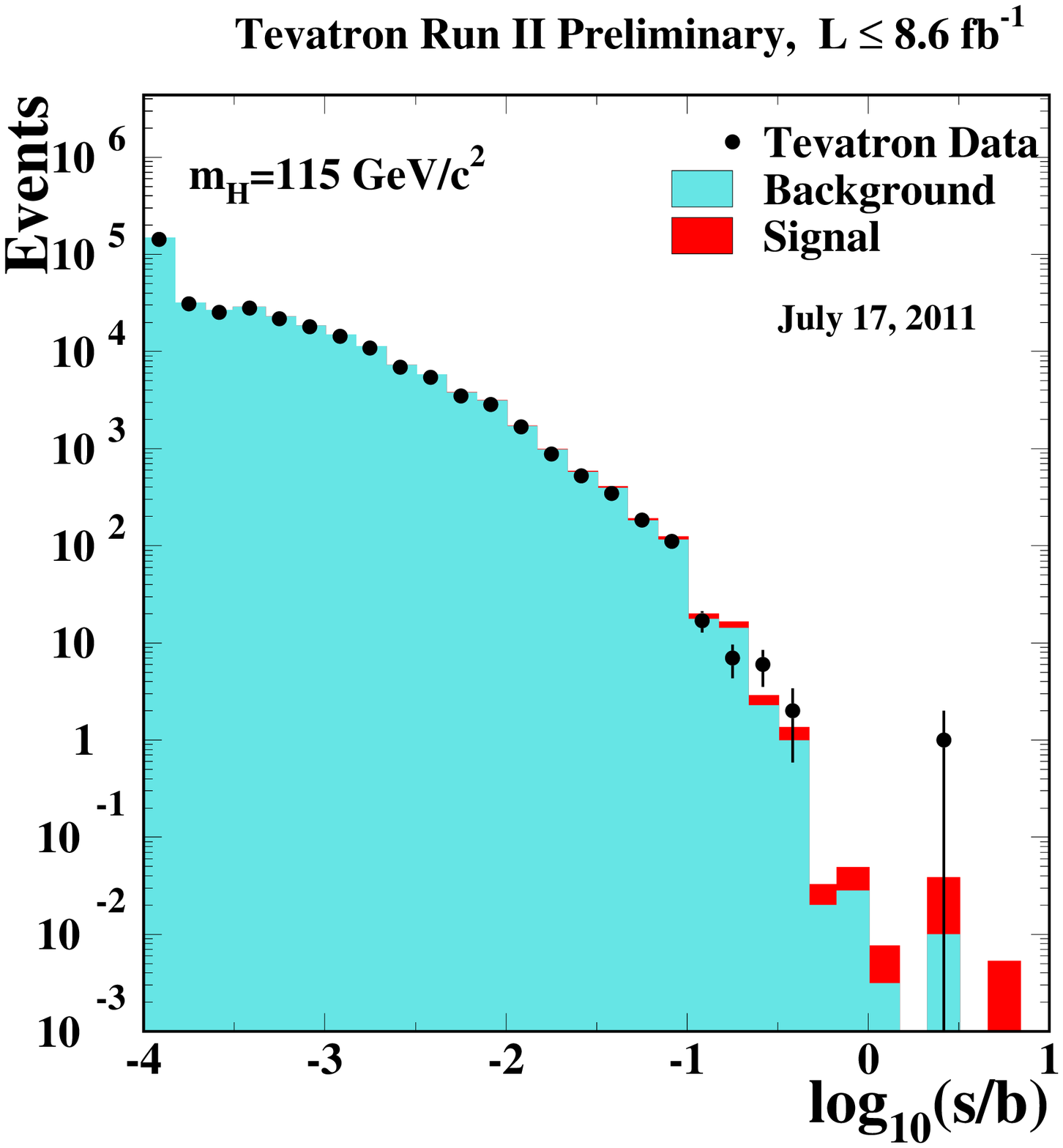}\includegraphics[width=0.4\textwidth]{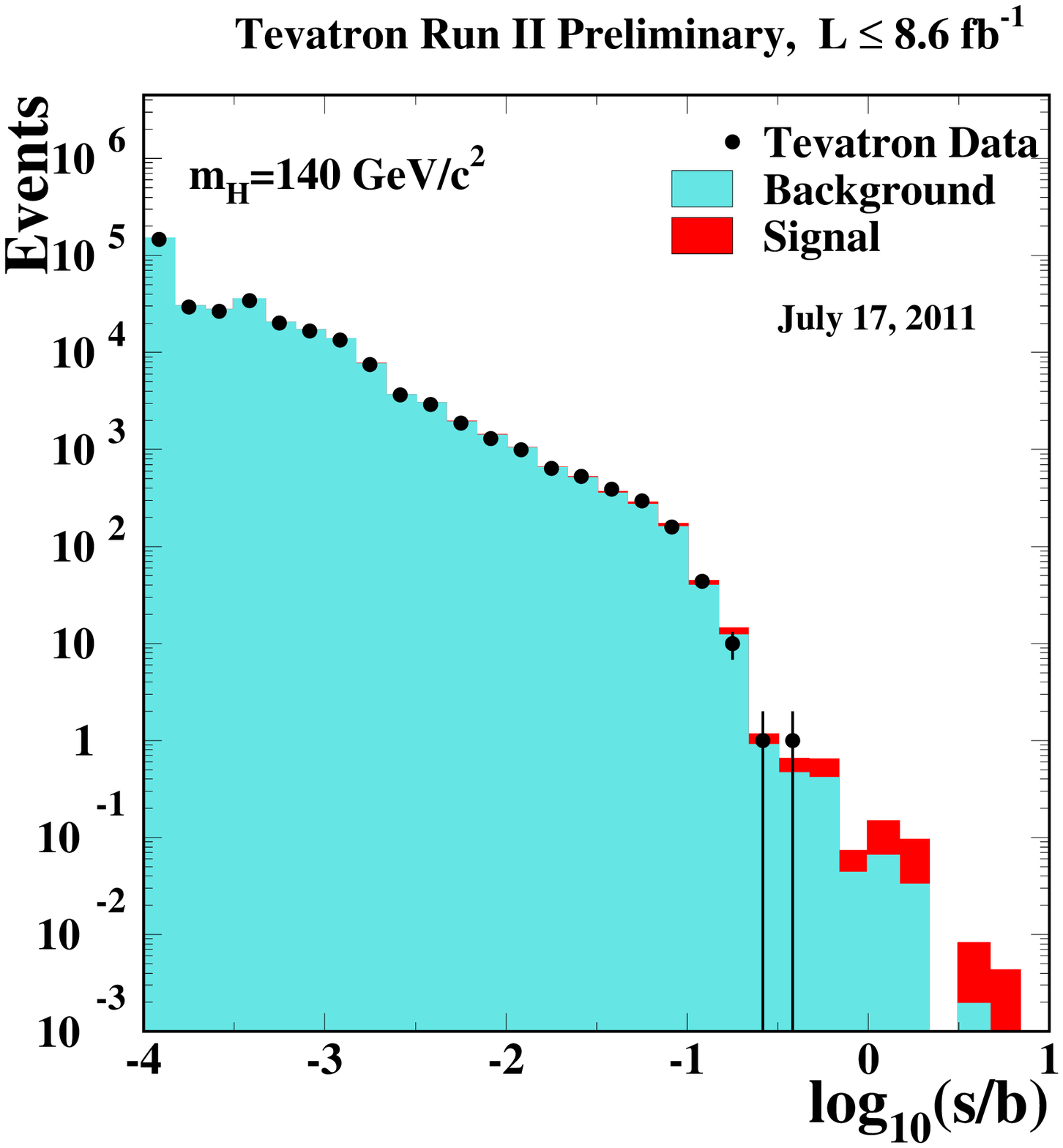}
\includegraphics[width=0.4\textwidth]{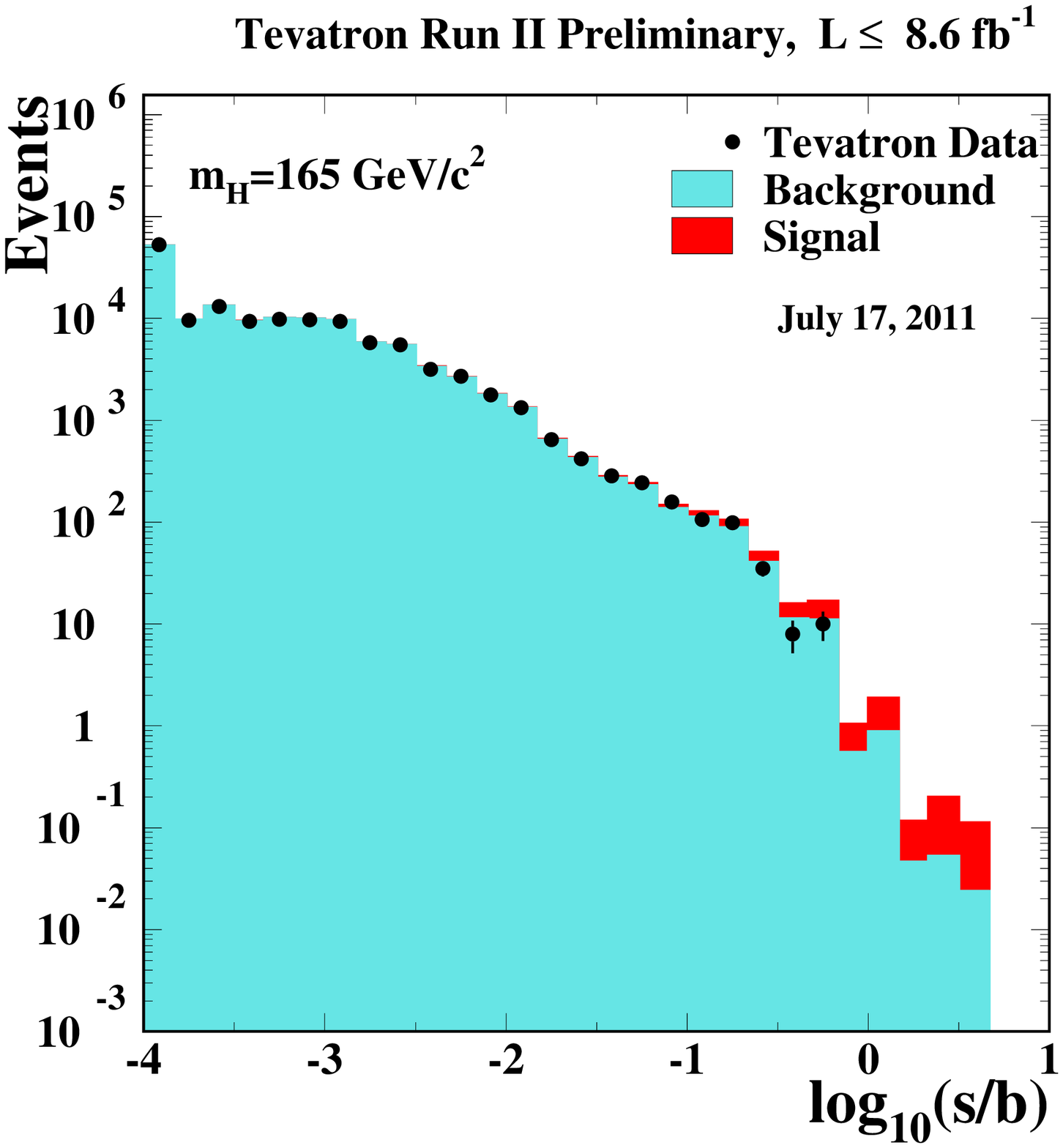}
 \caption{
 \label{fig:lnsb} Distributions of $\log_{10}(s/b)$, for the data from all
contributing channels from CDF and D0, for Higgs boson masses of 115, 140, and
165~GeV/$c^2$.  The data are shown with points, and the expected signal
is shown stacked on top of the backgrounds.  Underflows and overflows are
collected into the leftmost and rightmost bins. }
 \end{centering}
 \end{figure}

 \begin{figure}[t]
 \begin{centering}
 \includegraphics[width=0.4\textwidth]{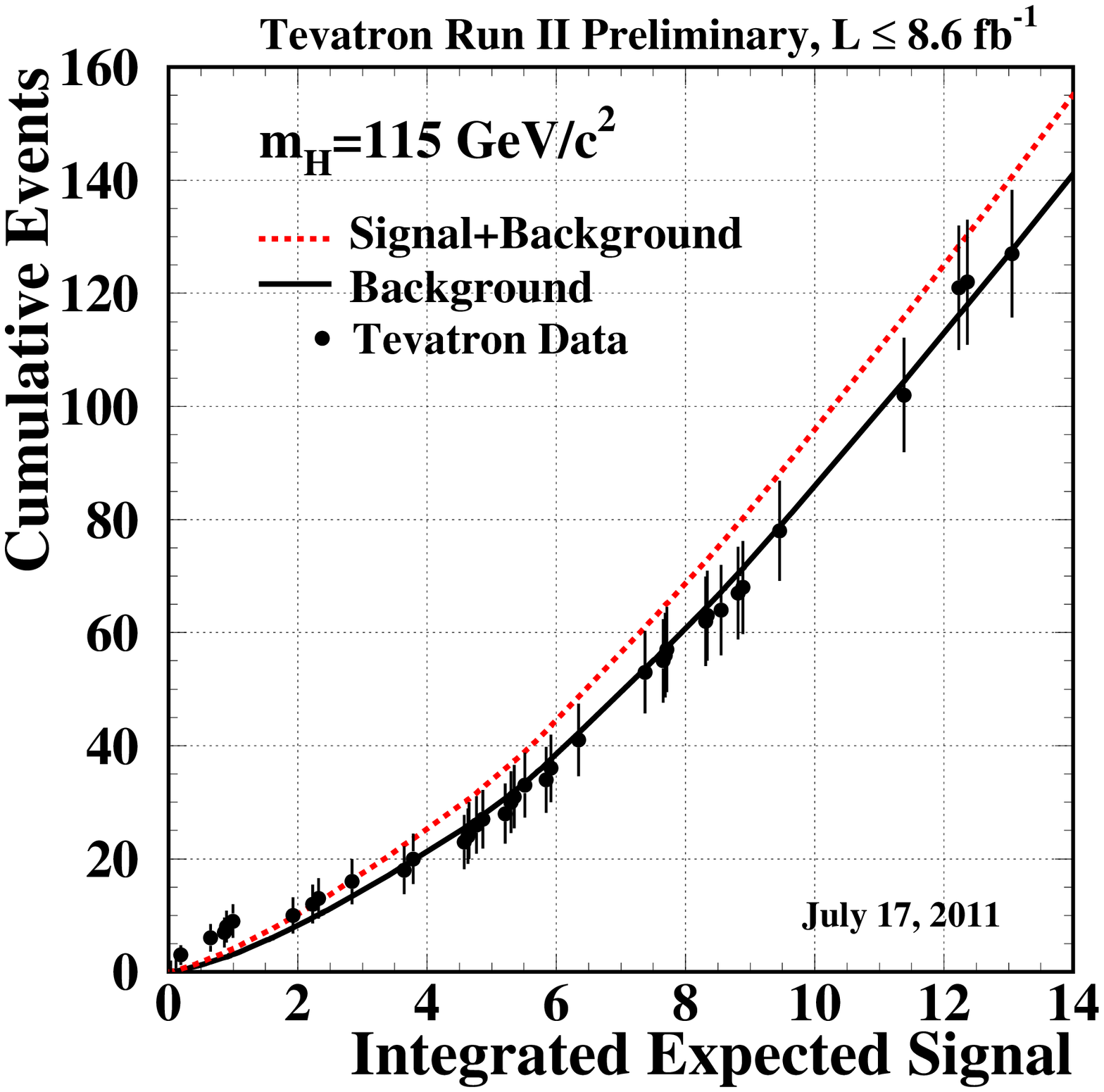}\includegraphics[width=0.4\textwidth]{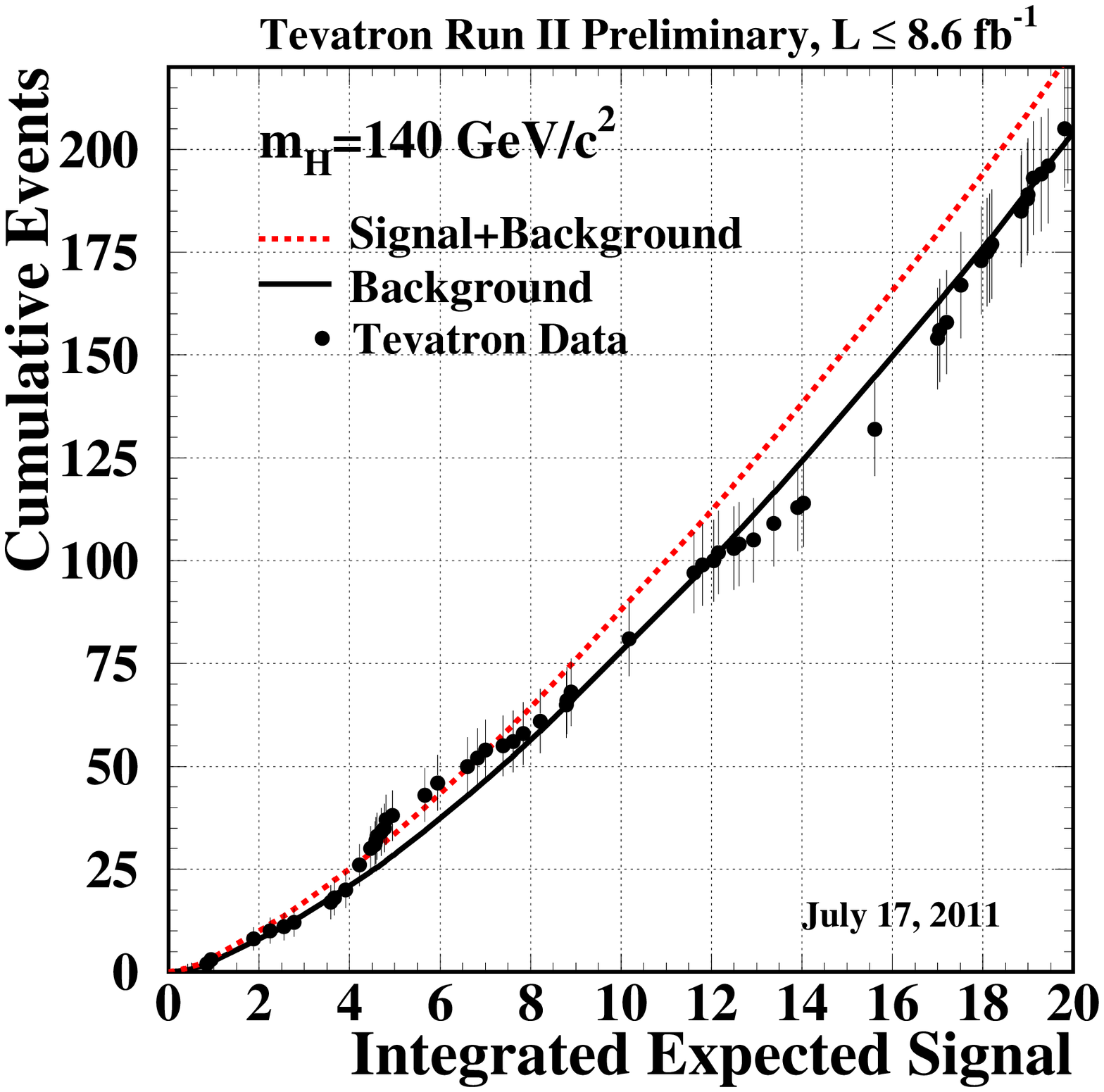}
\includegraphics[width=0.4\textwidth]{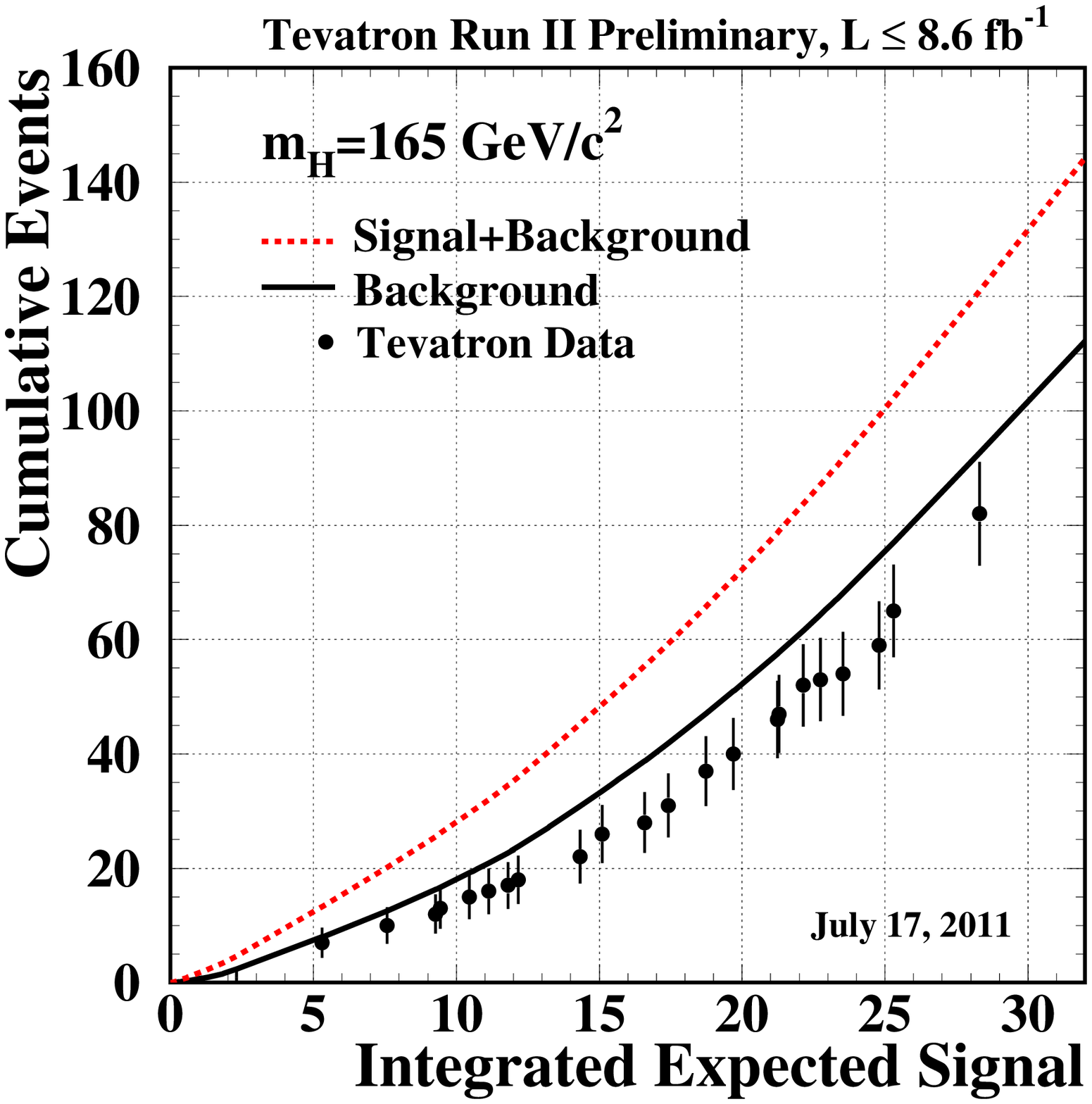}
 \caption{
 \label{fig:integ} Integrated distributions of $s/b$, starting at the high
$s/b$ side, for Higgs boson masses of 115, 140, and 165~GeV/$c^2$.  The total
signal+background and background-only integrals are shown separately, along
with the data sums.  Data are only shown for bins that have data events in
them.}
 \end{centering}
 \end{figure}

 \begin{figure}[t]
 \begin{centering}
 \includegraphics[width=0.45\textwidth]{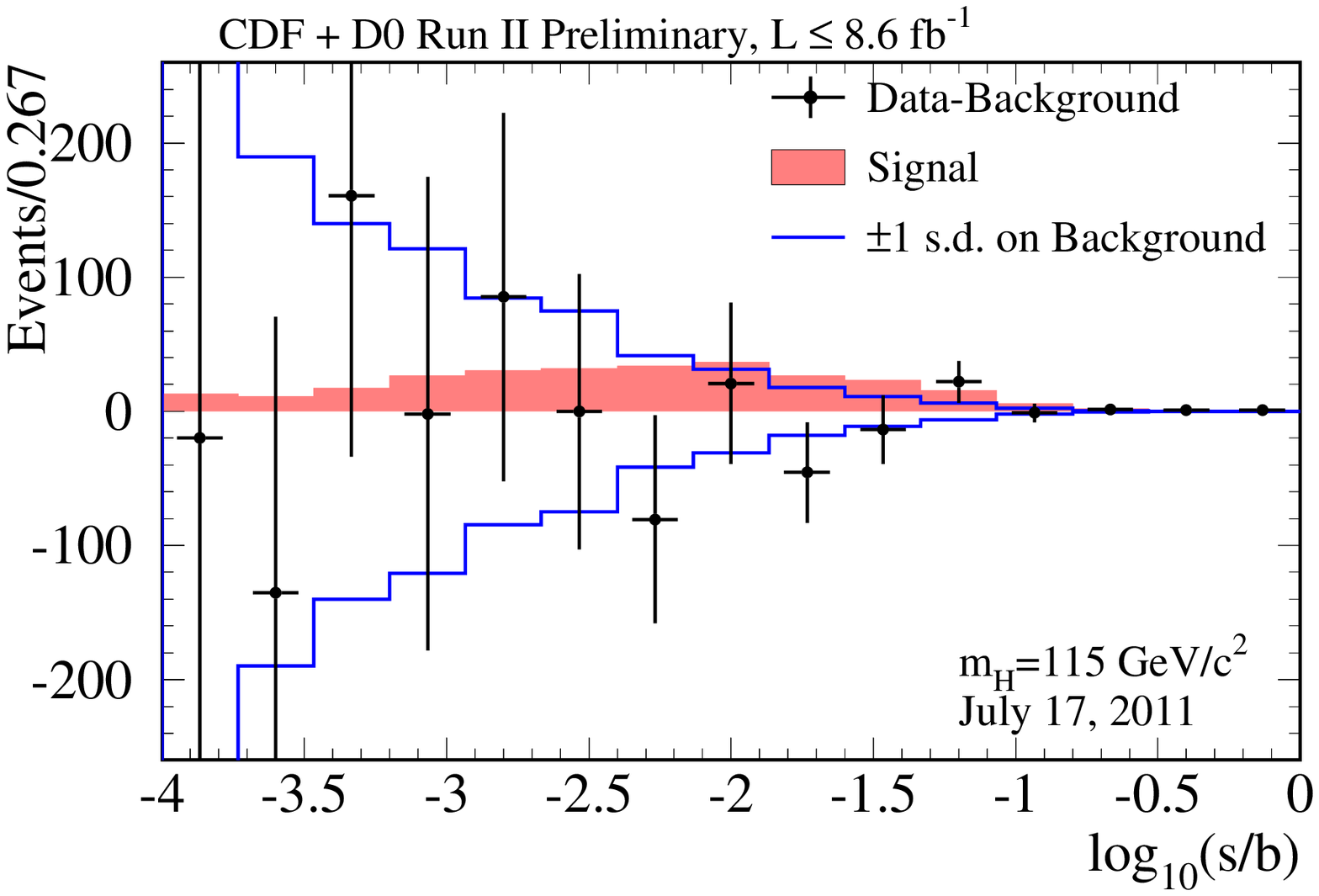}\includegraphics[width=0.45\textwidth]{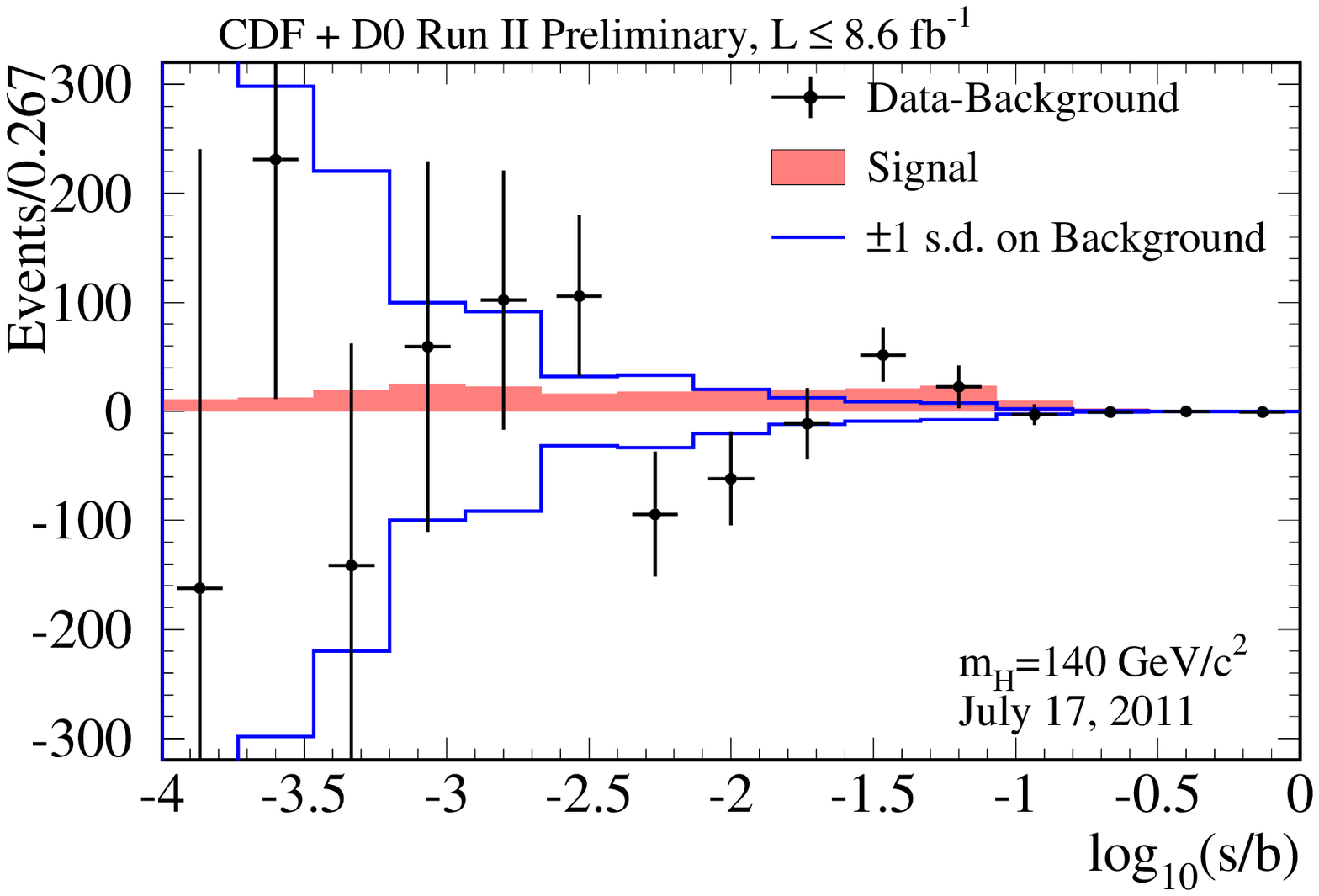}
\includegraphics[width=0.45\textwidth]{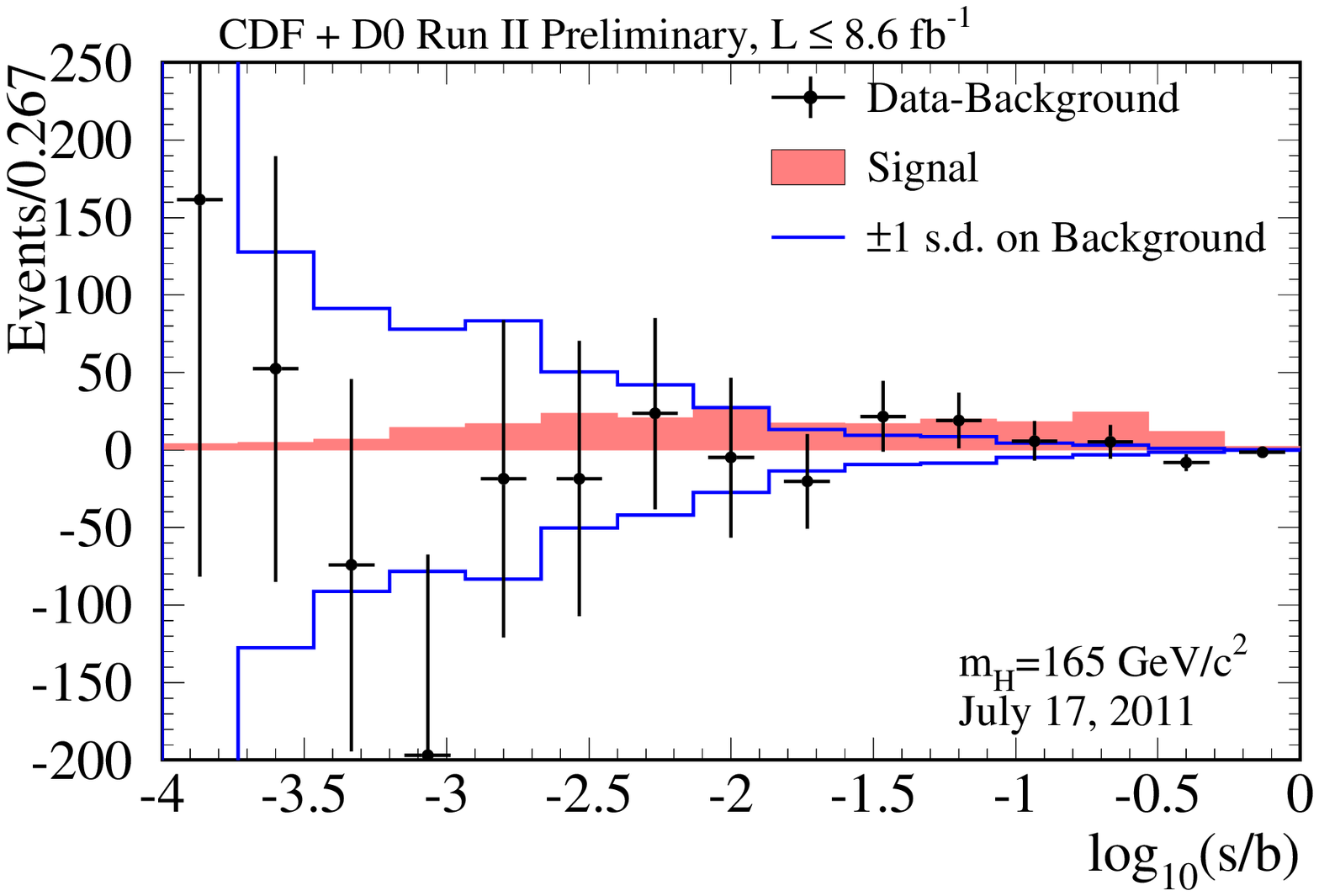}
 \caption{
 \label{fig:bgsub} Background-subtracted data distributions for all channels, summed in bins of $s/b$,
for Higgs boson masses of 115, 140, and 165~GeV/$c^2$.  The background has been fit, within its systematic
uncertainties, to the data.  The points with error bars indicate the background-subtracted data; the
sizes of the error bars are the square roots of the predicted background in each bin.  The unshaded
(blue-outline) histogram shows the systematic uncertainty on the best-fit background model, and the
shaded histogram shows the expected signal for a Standard Model Higgs boson.}
 \end{centering}
 \end{figure}

\section{Combining Channels} 

To gain confidence that the final result does not depend on the
details of the statistical formulation,
we perform two types of combinations, using
Bayesian and  Modified Frequentist approaches, which yield limits on the Higgs boson
production rate that agree within 10\% at each
value of $m_H$, and within 1\% on average.
Both methods rely on distributions in the final discriminants, and not just on
their single integrated values.  Systematic uncertainties enter on the
predicted number of signal and background events as well
as on the distribution of the discriminants in
each analysis (``shape uncertainties'').
Both methods use likelihood calculations based on Poisson
probabilities.

\subsection{Bayesian Method}

Because there is no experimental information on the production cross section for
the Higgs boson, in the Bayesian technique~\cite{CDFHiggs} we assign a flat prior
for the total number of selected Higgs boson events.  For a given Higgs boson mass, the
combined likelihood is a product of likelihoods for the individual
channels, each of which is a product over histogram bins:

\begin{equation}
{\cal{L}}(R,{\vec{s}},{\vec{b}}|{\vec{n}},{\vec{\theta}})\times\pi({\vec{\theta}})
= \prod_{i=1}^{N_C}\prod_{j=1}^{N_b} \mu_{ij}^{n_{ij}} e^{-\mu_{ij}}/n_{ij}!
\times\prod_{k=1}^{n_{np}}e^{-\theta_k^2/2}
\end{equation}

\noindent where the first product is over the number of channels
($N_C$), and the second product is over $N_b$ histogram bins containing
$n_{ij}$ events, binned in  ranges of the final discriminants used for
individual analyses, such as the dijet mass, neural-network outputs,
or matrix-element likelihoods.
 The parameters that contribute to the
expected bin contents are $\mu_{ij} =R \times s_{ij}({\vec{\theta}}) + b_{ij}({\vec{\theta}})$
for the
channel $i$ and the histogram bin $j$, where $s_{ij}$ and $b_{ij}$
represent the expected background and signal in the bin, and $R$ is a scaling factor
applied to the signal to test the sensitivity level of the experiment.
Truncated Gaussian priors are used for each of the nuisance parameters
$\theta_k$, which define the
sensitivity of the predicted signal and background estimates to systematic uncertainties. These
can take the form of uncertainties on overall rates, as well as the shapes of the distributions
used for combination.   These systematic uncertainties can be far larger
than the expected SM Higgs boson signal, and are therefore important in the calculation of limits.
The truncation is applied so that no prediction of any signal or background in any bin is negative.
The posterior density function is then integrated over all parameters (including correlations) except for $R$,
and a 95\% credibility level upper limit on $R$ is estimated
by calculating the value of $R$ that corresponds to 95\% of the area
of the resulting distribution.

\subsection{Modified Frequentist Method}

The Modified Frequentist technique relies on the ${\rm CL}_{\rm s}$ method, using
a log-likelihood ratio (LLR) as test statistic~\cite{DZHiggs}:
\begin{equation}
LLR = -2\ln\frac{p({\mathrm{data}}|H_1)}{p({\mathrm{data}}|H_0)},
\end{equation}
where $H_1$ denotes the test hypothesis, which admits the presence of
SM backgrounds and a Higgs boson signal, while $H_0$ is the null
hypothesis, for only SM backgrounds and 'data' is either an ensemble of pseudo-experiment
data constructed from the expected signal and backgrounds, or the
actual observed data.  The probabilities $p$ are
computed using the best-fit values of the nuisance parameters for each
pseudo-experiment, separately for each of the two hypotheses, and include the
Poisson probabilities of observing the data multiplied by Gaussian
priors for the values of the nuisance parameters.  This technique
extends the LEP procedure~\cite{pdgstats} which does not involve a
fit, in order to yield better sensitivity when expected signals are
small and systematic uncertainties on backgrounds are
large~\cite{pflh}.

The ${\rm CL}_{\rm s}$ technique involves computing two $p$-values, ${\rm CL}_{\rm s+b}$ and ${\rm CL}_{\rm b}$.
The latter is defined by
\begin{equation}
1-{\rm CL}_{\rm b} = p(LLR\le LLR_{\mathrm{obs}} | H_0),
\end{equation}
where $LLR_{\mathrm{obs}}$ is the value of the test statistic computed for the
data. $1-{\rm CL}_{\rm b}$ is the probability of observing a signal-plus-background-like outcome
without the presence of signal, i.e. the probability
that an upward fluctuation of the background provides  a signal-plus-background-like
response as observed in data.
The other $p$-value is defined by
\begin{equation}
{\rm CL}_{\rm s+b} = p(LLR\ge LLR_{\mathrm{obs}} | H_1),
\end{equation}
and this corresponds to the probability of a downward fluctuation of the sum
of signal and background in
the data.  A small value of ${\rm CL}_{\rm s+b}$ reflects inconsistency with  $H_1$.
It is also possible to have a downward fluctuation in data even in the absence of
any signal, and a small value of ${\rm CL}_{\rm s+b}$ is possible even if the expected signal is
so small that it cannot be tested with the experiment.  To minimize the possibility
of  excluding  a signal to which there is insufficient sensitivity
(an outcome  expected 5\% of the time at the 95\% C.L., for full coverage),
we use the quantity ${\rm CL}_{\rm s}={\rm CL}_{\rm s+b}/{\rm CL}_{\rm b}$.  If ${\rm CL}_{\rm s}<0.05$ for a particular choice
of $H_1$, that hypothesis is deemed to be excluded at the 95\% C.L. In an analogous
way, the expected ${\rm CL}_{\rm b}$, ${\rm CL}_{\rm s+b}$ and ${\rm CL}_{\rm s}$ values are computed from the median of the
LLR distribution for the background-only hypothesis.

Systematic uncertainties are included  by fluctuating the predictions for
signal and background rates in each bin of each histogram in a correlated way when
generating the pseudo-experiments used to compute ${\rm CL}_{\rm s+b}$ and ${\rm CL}_{\rm b}$.

\subsection{Systematic Uncertainties} 

Systematic uncertainties differ
between experiments and analyses, and they affect the rates and shapes of the predicted
signal and background in correlated ways.  The combined results incorporate
the sensitivity of predictions to  values of nuisance parameters,
and include correlations between rates and shapes, between signals and backgrounds,
and between channels within experiments and between experiments.
More on these issues can be found in the
individual analysis notes~\cite{cdfWH2J} through~\cite{dzHgg}.  Here we
consider only the largest contributions and correlations between and
within the two experiments.

\subsubsection{Correlated Systematics between CDF and D0}

The uncertainties on the measurements of the integrated luminosities are 6\%
(CDF) and 6.1\% (D0).
Of these values, 4\% arises from the uncertainty
on the inelastic \pp~scattering cross section, which is correlated
between CDF and D0.
CDF and D0 also share the assumed values and uncertainties on the production cross sections
for top-quark processes (\ttbar~and single top) and for electroweak processes
($WW$, $WZ$, and $ZZ$).  In order to provide a consistent combination, the values of these
cross sections assumed in each analysis are brought into agreement.  We use
$\sigma_{t\bar{t}}=7.04^{+0.24}_{-0.36}~{\rm (scale)}\pm 0.14{\rm (PDF)}\pm 0.30{\rm (mass)}$,
following the calculation of Moch and Uwer~\cite{mochuwer}, assuming
a top quark mass $m_t=173.0\pm 1.2$~GeV/$c^2$~\cite{tevtop09},
and using the MSTW2008nnlo PDF set~\cite{mstw2008}.  Other
calculations of $\sigma_{t\bar{t}}$ are similar~\cite{otherttbar}.

For single top, we use the NLL $t$-channel calculation of Kidonakis~\cite{kid1},
which has been updated using the MSTW2008nnlo PDF set~\cite{mstw2008}~\cite{kidprivcomm}.
For the $s$-channel process we use~\cite{kid2}, again based on the MSTW2008nnlo PDF set.
Both of the cross section values below are the sum of the single $t$ and single ${\bar{t}}$
cross sections, and both assume $m_t=173\pm 1.2$ GeV.
\begin{equation}
\sigma_{t-{\rm{chan}}} = 2.10\pm 0.027 {\rm{(scale)}} \pm 0.18 {\rm{(PDF)}}  \pm 0.045 {\rm{(mass)}}  {\rm {pb}}.
\end{equation}
\begin{equation}
\sigma_{s-{\rm{chan}}} = 1.05\pm 0.01 {\rm{(scale)}} \pm 0.06~{\rm{(PDF)}}  \pm 0.03~{\rm{(mass)}}~{\rm {pb}}.
\end{equation}
Other calculations of $\sigma_{\rm{SingleTop}}$ are
similar for our purposes~\cite{harris}.

MCFM~\cite{mcfm} has been used to compute the NLO cross sections for $WW$, $WZ$,
and $ZZ$ production~\cite{dibo}.  Using a scale choice $\mu_0=M_V^2+p_T^2(V)$ and
the MSTW2008 PDF set~\cite{mstw2008}, the cross section for inclusive $W^+W^-$
production is
\begin{equation}
\sigma_{W^+W^-} = 11.34^{+0.56}_{-0.49}~{\rm{(scale)}}~^{+0.35}_{-0.28} {\rm(PDF)} {\rm{pb}}
\end{equation}
and the cross section for inclusive $W^\pm Z$ production is
\begin{equation}
\sigma_{W^\pm Z} = 3.22^{+0.20}_{-0.17}~{\rm{(scale)}}~^{+0.11}_{-0.08}~{\rm(PDF)}~{\rm{pb}}
\end{equation}
For the $Z$, leptonic decays are used in the definition, with both $\gamma$
and $Z$ exchange.  The cross section quoted above involves the requirement
$75\leq m_{\ell^+\ell^-}\leq 105$~GeV for the leptons from the neutral current
exchange.  The same dilepton invariant mass requirement is applied to both
sets of leptons in determining the $ZZ$ cross section which is
\begin{equation}
\sigma_{ZZ} = 1.20^{+0.05}_{-0.04}~{\rm{(scale)}}~^{+0.04}_{-0.03}~{\rm(PDF)}~{\rm{pb}}
\end{equation}
For the diboson cross section calculations, $|\eta_{\ell}|<5$ for all calculations.
Loosening this requirement to include all leptons leads to $\sim$+0.4\% change in
the predictions.  Lowering the factorization and renormalization scales by a factor
of two increases the cross section, and raising the scales by a factor of two
decreases the cross section.  The PDF uncertainty has the same fractional impact on
the predicted cross section independent of the scale choice.  All PDF uncertainties
are computed as the quadrature sum of the twenty 68\% C.L. eigenvectors provided with
MSTW2008 (MSTW2008nlo68cl).

In many analyses, the dominant background yields are calibrated with data control
samples.  Since the methods of measuring the multijet (``QCD'') backgrounds differ
between CDF and D0, and even between analyses within the collaborations, there is
no correlation assumed between these rates.  Similarly, the large uncertainties on
the background rates for $W$+heavy flavor (HF) and $Z$+heavy flavor are considered
at this time to be uncorrelated.
The calibrations of fake leptons,
unvetoed $\gamma\rightarrow e^+e^-$ conversions, $b$-tag efficiencies and mistag
rates are performed by each collaboration using independent data samples and
methods, and are therefore also treated as uncorrelated.

\subsubsection{Correlated Systematic Uncertainties for CDF}
The dominant systematic uncertainties for the CDF analyses are shown in the
Appendix in Tables~\ref{tab:cdfsystwh2jet} and~\ref{tab:cdfsystwh3jet} for
the \WH\ channels, in Table~\ref{tab:cdfvvbb1} for the $WH,ZH\rightarrow\MET
b{\bar{b}}$ channels, in Tables~\ref{tab:cdfllbb1} and~\ref{tab:cdfllbb2}
for the $ZH\rightarrow\ell^+\ell^-b{\bar{b}}$ channels, in
Tables~\ref{tab:cdfsystww0}, \ref{tab:cdfsystww4}, and~\ref{tab:cdfsystww5}
for the $H \rightarrow W^+W^-\rightarrow \ell^{\prime \pm}\nu \ell^{\prime
\mp}\nu$ channels, in Table~\ref{tab:cdfsystwww} for the $WH \rightarrow
WWW \rightarrow\ell^{\prime \pm}\ell^{\prime \pm}$ and $WH\rightarrow
WWW \rightarrow \ell^{\pm}\ell^{\prime \pm} \ell^{\prime \prime \mp}$
channels, in Table~\ref{tab:cdfsystzww} for the $ZH \rightarrow ZWW
\rightarrow \ell^{\pm}\ell^{\mp} \ell^{\prime \pm}$ channels, In
Table~\ref{tab:cdfsystH4l} for the $H \rightarrow 4 \ell$ channel, in
Tables~\ref{tab:cdfsystttHLJ}, \ref{tab:cdfsysttthmetjets}, and
\ref{tab:cdfsysttthalljets} for the $t\bar{t}H \rightarrow W^+ b W^- \bar{b}
b\bar{b}$ channels, in Table~\ref{tab:cdfsysttautau} for the $H \rightarrow
\tau^+\tau^-$ channels, in Table~\ref{tab:cdfsystVtautau} for the $WH
\rightarrow \ell \nu \tau^+ \tau^-$ and $ZH \rightarrow \ell^+ \ell^- \tau^+
\tau^-$ channels, in Table~\ref{tab:cdfallhadsyst} for the $WH/ZH$ and VBF
$\rightarrow jjb{\bar{b}}$ channels, and in Table~\ref{tab:cdfsystgg}
for the $H \rightarrow \gamma \gamma$ channel.  Each source induces a
correlated uncertainty across all CDF channels' signal and background
contributions which are sensitive to that source.  For \hbb, the largest
uncertainties on signal arise from measured $b$-tagging efficiencies,
jet energy scale, and other Monte Carlo modeling.  Shape dependencies of
templates on jet energy scale, $b$-tagging, and gluon radiation (``ISR''
and ``FSR'') are taken into account for some analyses (see tables).
For \hww, the largest uncertainties on signal acceptance originate from
Monte Carlo modeling.  Uncertainties on background event rates vary
significantly for the different processes.  The backgrounds with the
largest systematic uncertainties are in general quite small. Such
uncertainties are constrained by fits to the nuisance parameters, and
they do not affect the result significantly.  Because the largest
background contributions are measured using data, these uncertainties
are treated as uncorrelated for the \hbb~channels.  The differences in
the resulting limits when treating the remaining uncertainties as either
correlated or uncorrelated is less than $5\%$.

\subsubsection{Correlated Systematic Uncertainties for D0 }
The dominant systematic uncertainties for the D0 analyses are shown in the Appendix,
in Tables~\ref{tab:d0systwh1}, \ref{tab:d0sysVHtau}, \ref{tab:d0vvbb}, \ref{tab:d0llbb1},
\ref{tab:d0systww}, \ref{tab:d0systwwtau}, \ref{tab:d0systwww}, \ref{tab:d0lvjj},
and \ref{tab:d0systgg}.  Each source induces a correlated
uncertainty across all D0 channels sensitive to that source. Wherever appropriate the
impact of systematic effects on both the rate and shape of the predicted signal and
background is included.  For the low mass, \hbb~analyses, significant sources of
uncertainty include the measured $b$-tagging rate and the normalization of the $W$
and $Z$ plus heavy flavor backgrounds. For the \hww and \vww\ analyses, significant
sources of uncertainty are the measured efficiencies for selecting leptons. For analyses
involving jets the determination of the jet energy scale, jet resolution and the
multijet background contribution are significant sources of uncertainty. Significant
sources for all analyses are the uncertainties on the luminosity and the cross sections
for the simulated backgrounds.  All systematic uncertainties arising from the same
source are taken to be correlated among the different backgrounds and between signal
and background.

 \begin{figure}[t]
 \begin{centering}
 \includegraphics[width=14.0cm]{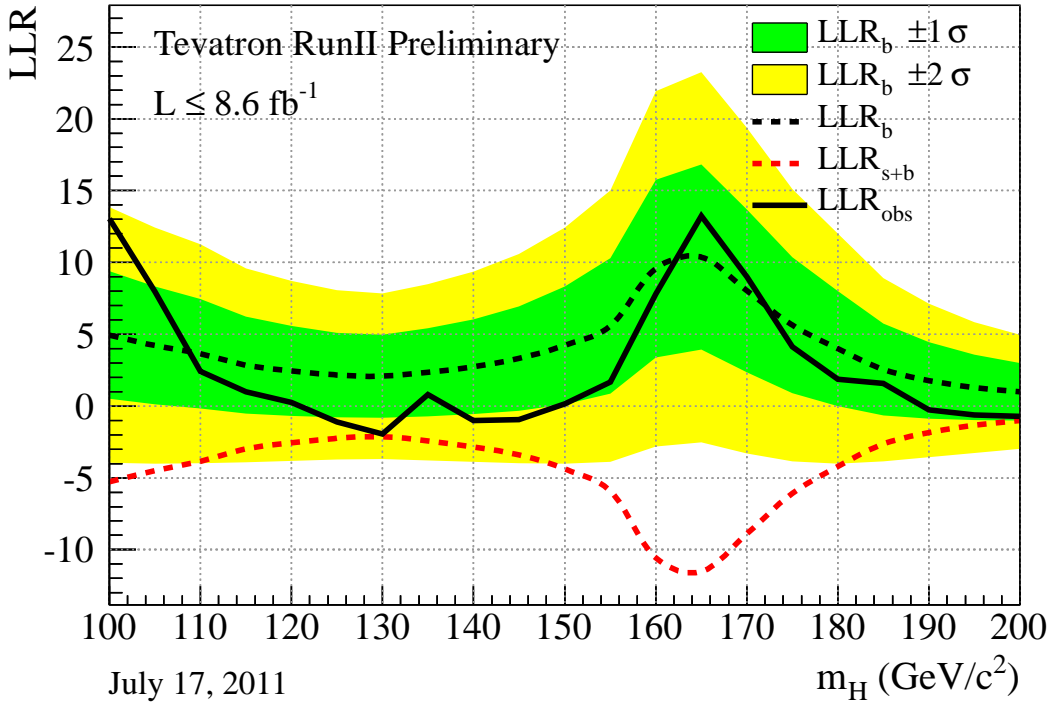}
 \caption{
 \label{fig:comboLLR} {
Distributions of the log-likelihood ratio (LLR) as a function of Higgs boson mass obtained with
the ${\rm CL}_{\rm s}$ method for the combination of all CDF and D0 analyses. The green and yellow
bands correspond to the regions enclosing 1-$\sigma$ and 2-$\sigma$ fluctuations of the background, respectively.
}}
 \end{centering}
 \end{figure}

\vspace*{1cm}
\section{Combined Results} 

Before extracting the combined limits we study the distributions of the
log-likelihood ratio (LLR) for different hypotheses to quantify the expected
sensitivity across the mass range tested.
Figure~\ref{fig:comboLLR} displays the LLR distributions for the combined
analyses as functions of $m_{H}$. Included are the median of the LLR distributions for the
background-only hypothesis (LLR$_{b}$), the signal-plus-background
hypothesis (LLR$_{s+b}$), and the observed value for the data (LLR$_{\rm{obs}}$).  The
shaded bands represent the one and two standard deviation ($\sigma$)
departures for LLR$_{b}$ centered on the median.
At $m_{H}=115$~GeV/$c^2$ a small excess in the data, at the 1- 2-sigma level, has the amplitude
expected from a Higgs boson of this mass. Table~\ref{tab:llrVals} lists the observed
and expected LLR values shown in Figure~\ref{fig:comboLLR}.

These distributions can be interpreted as follows:
The separation between the medians of the LLR$_{b}$ and LLR$_{s+b}$ distributions
provides a measure of the discriminating power of the search.  The sizes
of the one- and two-$\sigma$ LLR$_{b}$ bands indicate the width of the LLR$_{b}$
distribution, assuming no signal is truly present and only statistical fluctuations
and systematic effects are present.  The value of LLR$_{\rm{obs}}$ relative to
LLR$_{s+b}$ and LLR$_{b}$ indicates whether the data distribution appears to resemble
what we expect if a signal is present (i.e. closer to the LLR$_{s+b}$ distribution,
which is negative by construction) or whether it resembles the background expectation
more closely; the significance of any departures of LLR$_{\rm{obs}}$ from LLR$_{b}$
can be evaluated by the width of the LLR$_{b}$ bands.

Using the combination procedures outlined in Section III, we extract
limits on SM Higgs boson production $\sigma \times B(H\rightarrow X)$
in \pp~collisions at $\sqrt{s}=1.96$~TeV for $100\leq m_H \leq 200$ GeV/$c^2$.
To facilitate comparisons with the standard model and to accommodate
analyses with different degrees of sensitivity, we present our results
in terms of the ratio of obtained limits to the SM Higgs boson production
cross section, as a function of Higgs boson mass, for test masses for
which both experiments have performed dedicated searches in different
channels.  A value of the combined limit ratio which is less than or
equal to one indicates that that particular Higgs boson mass is excluded
at the 95\% C.L.

The combinations of results~\cite{CDFHiggs,DZHiggs} of each single
experiment, as used in this Tevatron combination, yield the following
ratios of 95\% C.L. observed (expected) limits to the SM cross section:
1.55~(1.49) for CDF and 1.83~(1.90) for D0 at $m_{H}=115$~GeV/$c^2$,
1.88~(1.55) for CDF and 2.42~(1.79) for D0 at $m_{H}=140$~GeV/$c^2$, and
0.75~(0.79) for CDF and 0.71~(0.87) for D0 at $m_{H}=165$~GeV/$c^2$.

The ratios of the 95\% C.L. expected and observed limit to the SM cross
section are shown in Figure~\ref{fig:comboRatio} for the combined CDF
and D0 analyses.  The observed and median expected ratios are listed
for the tested Higgs boson masses in Table~\ref{tab:ratios} for $m_{H}
\leq 150$~GeV/$c^2$, and in Table~\ref{tab:ratios-3} for $m_{H} \geq
155$~GeV/$c^2$, as obtained by the Bayesian and the ${\rm CL}_{\rm s}$
methods.  In the following summary we quote only the limits obtained
with the Bayesian method, which was decided upon {\it a priori}.
The corresponding limits and expected limits obtained using the
${\rm CL}_{\rm s}$ method are shown alongside the Bayesian limits in
the tables.  We obtain the observed (expected) values of
0.68~(0.96) at $m_{H}=105$~GeV/$c^2$, 1.17~(1.16) at $m_{H}=115$~GeV/$c^2$,
1.71~(1.16) at $m_{H}=140$~GeV/$c^2$,
1.08~(0.80) at $m_{H}=155$~GeV/$c^2$, 0.48~(0.57) at $m_{H}=165$~GeV/$c^2$,
0.91~(0.80) at $m_{H}=175$~GeV/$c^2$, and 1.31~(1.22) at $m_{H}=185$~GeV/$c^2$.

\begin{figure}[hb]
\begin{centering}
\includegraphics[width=16.5cm]{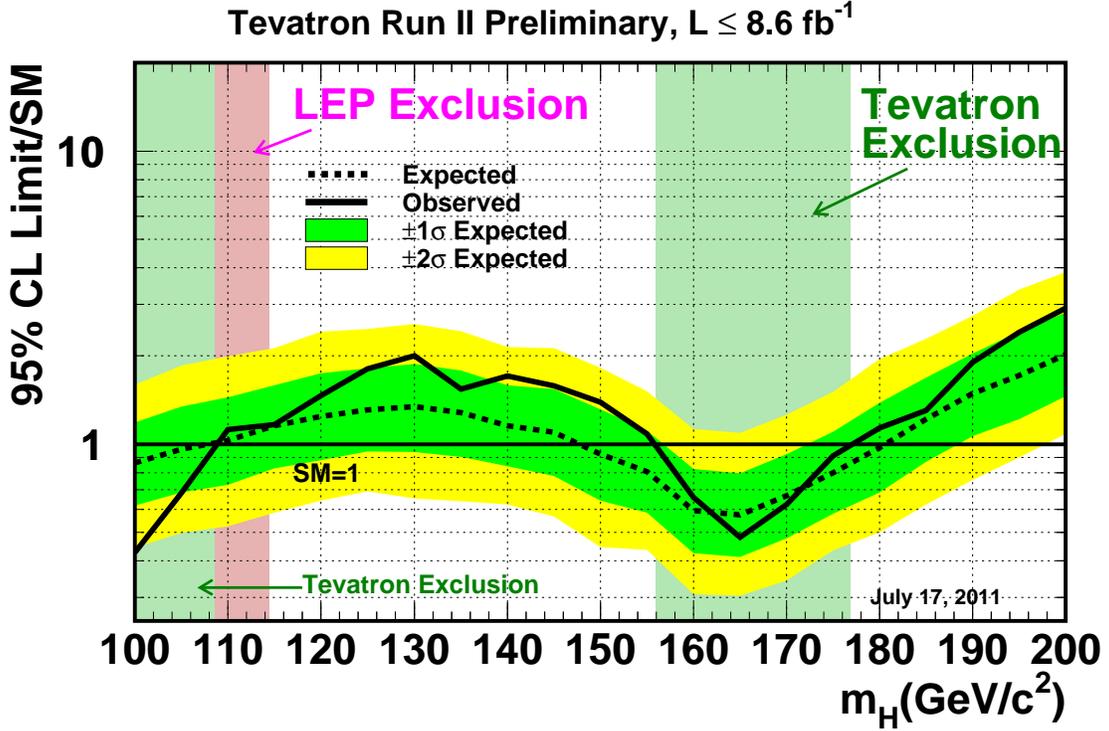}
\caption{
\label{fig:comboRatio}
Observed and expected (median, for the background-only hypothesis)
95\% C.L. upper limits on the ratios to the SM cross section, as
functions of the Higgs boson mass
for the combined CDF and D0 analyses.
The limits are expressed as a multiple of the SM prediction
for test masses (every 5 GeV/$c^2$)
for which both experiments have performed dedicated
searches in different channels.
The points are joined by straight lines
for better readability.
  The bands indicate the
68\% and 95\% probability regions where the limits can
fluctuate, in the absence of signal.
The limits displayed in this figure
are obtained with the Bayesian calculation.
}
\end{centering}
\end{figure}

\begin{table}[ht]
\caption{\label{tab:ratios} Ratios of median expected and observed 95\% C.L.
limit to the SM cross section for the combined CDF and D0 analyses as a function
of the Higgs boson mass in GeV/$c^2$, obtained with the Bayesian and with the ${\rm CL}_{\rm s}$ method.}
\begin{ruledtabular}
\begin{tabular}{lccccccccccc}\\
Bayesian       &  100 &  105 &  110 &  115 &  120 &  125 &  130 &  135 &  140 &  145 &  150 \\ \hline
Expected       & 0.86 & 0.96 & 1.03 & 1.16 & 1.24 & 1.31 & 1.35 & 1.28 & 1.16 & 1.10 & 0.93 \\
Observed       & 0.43 & 0.68 & 1.12 & 1.17 & 1.47 & 1.81 & 2.00 & 1.54 & 1.71 & 1.58 & 1.39 \\

\hline
\hline\\
${\rm CL}_{\rm s}$         &  100 &  105 &  110 &  115 &  120 &  125 &  130 &  135 &  140 &  145 &  150 \\ \hline
Expected       & 0.87 & 0.95 & 1.03 & 1.17 & 1.26 & 1.35 & 1.37 & 1.29 & 1.19 & 1.08 & 0.95 \\
Observed       & 0.44 & 0.70 & 1.13 & 1.22 & 1.57 & 1.92 & 2.02 & 1.59 & 1.77 & 1.65 & 1.32 \\
\end{tabular}
\end{ruledtabular}
\end{table}

\begin{table}[ht]
\caption{\label{tab:ratios-3}
Ratios of median expected and observed 95\% C.L.
limit to the SM cross section for the combined CDF and D0 analyses as a function
of the Higgs boson mass in GeV/$c^2$, obtained with the Bayesian and with the ${\rm CL}_{\rm s}$ method.}
\begin{ruledtabular}
\begin{tabular}{lccccccccccc}
Bayesian             &  155 &  160 &  165 &  170 &  175 &  180 &  185 &  190 &  195 &  200 \\ \hline
Expected             & 0.80 & 0.59 & 0.57 & 0.67 & 0.80 & 0.97 & 1.22 & 1.49 & 1.71 & 2.02 \\
Observed             & 1.08 & 0.66 & 0.48 & 0.62 & 0.91 & 1.14 & 1.31 & 1.90 & 2.41 & 2.91 \\
\hline
\hline\\
${\rm CL}_{\rm s}$               &  155 &  160 &  165 &  170 &  175 &  180 &  185 &  190 &  195 &  200 \\ \hline
Expected             & 0.82 & 0.61 & 0.58 & 0.67 & 0.81 & 0.98 & 1.24 & 1.50 & 1.77 & 2.04 \\
Observed             & 1.03 & 0.67 & 0.48 & 0.61 & 0.92 & 1.17 & 1.34 & 1.92 & 2.39 & 2.82 \\
\end{tabular}
\end{ruledtabular}
\end{table}

We also show in Figure~\ref{fig:comboCLS} and list in Table~\ref{tab:clsVals} the observed 1-${\rm CL}_{\rm s}$
and its expected distribution for the background-only hypothesis as a function of the Higgs boson mass.
This is directly interpreted as the level of exclusion of our search.  This
figure is obtained using the ${\rm CL}_{\rm s}$ method.  We also show in Figure~\ref{fig:comboCLS}
${\rm CL}_{\rm s}$ as a function of $m_H$, which on a logarithmic scale more clearly shows the strength of the
exclusion for values of $m_H$ for which we have a large sensitivity.  Figure~\ref{fig:comboCLSB} shows the
$p$-value ${\rm CL}_{\rm s+b}$ as a function of $m_H$ and also $1-{\rm CL}_{\rm s+b}$, as well as the expected
distributions in the absence of a Higgs boson signal.

We choose to use the intersections of piecewise linear interpolations of our observed and expected
rate limits in order to quote ranges of Higgs boson masses that are excluded and that are expected
to be excluded.  The sensitivities of our searches to Higgs bosons are smooth functions of the Higgs
boson mass and depend most strongly on the predicted cross sections and the decay branching ratios
(the decay $H\rightarrow W^+W^-$ is the dominant decay for the region of highest sensitivity).
We therefore use the linear interpolations to extend the results from the 5~GeV/$c^2$
mass grid investigated to points in between.  The regions
of Higgs boson masses excluded at the 95\% C.L. thus obtained are $156<m_{H}<177$~GeV/$c^{2}$
and $100<m_H<108$~GeV/$c^{2}$.  The expected exclusion region, given the current sensitivity,
is $148<m_{H}<180$~GeV/$c^{2}$ and $100<m_H<109$~GeV/$c^{2}$ (masses below $m_H<100$~GeV/$c^{2}$ were not studied).
The excluded region obtained by
finding the intersections of the linear interpolations of the observed $1-{\rm CL}_{\rm s}$ curve
shown in Figure~\ref{fig:comboCLS} is nearly identical to that obtained with the Bayesian
calculation.  As previously stated, we make the {\it a priori} choice to quote the exclusion
region using the Bayesian calculation.

We investigate the sensitivity and observed limits using CDF's and D0's searches for $H\rightarrow b{\bar{b}}$
taken in combination.
These channels contribute the most for values of $m_H$ below around 130 GeV/$c^2$. The contributing channels
for CDF are the
$WH\rightarrow \ell\nu b\bar{b}$ channels, the
$ZH\rightarrow \nu\bar{\nu} b\bar{b}$ channels, the
$ZH\rightarrow \ell^+\ell^- b\bar{b}$ channels, the
$WH+ZH+VBF\rightarrow jjb{\bar{b}}$ channels, and all of the
$t\bar{t}H$ channels.  The contributing channels for D0 are the
$WH\rightarrow \ell\nu b\bar{b}$ channels, the
$ZH\rightarrow \nu\bar{\nu} b\bar{b}$ channels, and the
$ZH\rightarrow \ell^+\ell^- b\bar{b}$ channels.  The result of this combination is shown in
Figure~\ref{fig:comboRatiobb}.

In summary, we combine all available CDF and D0 results on SM Higgs boson searches,
based on luminosities ranging from 4.0 to 8.6 fb$^{-1}$. Compared to our previous combination,
more data have been added to the existing
channels, additional channels have been included, and analyses have been further optimized to gain sensitivity.
The results presented here significantly extend the individual limits of each
collaboration and those obtained in our previous combination.  The sensitivity of our
combined search is sufficient to exclude a Higgs boson at high mass and is, in the absence of signal,
expected to
grow substantially in the future as more data are added and further improvements are
made to our analysis techniques.  We observe a small ($\approx 1\sigma$) excess in the range
$125<m_{H}<155$~GeV/$c^{2}$  which does not allow exclusion of a Higgs boson to as low a
mass as expected.  In addition, we combine the CDF and D0 analyses which seek specifically
the $H \rightarrow b\bar{b}$ decay, which dominates at the low end of the allowed mass range for the SM Higgs boson.
These are the search modes for which we expect Tevatron sensitivity to remain
competitive with the LHC experiments for several years to come.

 \begin{figure}[t]
 \begin{centering}
 \includegraphics[width=0.73\textwidth]{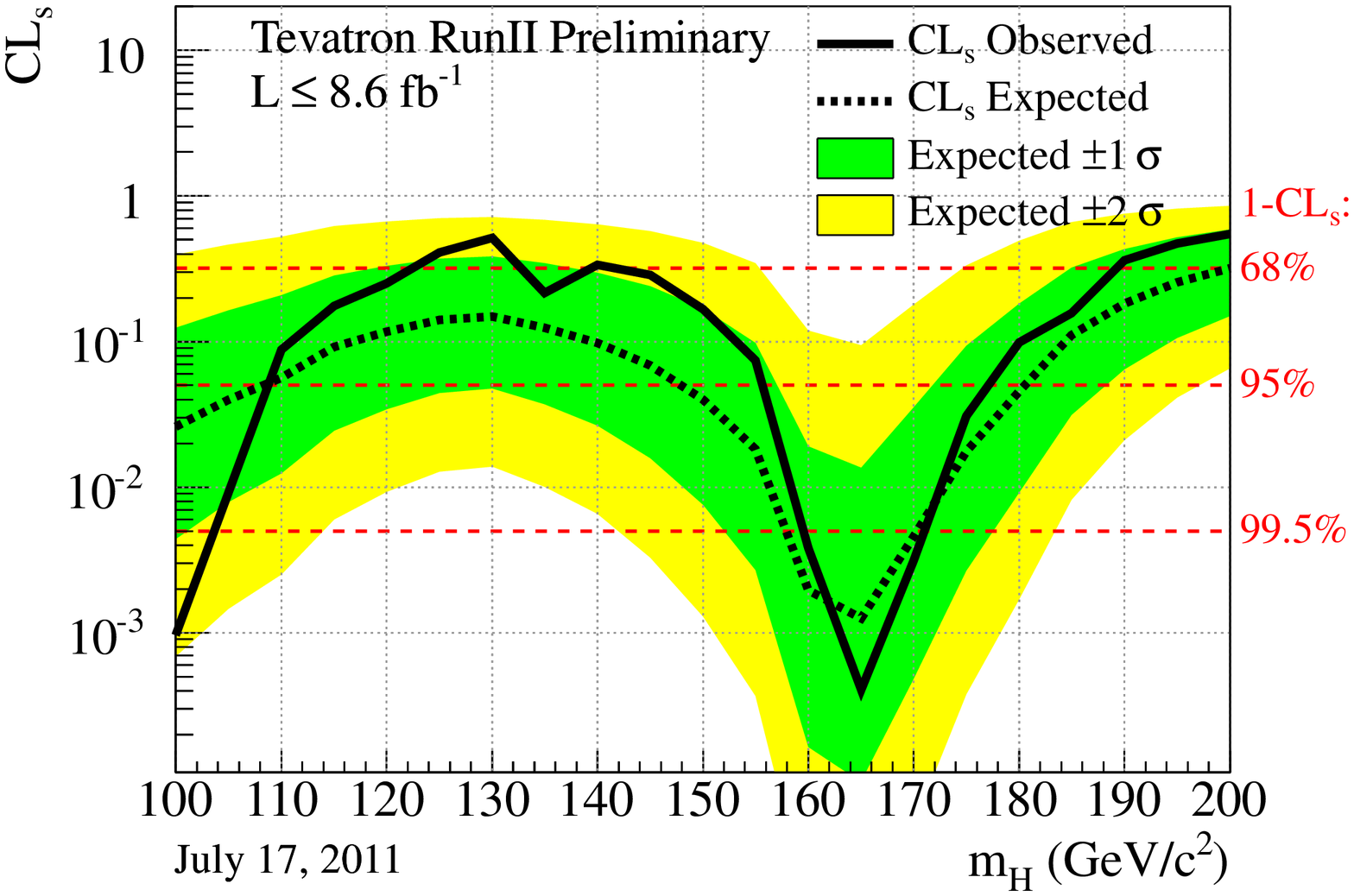} \\
(a) \\
\vspace{0.5cm}
 \includegraphics[width=0.7\textwidth]{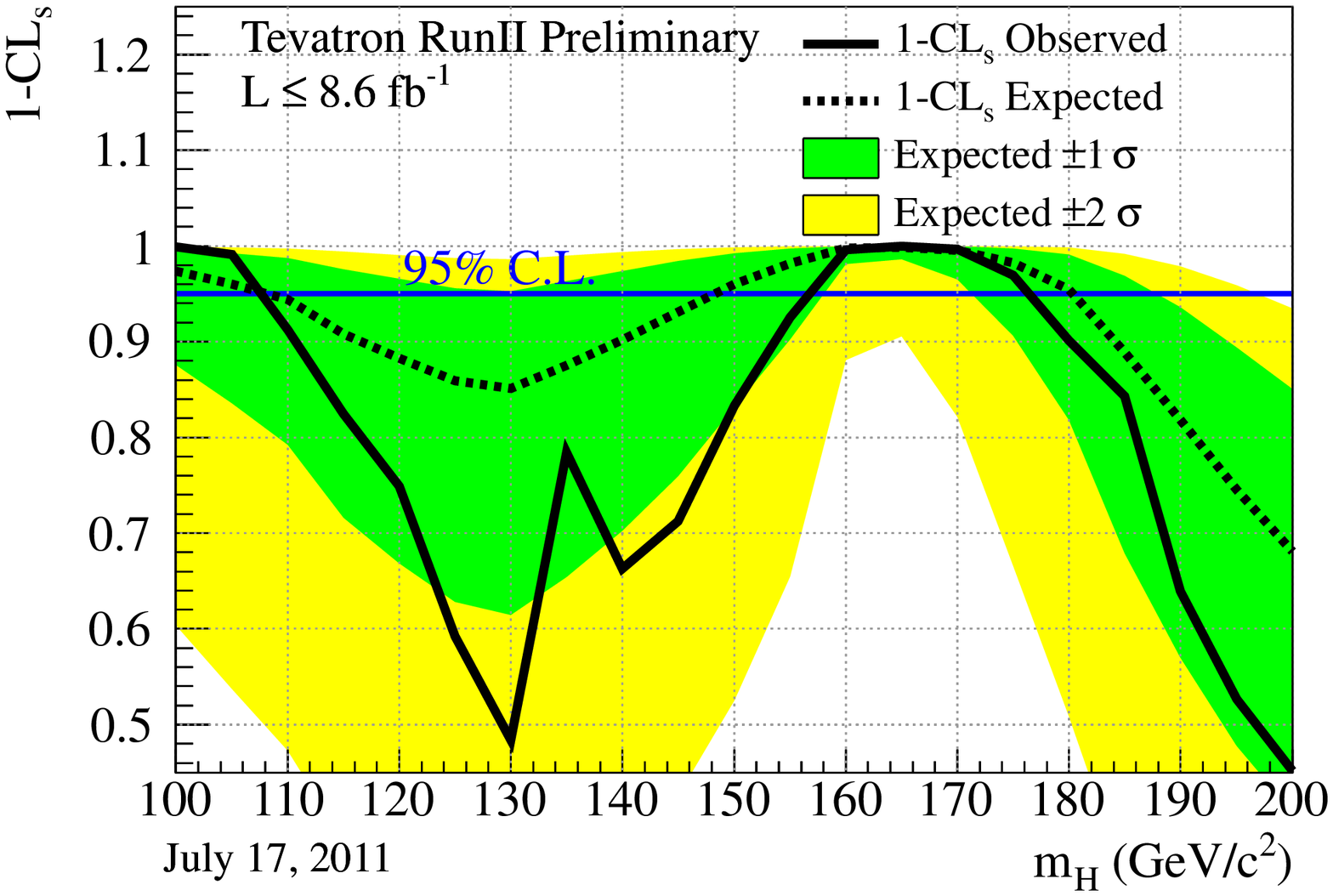} \\
(b)
 \caption{
 \label{fig:comboCLS}
 The exclusion strength ${\rm CL}_{\rm s}$ (a) and 1-${\rm CL}_{\rm s}$ (b) as functions of the Higgs boson mass
(in steps of 5 GeV/$c^2$), for the combination of the
 CDF and D0 analyses. The green and yellow
bands correspond to the regions enclosing 1-$\sigma$ and 2-$\sigma$ fluctuations of the background, respectively.}
 \end{centering}
 \end{figure}

 \begin{figure}[t]
 \begin{centering}
 \includegraphics[width=0.7\textwidth]{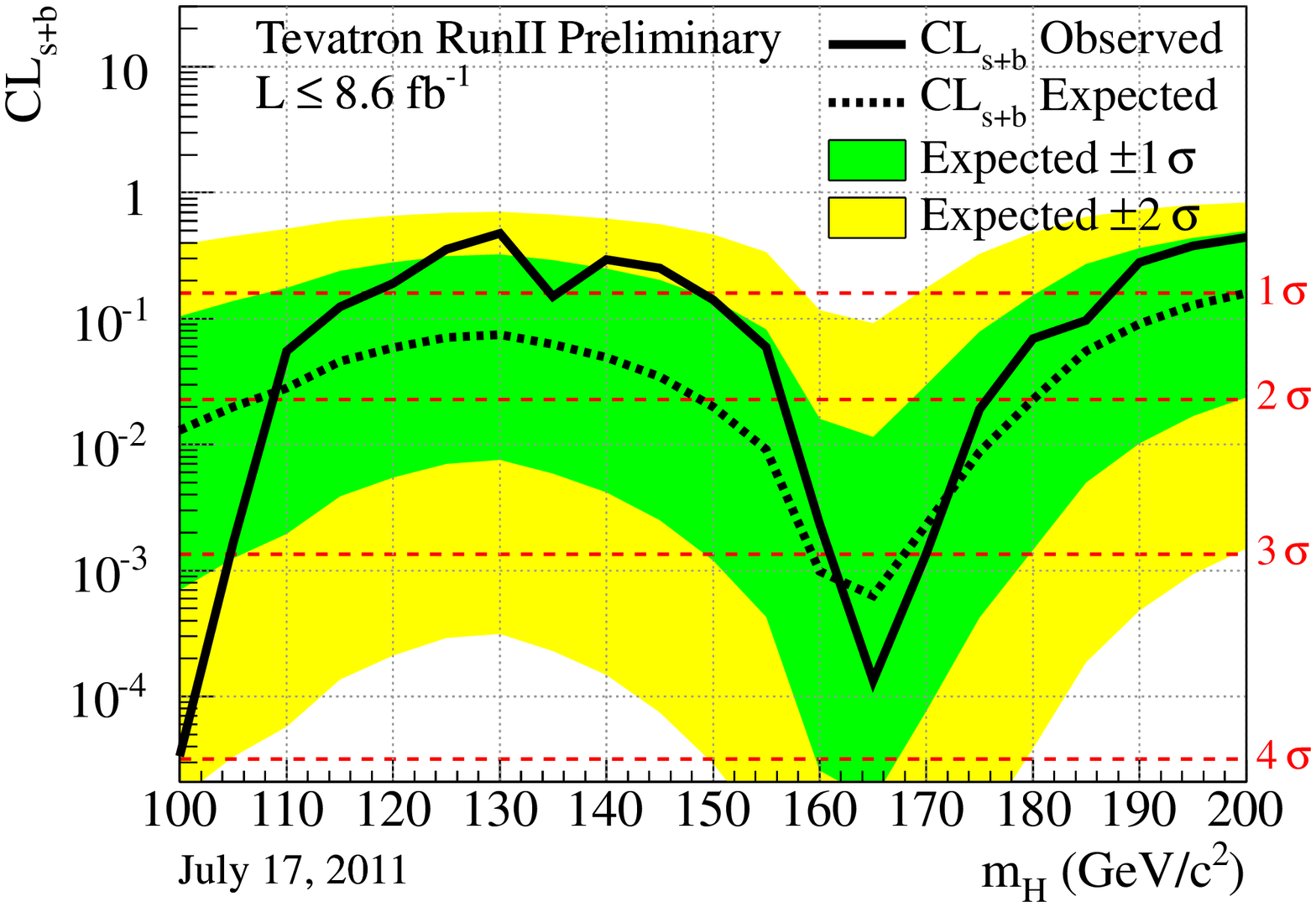} \\
(a) \\
\vspace{0.5cm}
 \includegraphics[width=0.7\textwidth]{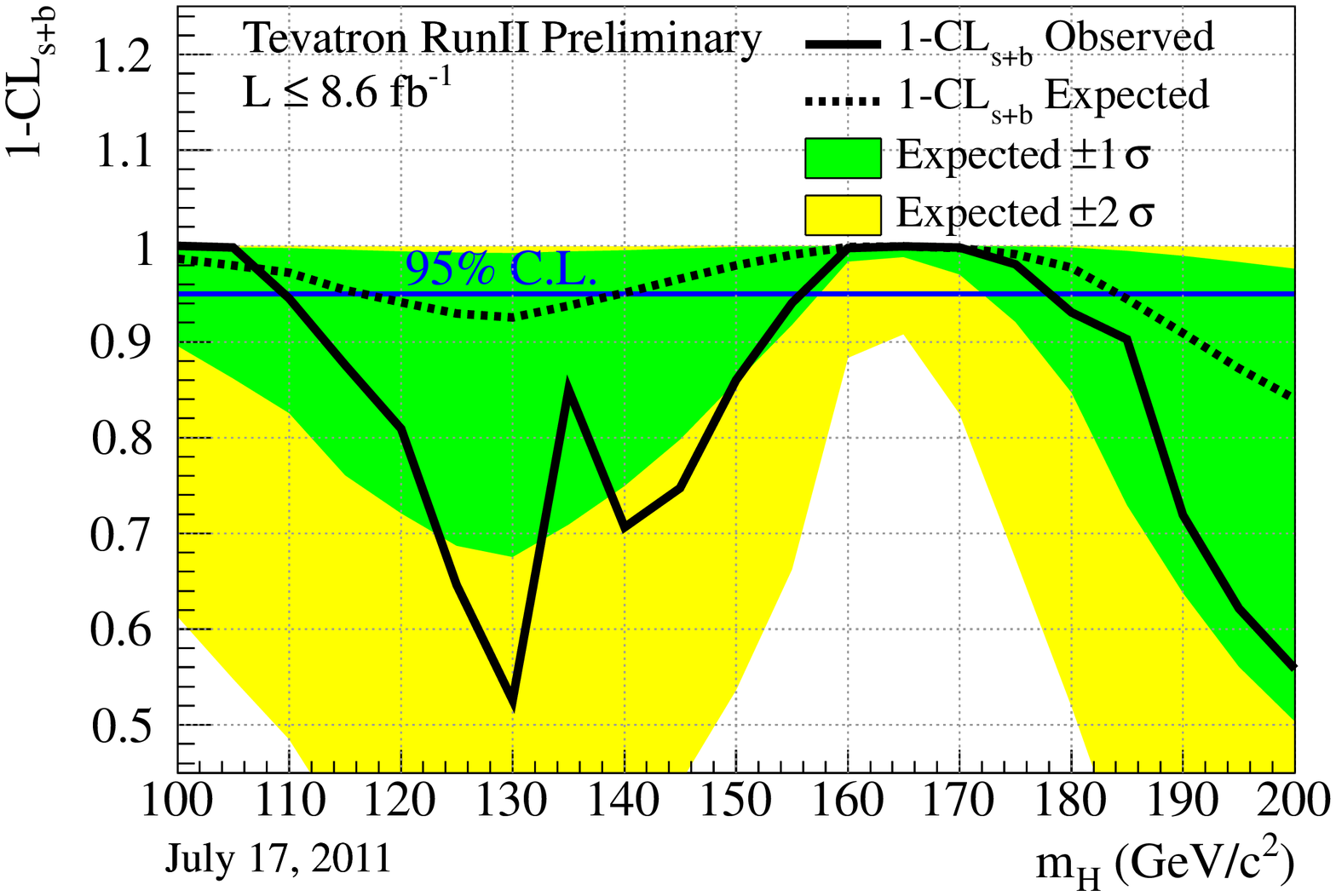} \\
(b)
 \caption{
 \label{fig:comboCLSB}
 The signal $p$-values ${\rm CL}_{\rm s+b}$ (a) and 1-${\rm CL}_{\rm s+b}$ (b) as functions of the Higgs boson mass
(in steps of 5 GeV/$c^2$),  for the combination of the
 CDF and D0 analyses. The green and yellow
bands correspond to the regions enclosing 1-$\sigma$ and 2-$\sigma$ fluctuations of the background, respectively.}
 \end{centering}
 \end{figure}

\begin{figure}[hb]
\begin{centering}
\includegraphics[width=16.5cm]{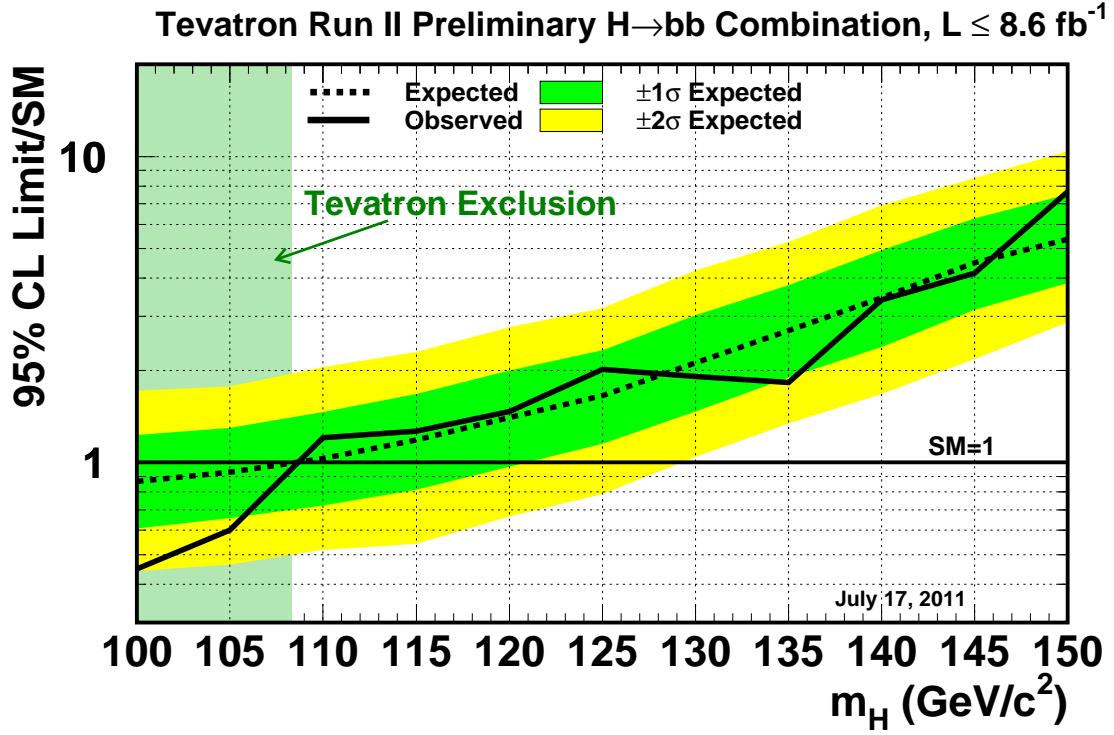}
\caption{
\label{fig:comboRatiobb}
Observed and expected (median, for the background-only hypothesis)
95\% C.L. upper limits on the ratios to the SM cross section,
as functions of the Higgs boson mass
for the combination of CDF and D0 analyses focusing on the $H
\rightarrow b\bar{b}$ decay channel.  The limits are expressed
as a multiple of the SM prediction for test masses (every 5
GeV/$c^2$) for which both experiments have performed dedicated
searches in different channels.
The points are joined by straight lines for better readability.
The bands indicate the 68\% and 95\% probability regions where
the limits can fluctuate, in the absence of signal.  The limits
displayed in this figure are obtained with the Bayesian calculation.
}
\end{centering}
\end{figure}

\newpage

\begin{table}[htpb]
\caption{\label{tab:llrVals} Log-likelihood ratio (LLR) values for the combined CDF + \Dzero Higgs boson search obtained using the {\rm CL}$_{S}$ method.}
\begin{ruledtabular}
\begin{tabular}{lccccccc}
$m_{H}$ (GeV/$c^2$ &  LLR$_{\rm{obs}}$ & LLR$_{S+B}^{\rm{med}}$ &
LLR$_{B}^{-2\sigma}$ & LLR$_{B}^{-1\sigma}$ & LLR$_{B}^{\rm{med}}$ &  LLR$_{B}^{+1\sigma}$ & LLR$_{B}^{+2\sigma}$ \\
\hline
100 & 13.03 & -5.26 & 13.85 & 9.40 & 4.95 & 0.50 & -3.95 \\
105 & 7.96 & -4.51 & 12.43 & 8.32 & 4.22 & 0.11 & -4.00 \\
110 & 2.43 & -3.84 & 11.28 & 7.46 & 3.64 & -0.17 & -3.99 \\
115 & 1.01 & -2.98 & 9.59 & 6.21 & 2.84 & -0.53 & -3.90 \\
120 & 0.24 & -2.55 & 8.71 & 5.58 & 2.45 & -0.68 & -3.81 \\
125 & -1.11 & -2.24 & 8.05 & 5.11 & 2.17 & -0.78 & -3.72 \\
130 & -1.94 & -2.13 & 7.85 & 4.97 & 2.08 & -0.80 & -3.69 \\
135 & 0.80 & -2.43 & 8.50 & 5.43 & 2.36 & -0.71 & -3.78 \\
140 & -1.01 & -2.84 & 9.35 & 6.05 & 2.74 & -0.57 & -3.88 \\
145 & -0.95 & -3.41 & 10.59 & 6.95 & 3.31 & -0.33 & -3.97 \\
150 & 0.14 & -4.38 & 12.44 & 8.33 & 4.22 & 0.11 & -4.00 \\
155 & 1.68 & -5.91 & 15.02 & 10.30 & 5.57 & 0.85 & -3.87 \\
160 & 7.82 & -10.57 & 21.93 & 15.74 & 9.56 & 3.38 & -2.81 \\
165 & 13.23 & -11.52 & 23.25 & 16.81 & 10.37 & 3.93 & -2.51 \\
170 & 9.00 & -8.88 & 19.36 & 13.69 & 8.03 & 2.36 & -3.31 \\
175 & 4.14 & -6.08 & 15.12 & 10.37 & 5.63 & 0.88 & -3.86 \\
180 & 1.86 & -4.20 & 12.00 & 8.00 & 4.00 & -0.00 & -4.00 \\
185 & 1.58 & -2.63 & 8.92 & 5.73 & 2.54 & -0.65 & -3.84 \\
190 & -0.27 & -1.85 & 7.13 & 4.46 & 1.78 & -0.89 & -3.56 \\
195 & -0.62 & -1.33 & 5.85 & 3.57 & 1.30 & -0.98 & -3.26 \\
200 & -0.72 & -1.02 & 4.97 & 2.98 & 0.99 & -1.00 & -2.99 \\
\hline
\end{tabular}
\end{ruledtabular}
\end{table}


\begin{table}[htpb]
\caption{\label{tab:clsVals} The observed and expected 1-{\rm CL}$_{\rm s}$ values as functions of $m_H$, for the combined
CDF and \Dzero Higgs boson searches.}
\begin{ruledtabular}
\begin{tabular}{lcccccc}
$m_H$ (GeV/$c^2$) & 1-{\rm CL}$_{\rm s}^{\rm{obs}}$ &
1-{\rm CL}$_{\rm s}^{-2\sigma}$ &
1-{\rm CL}$_{\rm s}^{-1\sigma}$ &
1-{\rm CL}$_{\rm s}^{\rm{median}}$ &
1-{\rm CL}$_{\rm s}^{+1\sigma}$ &
1-{\rm CL}$_{\rm s}^{+2\sigma}$ \\ \hline
100 & 0.999 & 0.999 & 0.996 & 0.974 & 0.876 & 0.604 \\
105 & 0.991 & 0.999 & 0.992 & 0.960 & 0.835 & 0.537 \\
110 & 0.912 & 0.997 & 0.988 & 0.944 & 0.792 & 0.473 \\
115 & 0.824 & 0.994 & 0.976 & 0.908 & 0.716 & 0.380 \\
120 & 0.749 & 0.991 & 0.966 & 0.883 & 0.669 & 0.332 \\
125 & 0.592 & 0.987 & 0.956 & 0.859 & 0.628 & 0.294 \\
130 & 0.484 & 0.986 & 0.953 & 0.851 & 0.614 & 0.281 \\
135 & 0.784 & 0.990 & 0.963 & 0.875 & 0.654 & 0.317 \\
140 & 0.663 & 0.993 & 0.974 & 0.902 & 0.702 & 0.365 \\
145 & 0.712 & 0.997 & 0.984 & 0.931 & 0.760 & 0.426 \\
150 & 0.833 & 0.999 & 0.992 & 0.960 & 0.832 & 0.526 \\
155 & 0.926 & 1.000 & 0.997 & 0.982 & 0.902 & 0.655 \\
160 & 0.996 & 1.000 & 1.000 & 0.998 & 0.981 & 0.881 \\
165 & 1.000 & 1.000 & 1.000 & 0.999 & 0.986 & 0.906 \\
170 & 0.997 & 1.000 & 1.000 & 0.995 & 0.965 & 0.821\\
175 & 0.969 & 1.000 & 0.997 & 0.982 & 0.906 & 0.666 \\
180 & 0.901 & 0.998 & 0.991 & 0.955 & 0.818 & 0.508 \\
185 & 0.843 & 0.992 & 0.969 & 0.889 & 0.679 & 0.341 \\
190 & 0.639 & 0.979 & 0.936 & 0.818 & 0.570 & 0.247 \\
195 & 0.527 & 0.959 & 0.894 & 0.745 & 0.478 & 0.184 \\
200 & 0.451 & 0.934 & 0.851 & 0.681 & 0.410 & 0.145 \\
\end{tabular}
\end{ruledtabular}
\end{table}

\begin{center}
{\bf Acknowledgments}  
\end{center}

We thank the Fermilab staff and the technical staffs of the
participating institutions for their vital contributions, and we
acknowledge support from the
DOE and NSF (USA);
CONICET and UBACyT (Argentina);
ARC (Australia);
CNPq, FAPERJ, FAPESP and FUNDUNESP (Brazil);
CRC Program and NSERC (Canada);
CAS, CNSF, and NSC (China);
Colciencias (Colombia);
MSMT and GACR (Czech Republic);
Academy of Finland (Finland);
CEA and CNRS/IN2P3 (France);
BMBF and DFG (Germany);
INFN (Italy);
DAE and DST (India);
SFI (Ireland);
Ministry of Education, Culture, Sports, Science and Technology (Japan);
KRF, KOSEF and World Class University Program (Korea);
CONACyT (Mexico);
FOM (The Netherlands);
FASI, Rosatom and RFBR (Russia);
Slovak R\&D Agency (Slovakia);
Ministerio de Ciencia e Innovaci\'{o}n, and Programa Consolider-Ingenio 2010 (Spain);
The Swedish Research Council (Sweden);
Swiss National Science Foundation (Switzerland);
STFC and the Royal Society (United Kingdom);
and
the A.P. Sloan Foundation (USA).


\clearpage
\newpage


\appendix
\appendixpage
\addappheadtotoc
\section{Systematic Uncertainties}


\begin{table}[h]
\begin{center}
\caption{\label{tab:cdfsystwh2jet} Systematic uncertainties on the signal and
background contributions for CDF's $WH\rightarrow\ell\nu b{\bar{b}}$ tight
double tag (TDT), loose double tag (LDT), looser double tag (LDTX), and single
tag (ST) 2 jet channels.  Systematic uncertainties are listed by name; see the
original references for a detailed explanation of their meaning and on how
they are derived.  Systematic uncertainties for $WH$ shown in this table are
obtained for $m_H=115$ GeV/$c^2$.  Uncertainties are relative, in percent, and
are symmetric unless otherwise indicated. Shape uncertainties are labeled with an "S".}
\vskip 0.1cm
{\centerline{CDF: tight and loose double-tag (TDT and LDT) $WH\rightarrow\ell\nu b{\bar{b}}$ channel relative uncertainties (\%)}}
\vskip 0.099cm
\begin{ruledtabular}
\begin{tabular}{lcccccc}\\
Contribution              & $W$+HF & Mistags & Top & Diboson & Non-$W$ & $WH$  \\ \hline
Luminosity ($\sigma_{\mathrm{inel}}(p{\bar{p}})$)
                          & 3.8      & 0       & 3.8 & 3.8     & 0       &    3.8   \\
Luminosity Monitor        & 4.4      & 0       & 4.4 & 4.4     & 0       &    4.4   \\
Lepton ID                 & 2.0-4.5      & 0       & 2.0-4.5   & 2.0-4.5       & 0       &    2.0-4.5   \\
Jet Energy Scale          & S      & 0       & S   & S       & 0       &    2(S)   \\
Mistag Rate               & 0      & 35     & 0   & 0       & 0       &    0   \\
$B$-Tag Efficiency          & 8.6      & 0       & 8.6 & 8.6     & 0       &    8.6   \\
$t{\bar{t}}$ Cross Section  & 0    & 0       & 10  & 0       & 0       &    0   \\
Diboson Rate              & 0      & 0       & 0   & 11.5    & 0       &    0   \\
Signal Cross Section      & 0      & 0       & 0   & 0       & 0       &    5 \\
HF Fraction in W+jets     &    45  & 0       & 0   & 0       & 0       &    0   \\
ISR+FSR+PDF               & 5.0-7.7      & 0       & 5.0-7.7   & 5.0-7.7       & 0       &    5.0-7.7 \\
$Q^2$                     & S          & 0        & 0    &   0      &   0      &  0   \\
QCD Rate                  & 0      & 0       & 0   & 0       & 40      &    0   \\
\end{tabular}
\end{ruledtabular}

\vskip 0.3cm
{\centerline{CDF: looser double-tag (LDTX) $WH\rightarrow\ell\nu b{\bar{b}}$ channel relative uncertainties (\%)}}
\vskip 0.099cm
\begin{ruledtabular}
\begin{tabular}{lcccccc}\\
Contribution              & $W$+HF & Mistags & Top & Diboson & Non-$W$ & $WH$  \\ \hline
Luminosity ($\sigma_{\mathrm{inel}}(p{\bar{p}})$)
                          & 3.8      & 0       & 3.8 & 3.8     & 0       &    3.8   \\
Luminosity Monitor        & 4.4      & 0       & 4.4 & 4.4     & 0       &    4.4   \\
Lepton ID                 & 2.0-4.5      & 0       & 2.0-4.5   & 2.0-4.5       & 0       &    2.0-4.5   \\
Jet Energy Scale          & S      &        & S   & S       &        &    2.2(S)   \\
Mistag Rate               & 0      & 36     & 0   & 0       & 0       &    0   \\
$B$-Tag Efficiency          & 13.6      & 0       & 13.6 & 13.6     & 0       &    13.6   \\
$t{\bar{t}}$ Cross Section  & 0    & 0       & 10  & 0       & 0       &    0   \\
Diboson Rate              & 0      & 0       & 0   & 11.5    & 0       &    0   \\
Signal Cross Section      & 0      & 0       & 0   & 0       & 0       &    5 \\
HF Fraction in W+jets     &    45  & 0       & 0   & 0       & 0       &    0   \\
ISR+FSR+PDF               & 4.9-19.5      & 0       & 4.9-19.5   & 4.9-19.5       & 0       &    4.9-19.5 \\
$Q^2$                     & S             &  0       &     0       &   0             & 0        & 0 \\
QCD Rate                  & 0      & 0       & 0   & 0       & 40      &    0   \\
\end{tabular}
\end{ruledtabular}

\vskip 0.3cm
{\centerline{CDF: single tag (ST) $WH\rightarrow\ell\nu b{\bar{b}}$ channel relative uncertainties (\%)}}
\vskip 0.099cm
\begin{ruledtabular}
\begin{tabular}{lcccccc}\\
Contribution              & $W$+HF & Mistags & Top & Diboson & Non-$W$ & $WH$  \\ \hline
Luminosity ($\sigma_{\mathrm{inel}}(p{\bar{p}})$)
                          & 3.8      & 0       & 3.8 & 3.8     & 0       &    3.8   \\
Luminosity Monitor        & 4.4      & 0       & 4.4 & 4.4     & 0       &    4.4   \\
Lepton ID                 & 2.0-4.5      & 0       & 2.0-4.5   & 2.0-4.5       & 0       &    2.0-4.5   \\
Jet Energy Scale          & S      & 0       & S   & S       & 0       &    2.3-4.7(S)   \\
Mistag Rate               & 0      & 35    & 0   & 0       & 0       &    0   \\
$B$-Tag Efficiency          & 4.3      & 0       & 4.3 & 4.3     & 0       &    4.3   \\
$t{\bar{t}}$ Cross Section  & 0    & 0       & 10  & 0       & 0       &    0   \\
Diboson Rate                & 0      & 0       & 0   & 11.5    & 0       &    0   \\
Signal Cross Section        & 0      & 0       & 0   & 0       & 0       &    5 \\
HF Fraction in W+jets       &    42  & 0       & 0   & 0       & 0       &    0   \\
ISR+FSR+PDF                 & 3.0-8.4      & 0       & 3.0-8.4   & 3.0-8.4       & 0       &    3.0-8.4 \\
$Q^2$                       &  S           &    0     &    0       &   0            & 0        &    0        \\
QCD Rate                    & 0      & 0       & 0   & 0       & 40      &    0   \\
\end{tabular}
\end{ruledtabular}

\end{center}
\end{table}


\begin{table}[t]
\begin{center}
\caption{\label{tab:cdfsystwh3jet} Systematic uncertainties on the signal and background
contributions for CDF's $WH\rightarrow\ell\nu b{\bar{b}}$ tight double tag (TDT), loose
double tag (LDT), and single tag (ST) 3 jet channels.  Systematic uncertainties are listed
by name; see the original references for a detailed explanation of their meaning and on how
they are derived.  Systematic uncertainties for $WH$ shown in this table are obtained for
$m_H=115$ GeV/$c^2$.  Uncertainties are relative, in percent, and are symmetric unless
otherwise indicated. Shape uncertainties are labeled with an "S". }

\vskip 0.1cm
{\centerline{CDF: tight and loose double-tag (TDT and LDT) $WH\rightarrow\ell\nu b{\bar{b}}$ channel relative uncertainties (\%)}}
\vskip 0.099cm
\begin{ruledtabular}
\begin{tabular}{lcccccc}\\
Contribution              & $W$+HF & Mistags & Top & Diboson & Non-$W$ & $WH$  \\ \hline
Luminosity ($\sigma_{\mathrm{inel}}(p{\bar{p}})$)
                          & 3.8      & 0       & 3.8 & 3.8     & 0       &    3.8   \\
Luminosity Monitor        & 4.4      & 0       & 4.4 & 4.4     & 0       &    4.4   \\
Lepton ID                 & 2      & 0       & 2   & 2       & 0       &    2   \\
Jet Energy Scale          & S      & 0       & S   & 0       & 0       &    13.5(S)   \\
Mistag Rate               & 0      & 9     & 0   & 0       & 0       &    0   \\
$B$-Tag Efficiency          & 8.4      & 0       & 8.4 & 8.4     & 0       &    8.4   \\
$t{\bar{t}}$ Cross Section  & 0    & 0       & 10  & 0       & 0       &    0   \\
Diboson Rate              & 0      & 0       & 0   & 10    & 0       &    0   \\
Signal Cross Section      & 0      & 0       & 0   & 0       & 0       &    10 \\
HF Fraction in W+jets     &    30  & 0       & 0   & 0       & 0       &    0   \\
ISR+FSR+PDF               & 21.4      & 0       & 21.4   & 21.4       & 0       &    21.4 \\
QCD Rate                  & 0      & 0       & 0   & 0       & 40      &    0   \\
\end{tabular}
\end{ruledtabular}

\vskip 0.3cm
{\centerline{CDF: single tag (ST) $WH\rightarrow\ell\nu b{\bar{b}}$ channel relative uncertainties (\%)}}
\vskip 0.099cm
\begin{ruledtabular}
\begin{tabular}{lcccccc}\\
Contribution              & $W$+HF & Mistags & Top & Diboson & Non-$W$ & $WH$  \\ \hline
Luminosity ($\sigma_{\mathrm{inel}}(p{\bar{p}})$)
                          & 3.8      & 0       & 3.8 & 3.8     & 0       &    3.8   \\
Luminosity Monitor        & 4.4      & 0       & 4.4 & 4.4     & 0       &    4.4   \\
Lepton ID                 & 2      & 0       & 2   & 2       & 0       &    2   \\
Jet Energy Scale          & S      & 0       & S   & 0       & 0       &    15.8(S)   \\
Mistag Rate               & 0      & 13.3    & 0   & 0       & 0       &    0   \\
$B$-Tag Efficiency          & 3.5      & 0       & 3.5 & 3.5     & 0       &    3.5   \\
$t{\bar{t}}$ Cross Section  & 0    & 0       & 10  & 0       & 0       &    0   \\
Diboson Rate              & 0      & 0       & 0   & 10    & 0       &    0   \\
Signal Cross Section      & 0      & 0       & 0   & 0       & 0       &    10 \\
HF Fraction in W+jets     &    30  & 0       & 0   & 0       & 0       &    0   \\
ISR+FSR+PDF               & 13.1      & 0       & 13.1   & 13.1       & 0       &    13.1 \\
QCD Rate                  & 0      & 0       & 0   & 0       & 40      &    0   \\
\end{tabular}
\end{ruledtabular}

\end{center}
\end{table}


\begin{table}[h]
\begin{center}
\caption{\label{tab:d0systwh1} Systematic uncertainties on the signal and background
contributions for D0's $WH\rightarrow\ell\nu b{\bar{b}}$ loose single (LST) and double tag
(LDT) channels.
Systematic uncertainties are listed by name, see the original
references for a detailed explanation of their meaning and on how they are derived.
Systematic uncertainties for $WH$ shown in this table are obtained for $m_H=115$ GeV/$c^2$.
Uncertainties are relative, in percent, and are symmetric unless otherwise indicated.
Shape uncertainties are labeled with an ``S'', and ``SH'' represents a shape-only uncertainty.}
\vskip 0.2cm
{\centerline{D0: single tag (ST) $WH \rightarrow\ell\nu b\bar{b}$ channel relative uncertainties (\%)}}
\vskip 0.099cm
\begin{ruledtabular}
\begin{tabular}{l c c c c c c c }\\
Contribution             &~WZ/WW~ & Wbb/Wcc& Wjj/Wcj& $~~~t\bar{t}~~~$ &single top&Multijet& ~~~WH~~~\\
\hline
Luminosity                &  6.1  &  6.1  &  6.1  &  6.1  &  6.1  &  0    &  6.1  \\
EM ID/Trigger eff.   (S)  & 1--5  & 2--4  &  2--4 & 1--2  & 1--2  &  0    &  2--3 \\
Muon Trigger eff. (S)     &  1--3 &  1--2 &  1--3 &  2--5 &  2--3 &  0    &  2--4 \\
Muon ID/Reco eff./resol.  &   4.1 &   4.1 &   4.1 &   4.1 &   4.1 &  0    &   4.1 \\
Jet ID/Reco eff.  (S)     &  2--5 &  1--2 &  1--3 &  3--5 &  2--4 &  0    &  2--4 \\
Jet Resolution    (S)     &  4--7 &  1--3 &  1--4 &  2--5 &  2--4 &  0    &  4--6 \\
Jet Energy Scale  (S)     &  4--7 &  2--5 &  2--5 &  2--5 &  2--4 &  0    &  2--5 \\
Vertex Conf. Jet  (S)     & 4--10 & 5--12 & 4--10 & 7--10 & 5--10 &  0    &  4--8 \\
$b$-tag/taggability (S)   & 1--4  &  1--2 & 3--7  &  3--5 &  1--2 &  0    &  1--2 \\
Heavy-Flavor K-factor     &  0    &    20 &     0 &  0    &  0    &  0    &  0    \\
Inst.-WH $e\nu b\bar{b}$ (S) & 1--2 & 2--4 & 1--3 & 1--2  &  1--3 &  -15  &  1--2 \\
Inst.-WH $\mu\nu b\bar{b} $  & 0  &   2.4 &   2.4 &  0    &  0    &  -20  &  0    \\
Cross Section             &     6 &     9 &     9 &    10 &   10  &  0    &     6 \\
Signal Branching Fraction &  0    &     0  &    0 &    0  &   0   &  0    &  1-9    \\
ALPGEN MLM pos/neg(S)     &  0    &   SH  &     0 &  0    &  0    &  0    &  0    \\
ALPGEN Scale (S)          &  0    &   SH  &    SH &  0    &  0    &  0    &  0    \\
Underlying Event (S)      &  0    &   SH  &     0 &  0    &  0    &  0    &  0    \\
PDF, reweighting          &  2    &  2    & 2     & 2     &  2    &  0    &  2    \\
\end{tabular}
\end{ruledtabular}
\vskip 0.5cm
{\centerline{D0: double tag (DT) $WH \rightarrow\ell\nu b\bar{b}$ channel relative uncertainties (\%)}}
\vskip 0.099cm
\begin{ruledtabular}
\begin{tabular}{ l c c c c c c c }   \\
Contribution  &~WZ/WW~&Wbb/Wcc&Wjj/Wcj&$~~~t\bar{t}~~~$&single top&Multijet& ~~~WH~~~\\
\hline
Luminosity                &  6.1  &  6.1  &  6.1  &  6.1  &  6.1  &  0    &  6.1  \\
EM ID/Trigger eff.   (S)  & 2--5  & 2--3  &  2--3 & 1--2  & 1--2  &  0    &  1--2 \\
Muon Trigger eff. (S)     &  2--4 &  1--2 &  1--2 &  2--4 &  1--3 &  0    &  2--5 \\
Muon ID/Reco eff./resol.  &   4.1 &   4.1 &   4.1 &   4.1 &   4.1 &  0    &   4.1 \\
Jet ID/Reco eff.  (S)     &  2--8 &  2--5 &  4--9 &  3--7 &  2--4 &  0    &  3--7 \\
Jet Resolution    (S)     &  4--7 &  2--7 &  2--7 &  2--9 &  2--4 &  0    &  4--6 \\
Jet Energy Scale  (S)     &  4--7 &  2--6 &  2--7 &  2--6 &  2--7 &  0    &  4--6 \\
Vertex Conf. Jet  (S)     & 4--10 & 5--12 & 4--10 & 7--10 & 5--10 &  0    &  4--6 \\
$b$-tag/taggability (S)   & 3--7  &  4--6 & 3--10 & 5--10 & 4--10 &  0    &  4--9 \\
Heavy-Flavor K-factor     &  0    &    20 &     0 &  0    &  0    &  0    &  0    \\
Inst.-WH $e\nu b\bar{b}$ (S) & 1--2 & 2--4 & 1--3 & 1--2  &  1--3 &  -15  &  1--2 \\
Inst.-WH $\mu\nu b\bar{b} $  & 0  &   2.4 &   2.4 &  0    &  0    &  -20  &  0    \\
Cross Section             &     6 &     9 &     9 &    10 &    10 &  0    &     6 \\
Signal Branching Fraction &  0    &     0  &    0 &    0  &   0   &  0    &  1-9    \\
ALPGEN MLM pos/neg(S)     &  0    &   SH  &     0 &  0    &  0    &  0    &  0    \\
ALPGEN Scale (S)          &  0    &   SH  &    SH &  0    &  0    &  0    &  0    \\
Underlying Event (S)      &  0    &   SH  &     0 &  0    &  0    &  0    &  0    \\
PDF, reweighting          &  2    &  2    & 2     & 2     &  2    &  0    &  2    \\
\end{tabular}
\end{ruledtabular}

\end{center}
\end{table}


\begin{table}
\begin{center}
\caption{\label{tab:d0sysVHtau}  Systematic uncertainties on the signal and background contributions for
D0's
$\tau \tau jj$ Run~IIb channel.
Systematic uncertainties for the
Higgs signal shown in this table are obtained for $m_H=135$ GeV/$c^2$.  Systematic uncertainties are listed
by name; see the original references for a detailed explanation of their meaning and on how they are derived.
Uncertainties are relative, in percent, and are symmetric unless otherwise indicated. A systematic is denoted
as flat if it affects the normalization only, and as a shape ``S'' uncertainty otherwise.}
\vskip 0.1cm
{\centerline{D0: $\mu \tau_{\rm{had}} jj$ Run IIb channel relative uncertainties (\%)}}
\vskip 0.099cm
\begin{ruledtabular}
\begin{tabular}{lccccccccc}\\
Contribution &$VH$ Signal &$VBF$ Signal &$ggH$ Signal  &$W+jets$ & $Z+jets$ & Top& diboson& Multijet\\ \hline

Luminosity (D0 specific)  & 4.1 & 4.1 & 4.1 & 4.1 & 4.1 & 4.1 &4.1 &- \\
Luminosity (Tevatron common)  & 4.6 & 4.6 & 4.6 & 4.6 & 4.6 & 4.6 &4.6 &- \\
$\mu$ ID & 2.9 & 2.9 & 2.9 & 2.9 & 2.9 & 2.9 &2.9 &- \\
$\mu$ trigger & 8.6 & 8.6 & 8.6 & 8.6 & 8.6 & 8.6 &8.6 &- \\
$\tau$ energy correction & 9.8 & 9.8 & 9.8 & 9.8 & 9.8 & 9.8 &9.8 &- \\
$\tau$ track efficiency & 1.4 & 1.4 & 1.4 & 1.4 & 1.4 & 1.4 &1.4 &- \\
$\tau$ selection by type & 12,4.2,7 & 12,4.2,7 & 12,4.2,7 & 12,4.2,7 & 12,4.2,7 & 12,4.2,7 &12,4.2,7 &- \\
Cross section & 6.2 &4.9& 33 &6.0&6.0&10.0&7.0&-\\
ggH Signal PDF &-&-&29&-&-&-&-&-\\
ggH $H p_T$ Reweighting (S)& 1.0 & 1.0 & 1.0 & 1.0 & 1.0 & 1.0 &1.0 &-\\
Signal Branching Fraction & 0-7.3 & 0-7.3 & 0-7.3 & - & - & - & - &-\\
Vertex confirmation for jets  & 4.0 & 4.0 & 4.0 & 4.0 & 4.0 & 4.0 &4.0 &- \\

Jet ID(S)  & $\sim$10 & $\sim$10 & $\sim$10 & $\sim$10& $\sim$10 & $\sim$10 &$\sim$10 &- \\
Jet Energy Resolution (S)  & $\sim$10 & $\sim$10 & $\sim$10 & $\sim$10& $\sim$10 & $\sim$10 &$\sim$10 &- \\
Jet energy Scale (S)  & $\sim$15 & $\sim$15 & $\sim$15 & $\sim$15& $\sim$15 & $\sim$15 &$\sim$15 &- \\

Jet pT  & 5.5 & 5.5 & 5.5 & 5.5 & 5.5 & 5.5 &5.5 &- \\
PDF reweighting  & 2 & 2 & 2 & 2 & 2 & 2 &2 &- \\
Multijet Normalization  & - & - & - & - & - & - &- &5.3 \\
Multijet Shape   & - & - & - & - & - & - &- &$\sim$15 \\

\end{tabular}
\end{ruledtabular}

\vskip 0.3cm
{\centerline{D0: $e \tau_{\rm{had}} jj$ Run IIb relative uncertainties (\%)}}
\vskip 0.099cm
\begin{ruledtabular}
\begin{tabular}{lccccccccc}\\
Contribution &$VH$ Signal &$VBF$ Signal &$ggH$ Signal  &$W+jets$ & $Z+jets$ & Top& diboson& Multijet\\
\hline
Luminosity (D0 specific)  & 4.1 & 4.1 & 4.1 & 4.1 & 4.1 & 4.1 &4.1 &- \\
Luminosity (Tevatron common)  & 4.6 & 4.6 & 4.6 & 4.6 & 4.6 & 4.6 &4.6 &- \\
EM ID & 4 & 4 & 4 & 4 & 4 & 4 &4 &- \\
e trigger & 2 & 2 & 2 & 2 & 2 & 2 &2 &- \\
$\tau$ energy correction & 9.8 & 9.8 & 9.8 & 9.8 & 9.8 & 9.8 &9.8 &- \\
$\tau$ track efficiency & 1.4 & 1.4 & 1.4 & 1.4 & 1.4 & 1.4 &1.4 &- \\
$\tau$ selection by type & 12,4.2,7 & 12,4.2,7 & 12,4.2,7 & 12,4.2,7 & 12,4.2,7 & 12,4.2,7 &12,4.2,7 &- \\
Cross section & 6.2 &4.9& 33 &6.0&6.0&10.0&7.0&-\\
ggH Signal PDF &-&-&29&-&-&-&-&-\\
ggH $H p_T$ Reweighting (S)& 1.0 & 1.0 & 1.0 & 1.0 & 1.0 & 1.0 &1.0 &-\\
Signal Branching Fraction & 0-7.3 & 0-7.3 & 0-7.3 & - & - & - & - &-\\
Vertex confirmation for jets  & 4.0 & 4.0 & 4.0 & 4.0 & 4.0 & 4.0 &4.0 &- \\

Jet ID(S)  & $\sim$10 & $\sim$10 & $\sim$10 & $\sim$10& $\sim$10 & $\sim$10 &$\sim$10 &- \\
Jet Energy Resolution (S)  & $\sim$10 & $\sim$10 & $\sim$10 & $\sim$10& $\sim$10 & $\sim$10 &$\sim$10 &- \\
Jet energy Scale (S)  & $\sim$15 & $\sim$15 & $\sim$15 & $\sim$15& $\sim$15 & $\sim$15 &$\sim$15 &- \\

Jet pT  & 5.5 & 5.5 & 5.5 & 5.5 & 5.5 & 5.5 &5.5 &- \\
PDF reweighting  & 2 & 2 & 2 & 2 & 2 & 2 &2 &- \\
Multijet Normalization  & - & - & - & - & - & - &- &4.7 \\
Multijet Shape   & - & - & - & - & - & - &- & $\sim$15 \\

\end{tabular}
\end{ruledtabular}

\end{center}
\end{table}


\begin{table}
\begin{center}
\caption{\label{tab:cdfvvbb1} Systematic uncertainties on the signal and background contributions for CDF's
$WH,ZH\rightarrow\MET b{\bar{b}}$ tight double tag (TDT), loose double tag (LDT), and single tag (ST) channels.
Systematic uncertainties are listed by name; see the original references for a detailed explanation of their
meaning and on how they are derived.  Systematic uncertainties for $ZH$ and $WH$ shown in this table are
obtained for $m_H=120$~GeV/$c^2$.  Uncertainties are relative, in percent, and are symmetric unless otherwise
indicated. Shape uncertainties are labeled with an "S".}
\vskip 0.1cm
{\centerline{CDF: tight double-tag (TDT) $WH,ZH\rightarrow\MET b{\bar{b}}$ channel relative uncertainties (\%)}}
\vskip 0.099cm
\begin{ruledtabular}
      \begin{tabular}{lccccccccc}\\
        Contribution & ZH & WH & Multijet & Mistags & Top Pair & S. Top  & Di-boson  & W + HF  & Z + HF \\\hline
        Luminosity       & 3.8 & 3.8 &     &  & 3.8 & 3.8 & 3.8     & 3.8     & 3.8     \\
        Lumi Monitor      & 4.4 & 4.4 &     &  & 4.4 & 4.4 & 4.4     & 4.4     & 4.4     \\
        Tagging SF        & 10.4& 10.4&      & & 10.4& 10.4& 10.4    & 10.4    & 10.4    \\
      Trigger Eff. (S)& 0.9 & 1.4 & 0.9 & & 0.9 & 1.6 & 2.0     & 1.8     & 1.2     \\
        Lepton Veto       & 2.0 & 2.0 &      & & 2.0 & 2.0 &2.0      & 2.0     & 2.0     \\
        PDF Acceptance    & 3.0 & 3.0 &    &   & 3.0 & 3.0 &3.0      & 3.0     & 3.0     \\
        JES (S)       & $^{+1.7}_{-1.8}$
                                  & $^{+2.4}_{-2.3}$
                                          & &
                                                  & $^{+0.0}_{-0.1}$
                                                          & $^{+2.5}_{-2.4}$
                                                                  & $^{+4.1}_{-4.5}$
                                                                             & $^{+4.3}_{-4.6}$
                                                                                          & $^{+8.8}_{-3.2}$    \\
        ISR/FSR               & \multicolumn{2}{c}{$^{+3.0}_{+3.0}$} &       &       &       &           &           &      \\
        Cross-Section     &  5  & 5 &   &    & 10 & 10 & 6    & 30      & 30      \\
        Multijet Norm.  (shape)   &   &    & 2.5 &  &      & &          &           &           \\
        Mistag (S) & & & & $^{+36.7}_{-30}$ & & & & &\\
      \end{tabular}
\end{ruledtabular}

\vskip 0.3cm
{\centerline{CDF: loose double-tag (LDT) $WH,ZH\rightarrow\MET b{\bar{b}}$ channel relative uncertainties (\%)}}
\vskip 0.099cm
 \begin{ruledtabular}
     \begin{tabular}{lccccccccc}\\
        Contribution & ZH & WH & Multijet & Mistags & Top Pair & S. Top  & Di-boson  & W + HF  & Z + HF \\\hline
        Luminosity       & 3.8  & 3.8  &   &  & 3.8  & 3.8  & 3.8      & 3.8      & 3.8     \\
        Lumi Monitor      & 4.4  & 4.4  &   &  & 4.4  & 4.4  & 4.4      & 4.4      & 4.4     \\
        Tagging SF        & 8.3 & 8.3 &   &  & 8.3 & 8.3 & 8.3     & 8.3     & 8.3     \\
      Trigger Eff. (S)& 1.2 & 1.7 & 1.6 & & 0.9 & 1.8 & 2.0     & 2.5     & 1.9     \\
        Lepton Veto       & 2.0  & 2.0  &    & & 2.0  & 2.0  &2.0       & 2.0      & 2.0     \\
        PDF Acceptance    & 3.0  & 3.0  &  &   & 3.0  & 3.0  & 3.0       & 3.0      & 3.0     \\
        JES (S)       & $^{+1.9}_{-1.9}$
                                   & $^{+2.4}_{-2.4}$
                                          & &
                                                          & $^{+3.0}_{-2.8}$
                                                  			& $^{-0.6}_{0.2}$
                                                                    & $^{+4.2}_{-4.2}$
                                                                                 & $^{+6.8}_{-5.9}$
                                                                                              & $^{+8.3}_{-3.1}$    \\
        ISR/FSR               & \multicolumn{2}{c}{$^{+2.4}_{-2.4}$} &    &   &       &       &           &           &      \\
        Cross-Section     &  5.0   & 5.0 &   &   & 10 & 10 & 6    & 30      & 30      \\
        Multijet Norm.  &       & & 1.6 &       & &          &           &           \\
        Mistag (S) & & & & $^{+65.2}_{-38.5}$ & & & & &\\
      \end{tabular}
\end{ruledtabular}

\vskip 0.3cm
{\centerline{CDF: single-tag (ST) $WH,ZH\rightarrow\MET b{\bar{b}}$ channel relative uncertainties (\%)}}
\vskip 0.099cm
\begin{ruledtabular}
      \begin{tabular}{lccccccccc}\\
        Contribution & ZH & WH & Multijet & Mistags & Top Pair & S. Top  & Di-boson  & W + HF  & Z + HF \\\hline
        Luminosity       & 3.8  & 3.8  &    & & 3.8  & 3.8  & 3.8      & 3.8      & 3.8     \\
        Lumi Monitor      & 4.4  & 4.4  &   & & 4.4  & 4.4  & 4.4      & 4.4      & 4.4     \\
        Tagging SF        & 5.2  & 5.2  &     & & 5.2  & 5.2  & 5.2      & 5.2      & 5.2     \\
      Trigger Eff. (S)& 1.2 & 1.7 & 1.6 & & 0.9 & 1.8 & 2.0     & 2.5     & 1.9     \\
        Lepton Veto       & 2.0  & 2.0  &  &   & 2.0  & 2.0  &2.0       & 2.0      & 2.0     \\
        PDF Acceptance    & 3.0  & 3.0  &   &  & 3.0  & 3.0  & 3.0       & 3.0      & 3.0     \\
        JES (S)       & $^{+2.6}_{-2.6}$
                                  & $^{+3.3}_{-3.1}$
                                          & &
                                                  		& $^{-0.8}_{+0.6}$
                                                          		& $^{+2.7}_{-2.8}$
                                                                  		& $^{+5.1}_{-5.1}$
                                                                            		 & $^{+8.2}_{-6.8}$
                                                                                          		& $^{+10.8}_{-3.4}$    \\
        ISR/FSR               & \multicolumn{2}{c}{$^{+2.0}_{-2.0}$} &       &       &       &           &           &      \\
        Cross-Section     &  5.0   & 5.0 &   &   & 10 & 10 & 6    & 30      & 30      \\
        Multijet Norm.  &       & & 0.7 &       & &          &           &           \\
        Mistag (S) & & & & $^{+17.9}_{-17.4}$ & & & & &\\
      \end{tabular}
\end{ruledtabular}

\end{center}
\end{table}


\begin{table}
\begin{center}
\caption{\label{tab:d0vvbb} Systematic uncertainties on the signal and background contributions for
D0's $ZH\rightarrow \nu \nu b{\bar{b}}$ loose single tag (LST) and double tag (LDT) channels.
Systematic uncertainties are listed by name; see the original references for a detailed explanation
of their meaning and on how they are derived.  Systematic uncertainties for $VH$ ($WH$+$ZH$) shown in this
table are obtained for $m_H=115$ GeV/$c^2$.  Uncertainties are relative, in percent, and are symmetric
unless otherwise indicated. Shape uncertainties are labeled with an ``S'', and ``SH'' represents
a shape only uncertainty.}

\vskip 0.1cm
{\centerline{D0: single tag (LST)~ $ZH \rightarrow \nu\nu b \bar{b}$ channel relative uncertainties (\%)}}
\vskip 0.099cm
\begin{ruledtabular}
\begin{tabular}{ l c c c c c c } \\
Contribution            & Top  & V+HF & V+LF & Diboson  & Total Bkgd & VH  \\
\hline
Jet ID/Reco Eff (S)             & 2.0  &  2.0   &  2.0   &  2.0     & 1.9  & 2.0  \\
Jet Energy Scale (S)            & 2.2  &  1.6   &  3.1   &  1.0     & 2.5  & 0.5  \\
Jet Resolution (S)              & 0.5  &  0.3   &  0.3   &  0.9     & 0.3  & 0.8  \\
Vertex Conf. / Taggability (S)  & 3.2  &  1.9   &  1.7   &  1.8     & 1.7  & 1.6  \\
b Tagging (S)                   & 1.1  &  0.8   &  1.8   &  1.2     & 1.3  & 3.2  \\
Lepton Identification           & 1.6  &  0.9   &  0.8   &  1.0     & 0.8  & 1.1  \\
Trigger                         & 2.0  &  2.0   &  2.0   &  2.0     & 1.9  & 2.0  \\
Heavy Flavor Fractions          & --   &  20.0  &  --    &  --      & 4.1  & --   \\
Cross Sections                  & 10.0 &  10.2  &  10.2  &  7.0     & 9.8  & 6.0  \\
Signal Branching Fractions      & --   &  --    & --     & --       & --   & 1-9   \\
Luminosity                      & 6.1  &  6.1   &  6.1   &  6.1     & 5.8  & 6.1  \\
Multijet Normalization         & --   &  --    &  --    &  --      & 1.3  & --   \\
ALPGEN MLM (S)                  & --   &  --    &  SH    &  --      & --   & --   \\
ALPGEN Scale (S)                & --   &  SH    &  SH    &  --      & --   & --   \\
Underlying Event (S)            & --   &  SH    &  SH    &  --      & --   & --   \\
PDF, reweighting (S)            & SH   &  SH    &  SH    &  SH      & SH   & SH   \\
Total uncertainty     & 12.8 &  23.6  &  12.9  &  10.1    & 12.3 & 9.8  \\

\end{tabular}
\end{ruledtabular}

\vskip 0.3cm
{\centerline{D0: double tag (LDT)~ $ZH \rightarrow \nu\nu b \bar{b}$ channel relative uncertainties (\%)}}
\vskip 0.099cm
\begin{ruledtabular}
\begin{tabular}{ l  c  c  c  c  c  c } \\
Contribution            & Top  & V+HF & V+LF & Diboson  & Total Bkgd & VH  \\
\hline
Jet ID/Reco Eff                 & 2.0  &  2.0   &  2.0   &  2.0     & 1.9  & 2.0  \\
Jet Energy Scale                & 2.1  &  1.6   &  3.4   &  1.2     & 2.2  & 0.2  \\
Jet Resolution                  & 0.7  &  0.4   &  0.5   &  1.5     & 0.5  & 0.7  \\
Vertex Conf. / Taggability      & 2.6  &  1.6   &  1.6   &  1.8     & 1.7  & 1.4  \\
b Tagging                       & 6.2  &  4.3   &  4.3   &  3.7     & 3.6  & 5.8  \\
Lepton Identification           & 2.0  &  0.9   &  0.8   &  0.9     & 1.0  & 1.1  \\
Trigger                         & 2.0  &  2.0   &  2.0   &  2.0     & 1.9  & 2.0  \\
Heavy Flavor Fractions          & --   &  20.0  &  --    &  --      & 8.0  & --   \\
Cross Sections                  & 10.0 &  10.2  &  10.2  &  7.0     & 9.8  & 6.0  \\
Signal Branching Fractions      & --   &  --    & --     & --       & --   & 1-9   \\
Luminosity                      & 6.1  &  6.1   &  6.1   &  6.1     & 5.8  & 6.1  \\
Multijet Normalilzation         & --   &  --    &  --    &  --      & 1.0  & --   \\
ALPGEN MLM pos/neg(S)           &  --  &  --    &  SH    &  --      & --   & --   \\
ALPGEN Scale (S)                &  --  &  SH    &  SH    &  --      & --   & --   \\
Underlying Event (S)            &  --  &  SH    &  SH    &  --      & --   & --   \\
PDF, reweighting (S)            &  SH  &  SH    &  SH    &  SH      & SH   & SH   \\
Total uncertainty       & 14.1 &  24.0  &  13.5  &  10.7    & 13.9 & 10.9 \\

\end{tabular}
\end{ruledtabular}
\end{center}
\end{table}

%


\begin{table}
\begin{center}

\caption{\label{tab:cdfllbb1} Systematic uncertainties on the signal and background contributions for CDF's
$ZH\rightarrow \mu^+\mu^-b{\bar{b}}$ single tag (ST), tight double tag (TDT), and loose double tag (LDT)
channels.  Systematic uncertainties
are listed by name; see the original references for a detailed explanation of their meaning and on how they
are derived.  Systematic uncertainties for $ZH$  shown in this table are obtained for $m_H=115$ GeV/$c^2$.
Uncertainties are relative, in percent, and are symmetric unless otherwise indicated. Shape uncertainties are labeled with an "S".}
\vskip 0.1cm
{\centerline{CDF: single tag (ST) $ZH \rightarrow \mu^+\mu^- b \bar{b}$ channel relative uncertainties (\%)}}
\vskip 0.099cm
\begin{ruledtabular}
\begin{tabular}{lccccccccc} \\
Contribution   & ~Fakes~ & ~~~$t\bar{t}$~~~  & ~~$WW$~~ & ~~$WZ$~~ & ~~$ZZ$~~  & ~$Z+b{\bar{b}}$~ & ~$Z+c{\bar{c}}$~& ~Mistags~ & ~~~$ZH$~~~ \\ \hline
Luminosity ($\sigma_{\mathrm{inel}}(p{\bar{p}})$)          &     &    3.8 &     3.8&3.8&3.8 &    3.8           &    3.8          &        &    3.8  \\
Luminosity Monitor        &     &    4.4 & 4.4 & 4.4 &      4.4 &    4.4           &    4.4          &       &    4.4  \\
Lepton ID    &     &    1 &    1& 1& 1 &      1           &    1          &       &    1  \\
Lepton Energy Scale    &     &    1.5 &      1.5 & 1.5& 1.5 &    1.5           &    1.5          &        &    1.5  \\
Fake Leptons       & 5    &   & &   &    &              &             &        &     \\
Mistag Rate  &   &   & & &  &  &  & $^{+13.6}_{-13.7}$ &  \\
Jet Energy Scale  (S)       &     &
    $^{+1.8}_{-1.9}$ & 
    $^{+18.7}_{-3.7}$ & 
    $^{+3.9}_{-4.3}$ & 
    $^{+4.3}_{-5.4}$ & 
    $^{+7.9}_{-6.7}$ & 
    $^{+7.9}_{-6.6}$ & 
       & 
    $^{+1.7}_{-2.6}$ \\ 

$b$-tag Rate       &     &    5.2 &      5.2 &5.2 & 5.2 &    5.2           &   5.2          &      &    5.2 \\
$t{\bar{t}}$ Cross Section         &     &   10&  &&      &              &             &        &     \\
Diboson Cross Section        &     &    & 6 & 6& 6    &              &             &        &     \\
$Z+$HF Cross Section      &        &  &&  &    &  40            & 40           &        &     \\
$ZH$ Cross Section    &     &    &   &&   &              &             &        &    5 \\
ISR/FSR           &     &    &     &&  &              &             &        &   1 \\
NN Trigger Model  &        & 5    & 5 & 5& 5   & 5              & 5             &      &   5     \\
\end{tabular}
\end{ruledtabular}
\end{center}
\end{table}
\begin{table}
\begin{center}

\vskip 0.3cm
{\centerline{CDF: tight double tag (TDT) $ZH \rightarrow \mu^+\mu^- b \bar{b}$ channel relative uncertainties (\%)}}
\vskip 0.099cm
\begin{ruledtabular}
\begin{tabular}{lccccccccc} \\
Contribution   & ~Fakes~ & ~~~$t\bar{t}$~~~  & ~~$WZ$~~ & ~~$ZZ$~~ & ~~$WW$~~  & ~$Z+b{\bar{b}}$~ & ~$Z+c{\bar{c}}$~& ~Mistags~ & ~~~$ZH$~~~ \\ \hline
Luminosity ($\sigma_{\mathrm{inel}}(p{\bar{p}})$)          &      &    3.8 &     3.8 & 3.8 & 3.8 &    3.8           &    3.8          &        &    3.8  \\
Luminosity Monitor        &      &    4.4 &      4.4 & 4.4 & 4.4 &    4.4           &    4.4          &        &    4.4  \\
Lepton ID    &      &    1 &    1& 1 & 1 &      1           &    1          &        &    1  \\
Lepton Energy Scale    &      &    1.5 &      1.5 & 1.5 & 1.5 &    1.5           &    1.5          &       &    1.5  \\
Fake Leptons       & 5    &       &   & &  &               &             &        &     \\
Mistag Rate  &   &  & & &   &  &  &  $^{+28.7}_{-25.1}$ &  \\
Jet Energy Scale  (S)       &      &
    $^{+1.6}_{-1.7}$ & 
    $^{+3.5}_{-3.7}$& 
    $^{+3.5}_{-3.7}$& 
    $^{+4.1}_{-4.4}$& 
    $^{+7.5}_{-3.8}$ & 
    $^{+6.6}_{-5.2}$ & 
       & 
    $^{+1.4}_{-2.3}$ \\ 

$b$-tag Rate       &      &    10.2 &      10.2 & 10.2 & 10.2 &    10.2         &   10.2          &       &    10.2 \\
$t{\bar{t}}$ Cross Section         &      &   10 &    & &    &              &              &         &     \\
Diboson Cross Section        &      &     & 6  & 6 & 6   &               &              &         &     \\
$Z+$HF Cross Section      &         &   & &  &     &  40            & 40           &         &      \\
$ZH$ Cross Section    &      &     &  & &     &               &              &         &    5 \\
ISR/FSR           &      &     &  & &      &               &              &         &   5 \\
NN Trigger Model  &         & 5    & 5 & 5 & 5  & 5              & 5             &       &   5     \\
\end{tabular}
\end{ruledtabular}

\end{center}
\end{table}


\begin{table}[t]
\begin{center}

\vskip 0.3cm
{\centerline{CDF: loose double tag (LDT) $ZH \rightarrow \mu^+\mu^- b \bar{b}$ channel relative uncertainties (\%)}}
\vskip 0.099cm
\begin{ruledtabular}
\begin{tabular}{lccccccccc} \\
Contribution   & ~Fakes~ & ~~~$t\bar{t}$~~~  & ~~$WW$~~ & ~~$WZ$~~ & ~~$ZZ$~~  & ~$Z+b{\bar{b}}$~ & ~$Z+c{\bar{c}}$~& ~Mistags~ & ~~~$ZH$~~~ \\ \hline
Luminosity ($\sigma_{\mathrm{inel}}(p{\bar{p}})$)          &     &    3.8 &     3.8& 3.8 & 3.8 &    3.8           &    3.8          &        &    3.8  \\
Luminosity Monitor        &     &    4.4 & 4.4 & 4.4 &      4.4 &    4.4           &    4.4          &       &    4.4  \\
Lepton ID    &     &    1 &    1 & 1 & 1 &      1           &    1          &       &    1  \\
Lepton Energy Scale    &     &    1.5 &      1.5 & 1.5 & 1.5 &    1.5           &    1.5          &        &    1.5  \\
Fake Leptons       & 5    &   & &   &    &              &             &        &     \\
Mistag Rate  &   &   & & &  &  &  &  $^{+27.2}_{-24.0}$ &  \\
Jet Energy Scale  (S)       &     &
    $^{+1.6}_{-1.8}$ & 
    $^{+3.5}_{-3.7}$ & 
    $^{+4.6}_{-7.6}$ & 
    $^{+4.0}_{-4.2}$ & 
    $^{+6.9}_{-5.9}$ & 
    $^{+7.8}_{-5.9}$ & 
       & 
    $^{+1.5}_{-2.4}$ \\ 

$b$-tag Rate       &     &    8.7 &      8.7 & 8.7 & 8.7 &    8.7           &   8.7        &       &    8.7  \\
$t{\bar{t}}$ Cross Section         &     &   10 &     & &   &              &             &        &     \\
Diboson Cross Section        &     &    & 6  & 6 & 6   &              &             &        &     \\
$Z+$HF Cross Section      &        &  & &  &    &  40           & 40         &        &     \\
$ZH$ Cross Section    &     &    &  & &    &              &             &        &    5 \\
ISR/FSR           &     &    &    & &   &              &             &        &   2 \\
NN Trigger Model  &        & 5    & 5  & 5 & 5 & 5              & 5             &      &   5     \\
\end{tabular}
\end{ruledtabular}

\end{center}
\end{table}


\begin{table}
\begin{center}

\caption{\label{tab:cdfllbb2} Systematic uncertainties on the signal and background contributions for CDF's
$ZH\rightarrow e^+e^-b{\bar{b}}$ single tag (ST), tight double tag (TDT), and loose double tag (LDT)
channels.  Systematic uncertainties
are listed by name; see the original references for a detailed explanation of their meaning and on how they
are derived.  Systematic uncertainties for $ZH$  shown in this table are obtained for $m_H=115$ GeV/$c^2$.
Uncertainties are relative, in percent, and are symmetric unless otherwise indicated. Shape uncertainties are labeled with an "S".}
\vskip 0.1cm
{\centerline{CDF: single tag (ST) $ZH \rightarrow e^+e^- b \bar{b}$ channel relative uncertainties (\%)}}
\vskip 0.099cm
\begin{ruledtabular}
\begin{tabular}{lccccccccc} \\
Contribution   & ~Fakes~ & ~~~Top~~~ & ~~$WW$~~ & ~~$WZ$~~ & ~~$ZZ$~~ & ~$Z+b{\bar{b}}$~ & ~$Z+c{\bar{c}}$~& ~$Z+$LF~ & ~~~$ZH$~~~ \\ \hline
Luminosity ($\sigma_{\mathrm{inel}}(p{\bar{p}})$)   & 0     &    3.8 &    3.8 &    3.8 &    3.8           &    3.8          & 3.8        &    0 & 3.8 \\
Luminosity Monitor        & 0     &    4.4 &    4.4 &    4.4 &    4.4           &   4.4          & 4.4        &    0  & 4.4\\
Trigger Emulation    & 0     &    1 &    1 &    1 &    1           &    1          & 1        &    0  & 1\\
Lepton ID    & 0     &    2 &    2 &    2 &    2           &    2         & 2        &    0  & 2\\
Lepton Energy Scale    & 0     &    3 &    3 &    3 &    3           &    3          & 3        &    0  & 3\\
$ZH$ Cross Section    & 0     &    0 &    0 &    0 &    0           &    0          & 0        &    0 & 5\\
Fake Leptons       & 50    & 0    & 0    & 0    & 0              & 0             & 0        & 0     & 0\\
B-Tag Efficiency      & 0     &    5.2 &    5.2 &    5.2 &    5.2           &  5.2          & 5.2        &    0  & 5.2\\
$t{\bar{t}}$ Cross Section         & 0     &   10 & 0    & 0    & 0              & 0             & 0        & 0     & 0\\
Diboson Cross Section        & 0     & 0   & 6   & 6    & 6              & 0             & 0        & 0     & 0\\
$\sigma(p{\bar{p}}\rightarrow Z+HF)$      & 0     & 0    & 0    & 0    &  40            & 40           & 40        & 0     & 0\\
ISR/FSR           & 0     & 0    & 0    & 0    & 0              & 0             & 0        &   0     & $4.0$\\
Mistag Rate (S)      & 0     & 0    & 0    & 0    & 0              & 0             & 0 &  $^{+13.9}_{-13.8}$     & 0     \\
Jet Energy Scale  (S)       & 0     &
  $^{+1.9}_{-2.5}$   & 
    $^{+19.6}_{-4.0}$   & 
  $^{+5.2}_{-6.2}$   & 
  $^{+5.3}_{-7.1}$   & 
  $^{+12.1}_{-11.1}$   & 
  $^{+4.1}_{-9.9}$   & 
  0   & 
  $^{+3.0}_{-4.3}$   \\ 
\end{tabular}
\end{ruledtabular}
\end{center}
\end{table}
\begin{table}
\begin{center}

\vskip 0.3cm
{\centerline{CDF: tight double tag (TDT) $ZH \rightarrow e^+e^- b \bar{b}$ channel relative uncertainties (\%)}}
\vskip 0.099cm
\begin{ruledtabular}
\begin{tabular}{lccccccccc} \\
Contribution   & ~Fakes~ & ~~~Top~~~  & ~~$WZ$~~ & ~~$ZZ$~~ & ~$Z+b{\bar{b}}$~ & ~$Z+c{\bar{c}}$~& ~$Z+$LF~ & ~~~$ZH$~~~ \\ \hline
Luminosity ($\sigma_{\mathrm{inel}}(p{\bar{p}})$)  & 0     &        3.8 &    3.8 &    3.8           &    3.8          & 3.8        &    0 & 3.8 \\
Luminosity Monitor        & 0     &    4.4 &    4.4 &    4.4        &   4.4          & 4.4        &    0  & 4.4\\
Trigger Emulation    & 0     &    1 &    1 &    1 &    1                     & 1        &    0  & 1\\
Lepton ID    & 0     &    2 &    2 &    2 &    2                & 2        &    0  & 2\\
Lepton Energy Scale    & 0     &    3 &    3 &    3           &    3          & 3        &    0  & 3\\
$ZH$ Cross Section    & 0     &    0 &    0 &    0            &    0          & 0        &    0 & 5\\
Fake Leptons       & 50    & 0    & 0    & 0                 & 0             & 0        & 0     & 0\\
B-Tag Efficiency      & 0     &    10.4 &     10.4  &       10.4            &   10.4           &  10.4         &    0  &  10.4 \\
$t{\bar{t}}$ Cross Section         & 0     &   10 & 0        & 0              & 0             & 0        & 0     & 0\\
Diboson Cross Section        & 0     & 0   & 6   & 6                & 0             & 0        & 0     & 0\\
$\sigma(p{\bar{p}}\rightarrow Z+HF)$      & 0       & 0    & 0    &  40            & 40           & 40        & 0     & 0\\
ISR/FSR           & 0     & 0    & 0    & 0    & 0                      & 0        &   0     & $4.0$\\
Mistag Rate (S)      & 0     & 0    & 0    & 0                  & 0             & 0 &  $^{+29.3}_{-25.4}$     & 0     \\
Jet Energy Scale  (S)       & 0     &
  $^{+1.4}_{-2.6}$   & 
  $^{+7.8}_{-3.1}$   & 
  $^{+3.4}_{-5.9}$   & 
  $^{+6.8}_{-6.6}$   & 
  $^{+1.0}_{-3.7}$   & 
  0   & 
  $^{+1.6}_{-2.7}$   \\ 
\end{tabular}
\end{ruledtabular}

\end{center}
\end{table}


\begin{table}[t]
\begin{center}

\vskip 0.3cm
{\centerline{CDF: loose double tag (LDT) $ZH \rightarrow e^+e^- b \bar{b}$ channel relative uncertainties (\%)}}
\vskip 0.099cm
\begin{ruledtabular}
\begin{tabular}{lccccccccc} \\
Contribution   & ~Fakes~ & ~~~Top~~~ & ~~$WW$~~ & ~~$WZ$~~ & ~~$ZZ$~~ & ~$Z+b{\bar{b}}$~ & ~$Z+c{\bar{c}}$~& ~$Z+$LF~ & ~~~$ZH$~~~ \\ \hline
Luminosity ($\sigma_{\mathrm{inel}}(p{\bar{p}})$)          & 0     &    3.8 &    3.8 &    3.8 &    3.8           &    3.8          & 3.8        &    0 & 3.8 \\
Luminosity Monitor     & 0     &    4.4 &    4.4 &    4.4 &    4.4           &   4.4          & 4.4        &    0  & 4.4\\
Trigger Emulation      & 0     &    1 &    1 &    1 &    1           &    1          & 1        &    0  & 1\\
Lepton ID              & 0     &    2 &    2 &    2 &    2           &    2         & 2        &    0  & 2\\
Lepton Energy Scale    & 0     &    3 &    3 &    3 &    3           &    3          & 3        &    0  & 3\\
$ZH$ Cross Section     & 0     &    0 &    0 &    0 &    0           &    0          & 0        &    0 & 5\\
Fake Leptons           & 50    & 0    & 0    & 0    & 0              & 0             & 0        & 0     & 0\\
B-Tag Efficiency       & 0     &    8.7 &    8.7 &    8.7 &    8.7           &  8.7          & 8.7        &    0  & 8.7\\
$t{\bar{t}}$ Cross Section         & 0     &   10 & 0    & 0    & 0              & 0             & 0        & 0     & 0\\
Diboson Cross Section        & 0     & 0   & 6   & 6    & 6              & 0             & 0        & 0     & 0\\
$\sigma(p{\bar{p}}\rightarrow Z+HF)$      & 0     & 0    & 0    & 0    &  40            & 40           & 40        & 0     & 0\\
ISR/FSR           & 0     & 0    & 0    & 0    & 0              & 0             & 0        &   0     & $4.0$\\
Mistag Rate (S)      & 0     & 0    & 0    & 0    & 0              & 0             & 0 &  $^{+25.5}_{-21.4}$     & 0     \\
Jet Energy Scale  (S)       & 0     &
  $^{+1.3}_{-2.3}$   & 
   0  & 
  $^{+7.5}_{-0.1}$   & 
  $^{+4.1}_{-4.4}$   & 
  $^{+8.2}_{-7.8}$   & 
  $^{+3.3}_{-5.5}$   & 
  0   & 
  $^{+2.1}_{-2.7}$   \\ 
\end{tabular}
\end{ruledtabular}

\end{center}
\end{table}


\begin{table}
\caption{\label{tab:d0llbb1}Systematic uncertainties on the contributions for
D0's $ZH\rightarrow \ell^+\ell^-b{\bar{b}}$ channels.
Systematic uncertainties are listed by name; see the original references for a
detailed explanation of their meaning and on how they
are derived.
Systematic uncertainties for $ZH$  shown in this table are obtained for $m_H=115$ GeV/$c^2$.
Uncertainties are relative, in percent, and are symmetric unless otherwise indicated. Shape uncertainties are
labeled with an ``S''. }
\vskip 0.8cm
{\centerline{D0: ~ $ZH \rightarrow \ell\ell b \bar{b}$ analyses relative uncertainties (\%)}}
\vskip 0.099cm
\begin{ruledtabular}
\begin{tabular}{  l  c  c  c  c  c  c  c  c }
Contribution              & Signal  & Multijet& $Z$+LF  &  $Z$\bb & $Z$\cc & Diboson & \ttbar\\ hline
Jet Energy Scale (S)      &  1.5  &         &  3.0   &  8.4   &  10   &  3.3   &  1.5  \\
Jet Energy Resolution (S) &  0.3  &         &  3.9   &  5.2   &  5.3  & 0.04   &  0.6  \\
Jet ID (S)                &  0.6  &         &  0.9   &  0.6   &  0.2  &  1.0   &  0.3  \\
Taggability (S)           &  5.1  &         &  5.2   &  7.2   &  7.3  &  6.9   &  6.5  \\
$Z p_T$ Model (S)         &       &         &  2.7   &  1.4   &  1.5  &        &       \\
HF Tagging Efficiency (S) &  4.9  &         &        &  5.0   &  9.4  &        &  5.2  \\
LF Tagging Efficiency (S) &       &         &   73   &        &       &  5.8   &       \\
$ee$ Multijet Shape (S)   &       &   53    &        &        &       &        &       \\
Multijet Normalization    &       &  20-50  &        &        &       &        &       \\
$Z$+jets Jet Angles (S)   &       &         &  1.7   &  2.7   &  2.8  &        &       \\
Alpgen MLM (S)            &       &         &  0.3   &        &       &        &       \\
Alpgen Scale (S)          &       &         &  0.4   &  0.2   &  0.2  &        &       \\
Underlying Event (S)      &       &         &  0.2   & 0.05   & 0.08  &        &       \\
Trigger (S)               & 0.4   &         &  0.03  &  0.2   &  0.3  &  0.3   &  0.4  \\
Cross Sections            & 6.2   &         &        &  20    &  20   &  7     &  10   \\
Signal Branching Fraction & 1-9   &         &        &        &       &        &       \\
Normalization             & 8     &         &  1.3   &  1.3   &  1.3  &  8     &  8    \\
PDFs                      & 0.55  &         &  1     &  2.4   &  1.1  &  0.66  &  5.9

\end{tabular}
\end{ruledtabular}
\vskip 0.8cm
{\centerline{D0: Double Tag (DT)~ $ZH \rightarrow \ell\ell b \bar{b}$ analysis relative uncertainties (\%)}}
\vskip 0.099cm
\begin{ruledtabular}
\begin{tabular}{  l  c c  c  c  c  c  c  c }  \\
Contribution             & Signal & Multijet& $Z$+LF  &  $Z$\bb & $Z$\cc & Diboson & \ttbar\\  \hline
Jet Energy Scale (S)     &  2.3  &         &  4.0   &  6.4   &  8.2   &  3.8   &  2.7  \\
Jet Energy Resolution(S) &  0.6  &         &  2.6   &  3.9   &  4.1   &  0.9   &  1.5  \\
JET ID (S)               &  0.8  &         &  0.7   &  0.3   &  0.2   &  0.7   &  0.4  \\
Taggability (S)          &  3.6  &         &  8.6   &  6.5   &  8.2   &  4.6   &  2.1  \\
$Z_{p_T}$ Model (S)      &       &         &  1.6   &  1.3   &  1.4   &        &       \\
HF Tagging Efficiency (S)&  0.8  &         &        &  1.3   &  3.2   &        &  0.7  \\
LF Tagging Efficiency (S)&       &         &   72   &        &        &  4.0   &       \\
$ee$ Multijet Shape (S)  &       &    59   &        &        &        &        &       \\
Multijet Normalization   &       &  20-50  &        &        &        &        &       \\
$Z$+jets Jet Angles (S)  &       &         &  2.0   &  1.5   &  1.5   &        &       \\
Alpgen MLM (S)           &       &         &  0.4   &        &        &        &       \\
Alpgen Scale (S)         &       &         &  0.2   &  0.2   &  0.2   &        &       \\
Underlying Event(S)      &       &         & 0.07   & 0.02   &  0.1   &        &       \\
Trigger (S)              &  0.4  &         &  0.3   &  0.2   &  0.1   &  0.2   &  0.5  \\
Cross Sections           &  6.2  &         &        & 20     & 20     & 7      & 10    \\
Signal Branching Fraction & 1-9   &         &        &        &       &        &       \\
Normalization            &  8    &         &  1.3   & 1.3    & 1.3    & 8      & 8     \\
PDFs                     & 0.55  &         &  1     & 2.4    & 1.1    & 0.66   & 5.9
\end{tabular}
\end{ruledtabular}
\end{table}


\clearpage


\begin{table}
\begin{center}
\caption{\label{tab:cdfsystww0} Systematic uncertainties on the signal and background contributions for CDF's
$H\rightarrow W^+W^-\rightarrow\ell^{\pm}\ell^{\prime \mp}$ channels with zero, one, and two or more associated
jets.  These channels are sensitive to gluon fusion production (all channels) and $WH, ZH$ and VBF production.
Systematic uncertainties are listed by name (see the original references for a detailed explanation of their
meaning and on how they are derived).  Systematic uncertainties for $H$ shown in this table are obtained for
$m_H=160$ GeV/$c^2$.  Uncertainties are relative, in percent, and are symmetric unless otherwise indicated.
The uncertainties associated with the different background and signal processed are correlated within individual
jet categories unless otherwise noted.  Boldface and italics indicate groups of uncertainties which are correlated
with each other but not the others on the line.}

\vskip 0.1cm
{\centerline{CDF: $H\rightarrow W^+W^-\rightarrow\ell^{\pm}\ell^{\prime \mp}$ with no associated jet channel relative uncertainties (\%)}}
\vskip 0.099cm
\begin{ruledtabular}
\begin{tabular}{lccccccccccc} \\
Contribution               &   $WW$     &  $WZ$         &  $ZZ$  &  $t\bar{t}$   &  DY    &  $W\gamma$   & $W$+jet & $gg\to H$ &  $WH$  &  $ZH$  &  VBF  \\ \hline
{\bf Cross Section :}      &        	&        	&        	&        &        	&        &         &           &        &        & 	 \\
Scale (Inclusive)          &        	&        	&        	&        &        	&        &         &    13.4   &        &        &       \\
Scale (1+ Jets)            &        	&        	&        	&        &        	&        &         &   -23.0   &        &        &       \\
Scale (2+ Jets)            &        	&        	&        	&        &        	&        &         &     0.0   &        &        &       \\
PDF Model                  &        	&        	&        	&        &        	&        &         &     7.6   &        &        &       \\
Total                      & {\it 6.0 } & {\it 6.0 }    & {\it 6.0 }    &   7.0  &  		& 	 & 	   & 	       & {\bf 5.0 } & {\bf 5.0 } & 10.0 \\
{\bf Acceptance :}         &        	&        	&        	&        &        	&        &         &           &        &        &       \\
Scale (jets)               & {\it 0.3 }&        	&        	&        &        	&        &         &           &        &        &       \\
PDF Model (leptons)        &            &       	&        	&        &              &  	 &         &     2.7   &        &        &       \\
PDF Model (jets)           & {\it 1.1 } &        	&        	&        &        	&        &         &     5.5   &        &        &       \\
Higher-order Diagrams      &            & {\it 10.0 }   & {\it 10.0 } 	& 10.0   &              & 10.0   &   	   &           & {\bf 10.0 } & {\bf 10.0 } & {\bf 10.0 } \\
$\MET$ Modeling            &  	 	&      	        &     	        &        & 19.5  	&	 &         &           &        &        &       \\
Conversion Modeling        &            &               &               &        &              & 10.0   &         &           &        &        &       \\
Jet Fake Rates             &        	&        	&        	&        &        	&        &         &           &        &        &       \\
(Low S/B)                  &        	&        	&        	&        &        	&        & 22.0    &           &        &        &       \\
(High S/B)                 &        	&        	&        	&        &        	&        & 26.0    &           &        &        &       \\
Jet Energy Scale           & {\it 2.6 } & {\it 6.1 }    & {\it 3.4 }    & {\it 26.0 } & {\it 17.5 } & {\it 3.1 } &   & {\it 5.0 } & {\it 10.5 } & {\it 5.0 } & {\it 11.5 } \\
Lepton ID Efficiencies     & {\it 3.8 } & {\it 3.8 } 	& {\it 3.8 } 	&{\it 3.8 }&{\it 3.8 }&  	 &         & {\it 3.8 } &  {\it 3.8 } &  {\it 3.8 } &  {\it 3.8 } \\
Trigger Efficiencies       & {\it 2.0 } & {\it 2.0 } 	& {\it 2.0 } 	&{\it 2.0 }&{\it 2.0 }&  	 &         & {\it 2.0 } &  {\it 2.0 } &  {\it 2.0 } &  {\it 2.0 } \\
{\bf Luminosity}           & {\it 3.8 } & {\it 3.8 } 	& {\it 3.8 } 	&{\it 3.8 }&{\it 3.8 }&  	 &         & {\it 3.8 } &  {\it 3.8 } &  {\it 3.8 } &  {\it 3.8 } \\
{\bf Luminosity Monitor}   & {\it 4.4 } & {\it 4.4 } 	& {\it 4.4 } 	&{\it 4.4 }&{\it 4.4 }&  	 &         & {\it 4.4 } &  {\it 4.4 } &  {\it 4.4 } &  {\it 4.4 } \\
\end{tabular}
\end{ruledtabular}

\vskip 0.3cm
{\centerline{CDF: $H\rightarrow W^+W^-\rightarrow\ell^{\pm}\ell^{\prime \mp}$ with one associated jet channel relative uncertainties (\%)}}
\vskip 0.099cm
\begin{ruledtabular}
\begin{tabular}{lccccccccccc} \\
Contribution               &   $WW$     &  $WZ$         &  $ZZ$  &  $t\bar{t}$   &  DY     & $W\gamma$   & $W$+jet &$gg \to H$&  $WH$  &  $ZH$  &  VBF \\ \hline
{\bf Cross Section :}      &        	&        	&        	&        &        	&        &         &           &        &        & 	 \\
Scale (Inclusive)          &        	&        	&        	&        &        	&        &         &     0.0   &        &        &       \\
Scale (1+ Jets)            &        	&        	&        	&        &        	&        &         &    35.0   &        &        &       \\
Scale (2+ Jets)            &        	&        	&        	&        &        	&        &         &   -12.7   &        &        &       \\
PDF Model                  &        	&        	&        	&        &        	&        &         &    17.3   &        &        &       \\
Total                      & {\it 6.0 } & {\it 6.0 }    & {\it 6.0 }    &   7.0  &  		& 	 & 	   & 	       & {\bf 5.0 } & {\bf 5.0 } & 10.0 \\
{\bf Acceptance :}         &        	&        	&        	&        &        	&        &         &        &        &        &        \\
Scale (jets)               &{\it -4.0 } &        	&        	&        &        	&        &         &       &        &        &        \\
PDF Model (leptons)        &            &               &               &        &              &  	 &	   &{\it 3.6 }&     &        &        \\
PDF Model (jets)           &{\it  4.7 }&        	&        	&        &        	&        &         & -6.3  &        &        &        \\
Higher-order Diagrams      &            & {\it 10.0 }  & {\it 10.0 }  & 10.0  &  	        & 10.0  &   	   &        &{\bf 10.0 }&{\bf 10.0 }&{\bf 10.0 }\\
$\MET$ Modeling            & 	 	&   	        &       	&        & 20.0  	& 	 &         &        &        &        &        \\
Conversion Modeling       &            &               &               &        &              & 10.0  &         &        &        &        &        \\
Jet Fake Rates             &        	&        	&        	&        &        	&        &         &        &        &        &        \\
(Low S/B)                  &        	&        	&        	&        &        	&        & 23.0   &        &        &        &        \\
(High S/B)                 &        	&        	&        	&        &        	&        & 29.0   &        &        &        &        \\
Jet Energy Scale &  {\it -5.5 }  & {\it -1.0 } &   {\it -4.3 }  &   {\it -13.0 } & {\it -6.5 } &   {\it -9.5 } &   & {\it -4.0 } & {\it -8.5 } & {\it -7.0 } & {\it -6.5 } \\
Lepton ID Efficiencies     & {\it 3.8 } & {\it 3.8 } 	& {\it 3.8 } 	&{\it 3.8 }&{\it 3.8 }& 	 &      &{\it 3.8 }&{\it 3.8 }&{\it 3.8 }&{\it 3.8 }\\
Trigger Efficiencies       & {\it 2.0 } & {\it 2.0 } 	& {\it 2.0 } 	&{\it 2.0 }&{\it 2.0 }& 	 &      &{\it 2.0 }&{\it 2.0 }&{\it 2.0 }&{\it 2.0 }\\
{\bf Luminosity}           & {\it 3.8 } & {\it 3.8 } 	& {\it 3.8 } 	&{\it 3.8 }&{\it 3.8 }& 	 &      &{\it 3.8 }&{\it 3.8 }&{\it 3.8 }&{\it 3.8 }\\
{\bf Luminosity Monitor}   & {\it 4.4 } & {\it 4.4 } 	& {\it 4.4 } 	&{\it 4.4 }&{\it 4.4 }& 	 &      &{\it 4.4 }&{\it 4.4 }&{\it 4.4 }&{\it 4.4 }\\
\end{tabular}
\end{ruledtabular}

\end{center}
\end{table}

\begin{table}
\begin{center}
\vskip 0.3cm
{\centerline{CDF: $H\rightarrow W^+W^-\rightarrow\ell^{\pm}\ell^{\prime \mp}$ with two or more associated jets channel relative uncertainties (\%)}}
\vskip 0.0999cm
\begin{ruledtabular}
\begin{tabular}{lccccccccccc} \\
Contribution               &  $WW$    &  $WZ$         &  $ZZ$   &  $t\bar{t}$  &  DY   &  $W\gamma$    & $W$+jet &$gg\to H$&  $WH$ &  $ZH$  &  VBF    \\ \hline
{\bf Cross Section :}      &        	&        	&        	&        &        	&        &         &           &        &        & 	 \\
Scale (Inclusive)          &        	&        	&        	&        &        	&        &         &     0.0   &        &        &       \\
Scale (1+ Jets)            &        	&        	&        	&        &        	&        &         &     0.0   &        &        &       \\
Scale (2+ Jets)            &        	&        	&        	&        &        	&        &         &    33.0   &        &        &       \\
PDF Model                  &        	&        	&        	&        &        	&        &         &    29.7   &        &        &       \\
Total                      & {\it 6.0 } & {\it 6.0 }    & {\it 6.0 }    &   7.0  &  		& 	 & 	   & 	       & {\bf 5.0 } & {\bf 5.0 } & 10.0 \\
{\bf Acceptance :}         &        	&        	&        	&        &        	&        &         &        &        &        &        \\
Scale (jets)               & {\it -8.2 }&        	&        	&        &        	&        &         &       &        &        &        	\\
PDF Model (leptons)        & 		 &  		&  		&	 &	        &  	 &      &{\it 4.8 }&	     &        &         \\
PDF Model (jets)           & {\it 4.2 }&        	&        	&        &        	&        &         & -12.3  &        &        &        	\\
Higher-order Diagrams      &  		& {\it 10.0 }  & {\it 10.0 }  & 10.0  &              & 10.0  & 	   &        &{\bf 10.0 }&{\bf 10.0 }&{\bf 10.0 }\\
$\MET$ Modeling            &  		&   	        &     	        &        & 25.5  	& 	 &         &        &        &        &     	\\
Conversion Modeling      &            &               &               &        &              & 10.0  &         &        &        &        &     	\\
Jet Fake Rates             &        	&        	&        	&        &        	&        & 28.0   &        &        &        &        	\\
Jet Energy Scale& {\it -14.8 } & {\it -12.9 } & {\it -12.1 }  & {\it -1.7 }  & {\it -29.2 } & {\it -22.0 } &  & {\it -17.0 } & {\it -4.0 } & {\it -2.3 } & {\it -4.0 } \\
$b$-tag Veto               &        	&        	&        	&  3.8  &        	&        &         &        &        &        &        	\\
Lepton ID Efficiencies     & {\it 3.8 } &  {\it 3.8 } &  {\it 3.8 }  &{\it 3.8 }&{\it 3.8 }&   	 &      &{\it 3.8 }&{\it 3.8 }&{\it 3.8 }&{\it 3.8 }\\
Trigger Efficiencies       & {\it 2.0 } &  {\it 2.0 } &  {\it 2.0 }  &{\it 2.0 }&{\it 2.0 }&  	 &      &{\it 2.0 }&{\it 2.0 }&{\it 2.0 }&{\it 2.0 }\\
{\bf Luminosity}           & {\it 3.8 } &  {\it 3.8 } &  {\it 3.8 }  &{\it 3.8 }&{\it 3.8 }&  	 &      &{\it 3.8 }&{\it 3.8 }&{\it 3.8 }&{\it 3.8 }\\
{\bf Luminosity Monitor}   & {\it 4.4 } &  {\it 4.4 } &  {\it 4.4 }  &{\it 4.4 }&{\it 4.4 }&  	 &      &{\it 4.4 }&{\it 4.4 }&{\it 4.4 }&{\it 4.4 }\\
\end{tabular}
\end{ruledtabular}
\end{center}
\end{table}

%
%

\begin{table}
\begin{center}
\caption{\label{tab:cdfsystww4} Systematic uncertainties on the signal and background contributions for CDF's low-$M_{\ell\ell}$
$H\rightarrow W^+W^-\rightarrow\ell^{\pm}\ell^{\prime \mp}$ channel with zero or one associated jets.  This channel is sensitive
to only gluon fusion production.  Systematic uncertainties are listed by name (see the original references for a detailed
explanation of their meaning and on how they are derived).  Systematic uncertainties for $H$ shown in this table are obtained
for $m_H=160$ GeV/$c^2$.  Uncertainties are relative, in percent, and are symmetric unless otherwise indicated.  The uncertainties
associated with the different background and signal processed are correlated within individual categories unless otherwise noted.
In these special cases, the correlated uncertainties are shown in either italics or bold face text.}
\vskip 0.1cm
{\centerline{CDF: low $M_{\ell\ell}$ $H\rightarrow W^+W^-\rightarrow\ell^{\pm}\ell^{\prime \mp}$ with zero or one associated jets channel relative uncertainties (\%)}}
\vskip 0.099cm
\begin{ruledtabular}
\begin{tabular}{lccccccccccc} \\
Contribution            & $WW$       & $WZ$       & $ZZ$       & $t\bar{t}$ & DY      & $W\gamma$  & $W$+jet(s) & $gg\to H$&  $WH$ &  $ZH$  &  VBF    \\ \hline
{\bf Cross Section :}      &        	&        	&        	&        &        	&        &         &           &        &        & 	 \\
Scale (Inclusive)          &        	&        	&        	&        &        	&        &         &     8.1   &        &        &       \\
Scale (1+ Jets)            &        	&        	&        	&        &        	&        &         &     0.0   &        &        &       \\
Scale (2+ Jets)            &        	&        	&        	&        &        	&        &         &    -5.1   &        &        &       \\
PDF Model                  &        	&        	&        	&        &        	&        &         &    10.5   &        &        &       \\
Total                      & {\it 6.0 } & {\it 6.0 }    & {\it 6.0 }    &   7.0  &  5.0		& 	 & 	   & 	       & {\bf 5.0 } & {\bf 5.0 } & 10.0 \\
{\bf Acceptance :}      &            &            &            &            &         &            &            &   &  &  &        \\
Scale (jets)            & {\it -0.4} &            &            &            &         &            &            &   &  &  &        \\
PDF Model (leptons)     & 	     &            &            &            &         &            &         & {\it 1.0 } &  &  & \\
PDF Model (jets)        &{\it 1.6 } &            &            &            &         &            &            & 2.1 &  &  &       \\
Higher-order Diagrams      &            & {\it 10.0 }  & {\it 10.0 }  & 10.0  & 10.0  	        &   &   	   &        &{\bf 10.0 }&{\bf 10.0 }&{\bf 10.0 }\\
Jet Energy Scale           & {\it 1.1 }  & {\it 2.2 } & {\it 2.0 } & {\it 13.5 } & {\it 6.4 } & {\it 1.3 } &            & {\it 2.4 } & {\it 9.2 }  & {\it 6.5 }  & {\it 7.8 } \\
Conversion Modeling      &            &            &            &            &         & 10.0        &            &     &  &  &      \\
Boson Radiation Model      &            &            &            &            & 25.0    &         &            &     &  &  &      \\
Jet Fake Rates          &            &            &            &            &         &            & 13.5        &           \\
Lepton ID Efficiencies     & {\it 3.8 } &  {\it 3.8 } &  {\it 3.8 } & {\it 3.8 } & {\it 3.8 } &  		  &         & {\it 3.8 } & {\it 3.8 } & {\it 3.8 } & {\it 3.8 } \\
Trigger Efficiencies       & {\it 2.0 } &  {\it 2.0 } &  {\it 2.0 } & {\it 2.0 } & {\it 2.0 } &  		  &         & {\it 2.0 } & {\it 2.0 } & {\it 2.0 } & {\it 2.0 } \\
{\bf Luminosity}           & {\it 3.8 } &  {\it 3.8 } &  {\it 3.8 } & {\it 3.8 } &   {\it 3.8 }    &          &        & {\it 3.8 } & {\it 3.8 } & {\it 3.8 } & {\it 3.8 }  \\
{\bf Luminosity Monitor}   & {\it 4.4 } &  {\it 4.4 } &  {\it 4.4 } & {\it 4.4 } &   {\it 4.4 }    &          &        & {\it 4.4 } & {\it 4.4 } & {\it 4.4 } & {\it 4.4 }  \\
\end{tabular}
\end{ruledtabular}
\end{center}
\end{table}


\begin{table}
\begin{center}
\caption{\label{tab:cdfsystww5} Systematic uncertainties on the signal and background contributions for CDF's
$H\rightarrow W^+W^-\rightarrow e^{\pm} \tau^{\mp}$ and $H\rightarrow W^+W^-\rightarrow \mu^{\pm} \tau^{\mp}$
channels.  These channels are sensitive to gluon fusion production, $WH, ZH$ and VBF production.  Systematic
uncertainties are listed by name (see the original references for a detailed explanation of their meaning
and on how they are derived).  Systematic uncertainties for $H$ shown in this table are obtained for
$m_H=160$ GeV/$c^2$.  Uncertainties are relative, in percent, and are symmetric unless otherwise indicated.
The uncertainties associated with the different background and signal processed are correlated within individual
categories unless otherwise noted.  In these special cases, the correlated uncertainties are shown in either
italics or bold face text.}
\vskip 0.1cm
{\centerline{CDF: $H\rightarrow W^+W^-\rightarrow e^{\pm} \tau^{\mp}$ channel relative uncertainties ( )}}
\vskip 0.099cm
\begin{ruledtabular}
\begin{tabular}{lccccccccccccccc} \\
Contribution                 & $WW$  & $WZ$ & $ZZ$ & $t\bar{t}$  & $Z\rightarrow\tau\tau$  & $Z\rightarrow\ell\ell$  & $W$+jet  & $W\gamma$  & $gg\to H$  & $WH$ & $ZH$ & VBF  \\ \hline
Cross section                & 6.0   & 6.0  & 6.0  & 10.0 &  5.0 &  5.0 &       &      &  10.3  &   5   &   5   &  10  \\
Measured W cross-section     &       &      &      &      &      &      & 12    &      &        &       &       &      \\
PDF Model                    & 1.6   & 2.3  & 3.2  & 2.3  &  2.7 &  4.6 & 2.2   & 3.1  &   2.5  & 2.0   &  1.9  & 1.8  \\
Higher order diagrams        & 10    & 10   & 10   & 10   &  10  &  10  &       & 10   &        & 10    &  10   & 10   \\
Conversion modeling          &       &      &      &      &      &      &       & 10   &        &       &       &      \\
Trigger Efficiency           & 0.5   & 0.6  & 0.6  & 0.6  & 0.7  & 0.5  & 0.6   & 0.6  &   0.5  & 0.5   &  0.6  & 0.5  \\
Lepton ID Efficiency         & 0.4   & 0.5  & 0.5  & 0.4  & 0.4  &  0.4 & 0.5   & 0.4  &   0.4  & 0.4   &  0.4  & 0.4  \\
$\tau$ ID Efficiency         & 1.0   & 1.3  & 1.9  & 1.3  & 2.1  &      &       & 0.3  &   2.8  &  1.6  &  1.7  & 2.8  \\
Jet into $\tau$ Fake rate    & 5.8   & 4.8  & 2.0  & 5.1  &      &  0.1 & 8.8   &      &        &  4.2  &  4.0  & 0.4  \\
Lepton into $\tau$ Fake rate & 0.2   & 0.1  & 0.6  & 0.2  &      &  2.3 &       & 2.1  &  0.15  & 0.06  &  0.15 & 0.11 \\
W+jet scale                  &       &      &      &      &      &      & 1.6   &      &        &       &       &      \\
MC Run dependence            & 2.6   & 2.6  & 2.6  &      &      &      & 2.6   &      &        &       &       &      \\
Luminosity                   & 3.8   & 3.8  & 3.8  & 3.8  & 3.8  & 3.8  & 3.8   & 3.8  &  3.8   & 3.8   &  3.8  & 3.8  \\
Luminosity Monitor           & 4.4   & 4.4  & 4.4  & 4.4  & 4.4  & 4.4  & 4.4   & 4.4  &  4.4   & 4.4   &  4.4  & 4.4  \\
\end{tabular}
\end{ruledtabular}

\vskip 0.3cm
{\centerline{CDF: $H\rightarrow W^+W^-\rightarrow \mu^{\pm} \tau^{\mp}$ channel relative uncertainties (\%)}}
\vskip 0.099cm
\begin{ruledtabular}
\begin{tabular}{lcccccccccccccc} \\
    Contribution                 & $WW$  & $WZ$ & $ZZ$ & $t\bar{t}$  & $Z\rightarrow\tau\tau$  & $Z\rightarrow\ell\ell$  & $W$+jet  & $W\gamma$  & $gg\to H$  & $WH$ & $ZH$ & VBF  \\ \hline
    Cross section                & 6.0  & 6.0  & 6.0  & 10.0 & 5.0 &  5.0 &     &      & 10.4  &   5   &   5   &  10  \\
    Measured W cross-section     &      &      &      &      &     &      & 12  &      &       &       &       &      \\
    PDF Model                    & 1.5  & 2.1  & 2.9  & 2.1  & 2.5 & 4.3  & 2.0 & 2.9  &  2.6  & 2.2   &  2.0  & 2.2  \\
    Higher order diagrams        & 10   & 10   & 10   & 10   &     &      &     & 11   &       & 10    &  10   & 10   \\
    Trigger Efficiency           & 1.3  & 0.7  & 0.7  & 1.1  & 0.9 & 1.3  & 1.0 & 1.0  &  1.3  & 1.3   &  1.2  & 1.3  \\
    Lepton ID Efficiency         & 1.1  & 1.4  & 1.4  & 1.1  & 1.2 & 1.1  & 1.4 & 1.3  &  1.0  & 1.0   &  1.0  & 1.0  \\
    $\tau$ ID Efficiency         & 1.0  & 1.2  & 1.4  & 1.6  & 1.9 &      &     &      &  2.9  &  1.6  &  1.7  & 2.8  \\
    Jet into $\tau$ Fake rate    & 5.8  & 5.0  & 4.4  & 4.4  &     & 0.2  & 8.8 &      &       &  4.5  &  4.2  & 0.4  \\
    Lepton into $\tau$ Fake rate & 0.06 & 0.05 & 0.09 & 0.04 &     & 1.9  &     & 1.2  & 0.04  & 0.02  &  0.02 & 0.04 \\
    W+jet scale                  &      &      &      &      &     &      & 1.4 &      &       &       &       &      \\
    MC Run dependence            & 3.0  & 3.0  & 3.0  &      &     &      & 3.0 &      &       &       &       &      \\
    Luminosity                   & 3.8  & 3.8  & 3.8  & 3.8  & 3.8 & 3.8  & 3.8 & 3.8  & 3.8   & 3.8   &  3.8  & 3.8  \\
    Luminosity Monitor           & 4.4  & 4.4  & 4.4  & 4.4  & 4.4 & 4.4  & 4.4 & 4.4  & 4.4   & 4.4   &  4.4  & 4.4  \\
\end{tabular}
\end{ruledtabular}
\end{center}
\end{table}


\begin{table}
\begin{center}
\caption{\label{tab:cdfsystwww} Systematic uncertainties on the signal and background contributions for
CDF's $WH\rightarrow WWW \rightarrow\ell^{\pm}\ell^{\prime \pm}$ channel with one or more associated
jets and $WH\rightarrow WWW \rightarrow \ell^{\pm}\ell^{\prime \pm} \ell^{\prime \prime \mp}$ channel.
These channels are sensitive to only $WH$ and $ZH$ production.  Systematic uncertainties are listed
by name (see the original references for a detailed explanation of their meaning and on how they are
derived).  Systematic uncertainties for $H$ shown in this table are obtained for $m_H=160$ GeV/$c^2$.
Uncertainties are relative, in percent, and are symmetric unless otherwise indicated.  The uncertainties
associated with the different background and signal processed are correlated within individual categories
unless otherwise noted.  In these special cases, the correlated uncertainties are shown in either italics
or bold face text.}
\vskip 0.1cm
{\centerline{CDF: $WH \rightarrow WWW \rightarrow\ell^{\pm}\ell^{\prime\pm}$ channel relative uncertainties (\%)}}
\vskip 0.099cm
\begin{ruledtabular}
\begin{tabular}{lccccccccc} \\
Contribution               & $WW$          &   $WZ$       &  $ZZ$        & $t\bar{t}$   &  DY          & $W\gamma$    & $W$+jet &  $WH$        &  $ZH$        \\ \hline
{\bf Cross Section}        &   {\it 6.0 } &  {\it 6.0 } &  {\it 6.0 } &        7.0  &        5.0  &              &     	 &  {\bf 5.0 } &  {\bf 5.0 } \\
Scale (Acceptance)         &  {\it -6.1 } &              &              &              &              &              &         &              &              \\
PDF Model (Acceptance)     &   {\it 5.7 } &  		  &  		 &  		&  	       &  	      &         &  	       & 	      \\
Higher-order Diagrams      &   		   & {\it 10.0 } & {\it 10.0 } &       10.0  &       10.0  &       10.0  &         & {\bf 10.0 } & {\bf 10.0 } \\
Conversion Modeling        &               &              &              &              &              &       10.0  &         &              &              \\
Jet Fake Rates             &               &              &              &              &              &              &  38.5  &              &              \\
Jet Energy Scale           & {\it -14.0 } & {\it -3.9 } & {\it -2.8 } & {\it -0.6 } & {\it -7.7 } & {\it -7.6 } &         & {\it -1.0 } & {\it -0.7 } \\
Charge Mismeasurement Rate &  {\it 40.0 } &              &              &            & {\it 40.0 } &              &         &              &              \\
Lepton ID Efficiencies     &   {\it 3.8 } &  {\it 3.8 } &  {\it 3.8 } & {\it 3.8 }  &  {\it 3.8 } &  	      &         &  {\it 3.8 } &  {\it 3.8 } \\
Trigger Efficiencies       &   {\it 2.0 } &  {\it 2.0 } &  {\it 2.0 } & {\it 2.0 }  &  {\it 2.0 } &  	      &         &  {\it 2.0 } &  {\it 2.0 } \\
{\bf Luminosity}           &   {\it 3.8 } &  {\it 3.8 } &  {\it 3.8 } & {\it 3.8 }  &  {\it 3.8 } &   	      &         &  {\it 3.8 } &  {\it 3.8 } \\
{\bf Luminosity Monitor}   &   {\it 4.4 } &  {\it 4.4 } &  {\it 4.4 } & {\it 4.4 }  &  {\it 4.4 } &   	      &         &  {\it 4.4 } &  {\it 4.4 } \\
\end{tabular}
\end{ruledtabular}

\vskip 0.3cm
{\centerline{CDF: $WH\rightarrow WWW \rightarrow \ell^{\pm}\ell^{\prime \pm} \ell^{\prime \prime \mp}$ channel relative uncertainties (\%)}}
\vskip 0.0999cm
\begin{ruledtabular}
\begin{tabular}{lccccccc} \\
Contribution                & $WZ$        & $ZZ$        & $Z\gamma$     & $t\bar{t}$  & Fakes       & $WH$         & $ZH$           \\ \hline
{\bf Cross Section}         & {\it 6.0 } &  {\it 6.0 } &    10.0     &       7.0  &             & {\bf 5.0 }  &   {\bf 5.0 }  \\
Higher-order Diagrams       & {\it 10.0 }& {\it 10.0 } &    15.0     &      10.0  &             & {\bf 10.0 } & {\bf 10.0 }    \\
Jet Energy Scale            &             &              &  {\it -2.7 }&             &             &              &                  \\
Jet Fake Rates              &             &              &              &             & 25.6       &              &                 \\
$b$-Jet Fake Rates          &             &              &              &  27.3      &             &              &                 \\
MC Run Dependence           &             &              &   5.0       &	      &             &  		   &                 \\
Lepton ID Efficiencies      & {\it 5.0 } &  {\it 5.0 } &              & {\it 5.0 } &             & {\it 5.0 }  &   {\it 5.0 }    \\
Trigger Efficiencies        & {\it 2.0 } &  {\it 2.0 } &              & {\it 2.0 } &             & {\it 2.0 }  &   {\it 2.0 }    \\
{\bf Luminosity}            & {\it 3.8 } &  {\it 3.8 } &              & {\it 3.8 } &             &  	 {\it 3.8 }&   {\it 3.8 }    \\
{\bf Luminosity Monitor}    & {\it 4.4 } &  {\it 4.4 } &              & {\it 4.4 } &             &  	 {\it 4.4 }&   {\it 4.4 }    \\
\end{tabular}
\end{ruledtabular}

\end{center}
\end{table}

\begin{table}
\begin{center}
\caption{\label{tab:cdfsystzww} Systematic uncertainties on the signal and background contributions for
CDF's $ZH\rightarrow ZWW \rightarrow \ell^{\pm}\ell^{\mp} \ell^{\prime \pm}$ channels with 1 jet and 2
or more jets.  These channels are sensitive to only $WH$ and $ZH$ production.  Systematic uncertainties
are listed by name (see the original references for a detailed explanation of their meaning and on how
they are derived).  Systematic uncertainties for $H$ shown in this table are obtained for $m_H=160$
GeV/$c^2$.  Uncertainties are relative, in percent, and are symmetric unless otherwise indicated.  The
uncertainties associated with the different background and signal processed are correlated within
individual categories unless otherwise noted.  In these special cases, the correlated uncertainties are
shown in either italics or bold face text.}
\vskip 0.1cm
{\centerline{CDF: $ZH\rightarrow ZWW \rightarrow \ell^{\pm}\ell^{\mp} \ell^{\prime \pm}$ with one associated jet channel relative uncertainties (\%)}}
\vskip 0.0999cm
\begin{ruledtabular}
\begin{tabular}{lccccccc} \\
Contribution                & $WZ$        & $ZZ$         & $Z\gamma$    & $t\bar{t}$  & Fakes       & $WH$         & $ZH$           \\ \hline
{\bf Cross Section}         & {\it 6.0 } &  {\it 6.0 } &   10.0      &       7.0  &             & {\bf 5.0 }  &   {\bf 5.0 }  \\
Higher-order Diagrams       & {\it 10.0 }& {\it 10.0 } &    15.0     &      10.0  &             & {\bf 10.0 } & {\bf 10.0 }    \\
Jet Energy Scale            & {\it -7.6 }& {\it -2.3 } &  {\it -5.3 }& {\it  9.4 }&             &  {\it -9.0 }& {\it  8.1 }    \\
Jet Fake Rates              &             &              &              &             & 24.8       &              &                 \\
$b$-Jet Fake Rates          &             &              &              &  42.0      &             &              &                 \\
MC Run Dependence           &             &              &   5.0       &	      &             &  		   &                 \\
Lepton ID Efficiencies      & {\it 5.0 } &  {\it 5.0 } &              & {\it 5.0 } &             & {\it 5.0 }  &   {\it 5.0 }   \\
Trigger Efficiencies        & {\it 2.0 } &  {\it 2.0 } &              & {\it 2.0 } &             &   {\it 2.0 }&   {\it 2.0 }   \\
{\bf Luminosity}            & {\it 3.8 } &  {\it 3.8 } &              & {\it 3.8 } &             &  	 {\it 3.8 }&   {\it 3.8 }    \\
{\bf Luminosity Monitor}    & {\it 4.4 } &  {\it 4.4 } &              & {\it 4.4 } &             &  	 {\it 4.4 }&   {\it 4.4\%}    \\
\end{tabular}
\end{ruledtabular}

\vskip 0.3cm
{\centerline{CDF: $ZH\rightarrow ZWW \rightarrow \ell^{\pm}\ell^{\mp} \ell^{\prime \pm}$ with two or more associated jets channel relative uncertainties (\%)}}
\vskip 0.0999cm
\begin{ruledtabular}
\begin{tabular}{lccccccc} \\
Contribution                & $WZ$        & $ZZ$         & $Z\gamma$    & $t\bar{t}$  & Fakes       & $WH$         & $ZH$           \\ \hline
{\bf Cross Section}         & {\it 6.0 } &  {\it 6.0 } &   10.0      &       7.0  &             & {\bf 5.0 }  &   {\bf 5.0 }  \\
Higher-order Diagrams       & {\it 10.0 }& {\it 10.0 } &    15.0     &      10.0  &             & {\bf 10.0 } & {\bf 10.0 }    \\
Jet Energy Scale            &{\it -17.8 }& {\it -13.1 }& {\it -18.2 }& {\it -3.6 }&             & {\it -15.4 }& {\it -4.9 }    \\
Jet Fake Rates              &             &              &              &             & 25.6       &              &                 \\
$b$-Jet Fake Rates          &             &              &              &  22.2      &             &              &                 \\
MC Run Dependence           &             &              &   5.0       &	      &             &  		   &                 \\
Lepton ID Efficiencies      & {\it 5.0 } &  {\it 5.0 } &              & {\it 5.0 } &             & {\it 5.0 }  &   {\it 5.0 }   \\
Trigger Efficiencies        & {\it 2.0 } &  {\it 2.0 } &              & {\it 2.0 } &             &   {\it 2.0 }&   {\it 2.0 }   \\
{\bf Luminosity}            & {\it 3.8 } &  {\it 3.8 } &              & {\it 3.8 } &             &  	 {\it 3.8 }&   {\it 3.8 }    \\
{\bf Luminosity Monitor}    & {\it 4.4 } &  {\it 4.4 } &              & {\it 4.4 } &             &  	 {\it 4.4 }&   {\it 4.4 }    \\
\end{tabular}
\end{ruledtabular}

\end{center}
\end{table}


\begin{table}
\begin{center}
\caption{\label{tab:d0systww} Systematic uncertainties on the signal and background contributions for D0's
$H\rightarrow WW \rightarrow\ell^{\pm}\ell^{\prime \mp}$ channels.  Systematic uncertainties are listed by
name; see the original references for a detailed explanation of their meaning and on how they are derived.
Shape uncertainties are labeled with the ``S'' designation. Systematic uncertainties shown in this table are obtained for the $m_H=165$ GeV/c$^2$ Higgs selection.
Uncertainties are relative, in percent, and are symmetric unless otherwise indicated.}
\vskip 0.1cm
{\centerline{D0: $H\rightarrow WW \rightarrow\ell^{\pm}\ell^{\prime \mp}$ channels relative uncertainties (\%)}}
\vskip 0.099cm
\begin{ruledtabular}
%
\begin{tabular}{ l  c  c  c  c  c  c  c  c  c}  \\
Contribution & Diboson & ~~$Z/\gamma^* \rightarrow \ell\ell$~~&$~~W+jet/\gamma$~~ &~~~~$t\bar{t}~~~~$    & ~~Multijet~~  & ~~~~$gg \rightarrow H$~~~~ & ~~~~$qq \rightarrow qqH$~~~~ & ~~~~$VH$~~~~     \\
\hline
Luminosity/Normalization    &  6    &   6   & 6   & 6    & 20-30   &   6           & 6   & 6      \\
Jet-bin Normalization:      & --    & 2-15  & 20     & --    & --      & --    & -- & -- \\
Cross Section (scale/PDF)   &  7    &   6   & 6   & 10   & --   &   13-33/7.6-30  & 4.9   & 6.1    \\
Signal Branching Fraction   & --    & --    & --  & --   & --   &  0-7.3        & 0-7.3 &  0-7.3 \\
PDF                         &2.5    &   2.5 & 2.5 & 2.5  & --   &   --          & --  & --    \\
EM Identification           &2.5    &   2.5 & 2.5 & 2.5  & --   &   2.5         & 2.5 & 2.5        \\
EM Resolution (each)        & 2     & 2     & 2   & 2    & --   & 2  & 2   &2    \\
Muon Identification         &  4    &  4    & 4   & 4    & --   &   4           &4      &4      \\
Vertex Confirmation (S)     &2-6    &  1-7  & 1-6 & 1-8  & --   &   1-8         &1-8  &1-8\\
Jet identification (S)      &2-5    &  2-5  & 2-5 & 2-5  & --   &  2-5          &  2-5 &  2-5  \\
Jet Energy Scale (S)        &2-3    &  1-4  & 1-8 & 1-4  & --   &   1-10        &   1-10 &   1-10  \\
Jet Energy Resolution(S)    &1-4    &  1-4  & 1-12 & 1-3 & --   &   1-12        &   1-12  &   1-12  \\
B-tagging                   &10     &   10  & 10   & 5   & --   &   10           &   10  &   10  \\
\end{tabular}
\end{ruledtabular}

\end{center}
\end{table}


\begin{table}
\begin{center}
\caption{\label{tab:d0systwwtau} Systematic uncertainties on the signal and background contributions for D0's
$H\rightarrow W^+ W^- \rightarrow \mu\nu \tau_{\rm{had}}\nu $ channel.  Systematic uncertainties are listed by
name; see the original references for a detailed explanation of their meaning and on how they are derived.
Shape uncertainties are labeled with the shape designation (S). Systematic uncertainties shown in this table are obtained for the $m_H=165$ GeV/c$^2$ Higgs selection.
Uncertainties are relative, in percent, and are symmetric unless otherwise indicated.}
\vskip 0.1cm
{\centerline{D0: $H\rightarrow W^+ W^- \rightarrow \mu\nu \tau_{\rm{had}}\nu $ channel relative uncertainties (\%)}}
\vskip 0.099cm
\begin{ruledtabular}
%
\begin{tabular}{ l  c  c  c  c  c  c  c  c  c}  \\
Contribution       & Diboson    & ~~$Z/\gamma^* \rightarrow \ell\ell$~~ & ~~$W$+$\rm{jets}$~~ &
~~~~$t\bar{t}$~~~~ & ~~Multijet~~ & ~~~~$gg \rightarrow H$~~~~ &
~~~~$qq \rightarrow qqH$~~~~ & ~~~~$VH$~~~~\\
\hline
Luminosity ($\sigma_{\rm{inel}}(p\bar{p})$) &4.6   & 4.6   & -  & 4.6    &    -   &   4.6 &   4.6 &   4.6    \\
Luminosity Monitor    &  4.1  & 4.1           & -          & 4.1         &-        &   4.1 &   4.1 &   4.1    \\
Trigger     &  5.0            &5.0             & -          & 5.0        &-        &   5.0  &   5.0 &   5.0  \\
Lepton ID    &  3.7   &3.7           & -           & 3.7              &-        &   3.7 &   3.7 &   3.7    \\
EM veto        &  5.0         &-         & -         & 5.0        &-        &   5.0  &   5.0 &   5.0   \\
Tau Energy Scale (S)    &  1.0        &1.1          & -        & $<$1     &-        &   $<$1 &   $<$1 &   $<$1   \\
Jet Energy Scale (S)    &  8.0     &  $<$1    & -       & 1.8     &-        &   2.5 &   2.5 &   2.5     \\
Jet identification (S)  &  $<$1       & $<$1       & -         & 7.5        &-         &   5.0 &   5.0 &   5.0    \\
Multijet  (S)  ~~~~~    &  -      & -    & -     & -       &20-50    &   -    &   - &   -  \\
Cross Section (scale/PDF)     &  7.0       & 4.0      & -      & 10       & -        &   7/8 & 4.9 & 6.1    \\
Signal Branching Fraction & -  & -        & -      &-         &-         &0-7.3  &0-7.3 &0-7.3 \\
Modeling    ~~~~~       &  1.0     &-      & 10    & -      &-        &   3.0  &   3.0 &   3.0   \\

\end{tabular}
\end{ruledtabular}
\end{center}
\end{table}

\begin{table}
\begin{center}
\caption{\label{tab:d0systwww} Systematic uncertainties on the signal and background contributions for D0's
$WH \rightarrow WWW \rightarrow\ell^{\prime \pm}\ell^{\prime \pm}$ channel.  Systematic uncertainties are
listed by name; see the original references for a detailed explanation of their meaning and on how they are
derived. Shape uncertainties are labeled with the ``S'' designation.  Systematic uncertainties for signal
shown in this table are obtained for $m_H=165$ GeV/$c^2$.  Uncertainties are relative, in percent, and are
symmetric unless otherwise indicated.}
\vskip 0.1cm
{\centerline{D0: $VH \rightarrow\ell^{\pm}\ell^{\prime\pm} + X $ Run~IIa channel relative uncertainties (\%)}}
\vskip 0.099cm
\begin{ruledtabular}
\begin{tabular}{ l  c  c  c  c c } \\
Contribution		&WZ/ZZ	&W+jet	&ChargeFlip	&Multijet 	& $VH \rightarrow \ell\ell X$	\\
\hline
Cross section		& 7		& 6		& 0		& 0			& 5		\\
Normalization		& 4.7		& 4.7		& 0		& 0			& 4.7		\\
Signal Branching Fraction     &-         &-           & -     &-          &0-7.3 \\
JetID/JES		  	& 2		& 2		& 0		& 0			& 2		\\
Jet-Lepton Fake	& 0		& 17-26	& 0		& 0			& 0		\\
Instrumental ($ee$) 	& 0		& 0		& 30		& 42			& 0		\\
Instrumental ($e\mu$ & 0		& 0		& 0		& 28			& 0		\\
Instrumental ($\mu\mu$) & 0	& 0		& 27		& 42			& 0		\\
Instrumental Model	& -		& -		& S	& S		& -		\\
\end{tabular}
\end{ruledtabular}

\vskip 0.3cm
{\centerline{D0: $VH \rightarrow\ell^{\pm}\ell^{\prime\pm} + X $ Run~IIb channel relative uncertainties (\%)}}
\vskip 0.099cm
\begin{ruledtabular}
\begin{tabular}{ l  c  c  c  c c } \\
Contribution 		&WZ/ZZ	&W+jet	&ChargeFlip	&Multijet 		& $VH \rightarrow \ell\ell X$	 \\
\hline
Cross section		& 7		& 6		& 0		& 0			& 5		\\
Normalization		& 4.7		& 4.7		& 0		& 0			& 4.7		\\
Signal Branching Fraction     &-         &-           & -     &-          &0-7.3 \\
JetID/JES		  	& 2		& 2		& 0		& 0			& 2		\\
Jet-Lepton Fake	& 0		& 20-32	& 0		& 0			& 0		\\
Instrumental ($ee$) 	& 0		& 0		& 15		& 30			& 0		\\
Instrumental ($e\mu$ & 0		& 0		& 0		& 18			& 0		\\
Instrumental ($\mu\mu$) & 0	& 0		& 11		& 29			& 0		\\
Instrumental Model	& -		& -		& S	& S		& -		\\
\end{tabular}
\end{ruledtabular}
\end{center}
\end{table}

\begin{table*}
\begin{center}
\caption{\label{tab:d0lvjj}
Systematic uncertainties on the signal and background contributions for D0's
 $H\rightarrow W W^{*} \rightarrow \ell\nu jj$ electron and muon channels.  Systematic uncertainties are listed
 by name; see the original references for a detailed explanation of their meaning and on how they are
 derived.
Signal uncertainties are shown for $m_H=160$ GeV/$c^2$ for all channels except for $WH$,
shown for $m_H=115$ GeV/$c^2$.  Those affecting the shape of
the RF discriminant are indicated with ``Y.''
Uncertainties are listed as relative changes in normalization,
in percent, except for those also marked by ``S,'' where
the overall normalization is constant, and the value given
denotes the maximum percentage change from nominal in any region of the
distribution.}

\vskip 0.1cm
{\centerline{D0: $H\rightarrow W W^{*} \rightarrow \ell\nu jj$ Run~II channel relative uncertainties (\%)}}
\vskip 0.099cm
\begin{ruledtabular}
\begin{tabular}{llccccccl}

Contribution & Shape & $W$+jets & $Z$+jets & Top & Diboson & $gg\to H$ & $qq\to qqH$ & $WH$ \\ \hline
Jet energy scale & Y & $\binom{+6.7}{-5.4}^S$ & $<0.1$ & $\pm$0.7 & $\pm$3.3 & $\binom{+5.7}{-4.0}$ & $\pm$1.5 &$\binom{+2.7}{-2.3}$  \\
Jet identification & Y & $\pm 6.6^S$ & $<0.1$ & $\pm$0.5 & $\pm$3.8  & $\pm$1.0 & $\pm$1.1 & $\pm$1.0 \\
Jet resolution & Y & $\binom{+6.6}{-4.1}^S$ & $<0.1$ & $\pm$0.5 & $\binom{+1.0}{-0.5}$ & $\binom{+3.0}{-0.5}$ & $\pm 0.8$ & $\pm 1.0$ \\
Association of jets with PV & Y & $\pm 3.2^S$ & $\pm 1.3^S$ & $\pm$1.2 & $\pm$3.2 & $\pm$2.9 & $\pm$2.4 & $\binom{+0.9}{-0.2}$ \\
Luminosity & N & n/a & n/a & $\pm$6.1 & $\pm$6.1 & $\pm$6.1 & $\pm$6.1 &  $\pm$6.1 \\
Muon trigger  & Y & $\pm 0.4^S$ & $<0.1$ & $<0.1$ & $<0.1$ & $<0.1$ & $<0.1$ &  $<0.1$ \\
Electron identification & N & $\pm$4.0  & $\pm$4.0  & $\pm$4.0  & $\pm$4.0  & $\pm$4.0  & $\pm$4.0  & $\pm$4.0 \\
Muon identification  & N & $\pm$4.0  & $\pm$4.0  & $\pm$4.0  & $\pm$4.0  & $\pm$4.0  & $\pm$4.0  & $\pm$4.0  \\
ALPGEN tuning & Y & $\pm 1.1^S$ & $\pm 0.3^S$ & n/a & n/a & n/a & n/a & n/a \\
Cross Section & N & $\pm$6 & $\pm$6 &  $\pm$10 & $\pm$7 & $\pm$10 & $\pm$10 & $\pm$6 \\
Heavy-flavor fraction  & Y & $\pm$20 & $\pm$20 & n/a & n/a & n/a & n/a & n/a  \\
Signal Branching Fraction & N & n/a &n/a &n/a& n/a & 0-7.3 & 0-7.3 &  0-7.3 \\
PDF & Y & $\pm 2.0^S$ & $\pm 0.7^S$ & $<0.1^S$ & $<0.1^S$ & $<0.1^S$ & $<0.1^S$ & $<0.1^S$ \\
 &  &  &  &  &  &  &  &  \\
 &  & \multicolumn{ 3}{c}{Electron channel} & \multicolumn{ 3}{c}{Muon channel} &  \\
Multijet Background & Y  & \multicolumn{ 3}{c}{$\pm$6.5} & \multicolumn{ 3}{c}{$\pm$26} &  \\
\end{tabular}
\end{ruledtabular}
\end{center}
\end{table*}


\begin{table}
\begin{center}
\caption{\label{tab:cdfsystH4l} Systematic uncertainties on the signal and background contributions for CDF's
$H\rightarrow \ell^{\pm}\ell^{\mp}\ell^{\prime \pm}\ell^{\prime \mp}$ channel. This channel is sensitive to
gluon fusion production and $WH$, $ZH$ and VBF production.  Systematic uncertainties are listed by name (see
the original references for a detailed explanation of their meaning and on how they are derived). Uncertainties
are relative, in percent, and are symmetric unless otherwise indicated.  The uncertainties associated with the
different background and signal processed are correlated unless otherwise noted.  Boldface and italics indicate
groups of uncertainties which are correlated with each other but not the others on the line.}

\vskip 0.1cm
{\centerline{CDF: $H\rightarrow \ell^{\pm}\ell^{\mp}\ell^{\prime \pm}\ell^{\prime \mp}$ channel relative uncertainties (\%)}}
\vskip 0.099cm
\begin{ruledtabular}
\begin{tabular}{lcccccc} \\
Contribution               &   $ZZ$     &  $Z(/\gamma^*)$+jets  &  $gg \to H$ &     $WH$    &    $ZH$    & VBF       \\ \hline
{\bf Cross Section :}      &            &                       &             &             &            &           \\
Scale                      &            &                       &   7.0       &             &            &           \\
PDF Model                  &            &                       &   7.7       &             &            &           \\
Total                      & {\it 10.0} &                       &             &   {\bf5.0}  &  {\bf5.0}  & 10.0      \\
$\mathcal{BR}(H\to VV)$    &            &                       &   3.0       &     3.0     &    3.0     & 3.0       \\
{\bf Acceptance :}         &            &                       &             &             &            &           \\
PDF Model                  &     2.7    &                       &             &             &            &           \\
Higher-order Diagrams      &     2.5    &                       &             &             &            &           \\
Jet Fake Rates             &            &     50.0              &             &             &            &           \\
Lepton ID Efficiencies     & {\it 3.6}  &                       &{\it 3.6 }   &   {\it 3.6} &  {\it 3.6} & {\it 3.6} \\
Trigger Efficiencies       & {\it 0.4}  &                       &{\it 0.4 }   &   {\it 0.4} &  {\it 0.4} & {\it 0.4} \\
{\bf Luminosity}           & {\it 3.8}  &                       &{\it 3.8 }   &   {\it 3.8} &  {\it 3.8} & {\it 3.8} \\
{\bf Luminosity Monitor}   & {\it 4.4}  &                       &{\it 4.4 }   &   {\it 4.4} &  {\it 4.4} & {\it 4.4} \\
\end{tabular}
\end{ruledtabular}

\end{center}
\end{table}



\clearpage\newpage


\begin{table}
\begin{center}

\caption{\label{tab:cdfsystttHLJ} Systematic uncertainties on the
signal and background contributions for CDF's $t\bar{t}H \to
\ell+$jets channels.  Systematic uncertainties are listed by name; see
the original references for a detailed explanation of their meaning
and on how they are derived.  Systematic uncertainties for $t\bar{t}H$
shown in this table are obtained for $m_H=115$ GeV/$c^2$.
Uncertainties are relative, in percent, and are symmetric unless
otherwise indicated.}
\begin{small}

\vskip 0.1cm
{\centerline{CDF: $t\bar{t}H$ $\protect \ell+\raisebox{.3ex}{$\not$}E_{T}$+4 jets relative uncertainties (\%)}}
\vskip 0.099cm
\begin{tabular}{lcccccccccc}\hline\hline
                    & \multicolumn{2}{c}{1 tight, 1 loose}& \multicolumn{2}{c}{1 tight, $\ge2$ loose}& \multicolumn{2}{c}{2 tight, 0 loose}& \multicolumn{2}{c}{2 tight, $\ge1$ loose}& \multicolumn{2}{c}{$\ge3$ tight, $\ge0$ loose} \\
Contribution              & $t\bar{t}$       & $t\bar{t}H$      & $t\bar{t}$       & $t\bar{t}H$           & $t\bar{t}$       & $t\bar{t}H$      & $t\bar{t}$       & $t\bar{t}H$           & $t\bar{t}$       & $t\bar{t}H$                 \\ \hline \noalign{\smallskip}
$t\bar{t}$ Cross Section  &                  & 10               &                  & 10                    &                  & 10               &                  & 10                    &                  & 10                          \\
$t\bar{t}H$ Cross Section & 10               &                  & 10               &                       & 10               &                  & 10               &                       & 10               &                             \\
Luminosity ($\sigma_{\mathrm{inel}}(p\bar{p})$)& 3.8 & 3.8      & 3.8              & 3.8                   & 3.8              & 3.8              & 3.8              & 3.8                   & 3.8              & 3.8                         \\
Luminosity Monitor        & 4.4              & 4.4              & 4.4              & 4.4                   & 4.4              & 4.4              & 4.4              & 4.4                   & 4.4              & 4.4                         \\[1.1ex]
$B$-Tag Efficiency        & $^{+1.4}_{-2.5}$ & $^{-2.9}_{-2.0}$ & $^{+3.3}_{-1.5}$ & $^{+0.3}_{+0.3}$      & $^{+7.3}_{-9.4}$ & $^{+6.7}_{-2.0}$ & $^{+8.3}_{-8.8}$ & $^{+7.0}_{-7.7}$      & $^{+11}_{-12}$   & $^{+11}_{-16}$              \\[2.2ex]
Mistag Rate               & $^{+1.7}_{-2.0}$ & $^{-0.4}_{-1.5}$ & $^{+10}_{-11}$   & $^{-1.1}_{-5.7}$      & $^{-1.2}_{+2.7}$ & $^{+2.7}_{+3.7}$ & $^{+7.6}_{-7.4}$ & $^{+1.7}_{+2.4}$      & $^{+3.3}_{-5.1}$ & $^{+1.6}_{+0.2}$            \\[2.2ex]
Jet Energy Scale          & $^{+3.8}_{-5.1}$ & $^{-13}_{+6.7}$  & $^{+2.5}_{-4.5}$ & $^{0.0}_{0.0}$        & $^{+4.2}_{-4.8}$ & $^{-5.9}_{+5.9}$ & $^{+2.5}_{-3.8}$ & $^{-12}_{0.0}$        & $^{+3.3}_{-4.4}$ & $^{-12}_{0.0}$              \\[2.2ex]
ISR+FSR+PDF               & $^{-1.8}_{-1.0}$ & $^{-0.1}_{+0.1}$ & $^{-1.3}_{+2.3}$ & $^{-0.5}_{+0.5}$      & $^{-3.8}_{-1.3}$ & $^{+0.2}_{-0.2}$ & $^{-4.4}_{-1.1}$ & $^{+0.0}_{-0.0}$      & $^{-2.9}_{-3.5}$ & $^{-0.2}_{+0.2}$            \\[2.2ex]\hline\hline
\end{tabular}

\vspace*{1cm}

{\centerline{CDF: $t\bar{t}H$ $\protect \ell+\raisebox{.3ex}{$\not$}E_{T}$+5 jets relative uncertainties (\%)}}
\vskip 0.099cm
\begin{tabular}{lcccccccccc}\hline\hline
                    & \multicolumn{2}{c}{1 tight, 1 loose}& \multicolumn{2}{c}{1 tight, $\ge2$ loose}& \multicolumn{2}{c}{2 tight, 0 loose}& \multicolumn{2}{c}{2 tight, $\ge1$ loose}& \multicolumn{2}{c}{$\ge3$ tight, $\ge0$ loose} \\
Contribution              & $t\bar{t}$       & $t\bar{t}H$      & $t\bar{t}$       & $t\bar{t}H$           & $t\bar{t}$       & $t\bar{t}H$      & $t\bar{t}$       & $t\bar{t}H$           & $t\bar{t}$       & $t\bar{t}H$                 \\ \hline \noalign{\smallskip}
$t\bar{t}$ Cross Section  &                  & 10               &                  & 10                    &                  & 10               &                  & 10                    &                  & 10                          \\
$t\bar{t}H$ Cross Section & 10               &                  & 10               &                       & 10               &                  & 10               &                       & 10               &                             \\
Luminosity ($\sigma_{\mathrm{inel}}(p\bar{p})$)& 3.8 & 3.8      & 3.8              & 3.8                   & 3.8              & 3.8              & 3.8              & 3.8                   & 3.8              & 3.8                         \\
Luminosity Monitor        & 4.4              & 4.4              & 4.4              & 4.4                   & 4.4              & 4.4              & 4.4              & 4.4                   & 4.4              & 4.4                         \\[1.1ex]
$B$-Tag Efficiency        & $^{+1.8}_{-3.5}$ & $^{-0.4}_{+2.7}$ & $^{+4.5}_{-4.1}$ & $^{-1.3}_{-1.6}$      & $^{+8.2}_{-6.8}$ & $^{+2.5}_{-5.0}$ & $^{+9.7}_{-7.7}$ & $^{+5.9}_{-5.5}$      & $^{+11}_{-16}$   & $^{+9.9}_{-13}$             \\[2.2ex]
Mistag Rate               & $^{+1.3}_{-2.9}$ & $^{-7.5}_{+1.8}$ & $^{+18}_{-8.9}$  & $^{+4.3}_{-6.6}$      & $^{-0.2}_{+2.6}$ & $^{-2.0}_{+1.0}$ & $^{+8.2}_{-8.7}$ & $^{+2.5}_{-2.2}$      & $^{+8.1}_{-3.4}$ & $^{+1.3}_{-0.5}$            \\[2.2ex]
Jet Energy Scale          & $^{+19}_{-16}$   & $^{+7.5}_{-7.5}$ & $^{+17}_{-15}$   & $^{+7.1}_{-14}$       & $^{+18}_{-17}$   & $^{+7.0}_{-4.7}$ & $^{+16}_{-16}$   & $^{+6.7}_{-3.3}$      & $^{+15}_{-15}$   & $^{-2.7}_{-8.1}$            \\[2.2ex]
ISR+FSR+PDF               & $^{+10}_{-1.2}$  & $^{-0.0}_{+0.0}$ & $^{+14}_{-1.0}$  & $^{-0.2}_{+0.2}$      & $^{+8.2}_{-6.5}$ & $^{+0.0}_{-0.0}$ & $^{+12}_{-5.1}$  & $^{-2.1}_{+2.1}$      & $^{+14}_{-2.0}$  & $^{-1.9}_{+1.9}$            \\[2.2ex]\hline\hline
\end{tabular}
\end{small}

\end{center}
\end{table}


\begin{table}[h]
\begin{center}
\caption{\label{tab:cdfsysttthmetjets} Systematic uncertainties on the
signal and background contributions for CDF's $t\bar{t}H$ 2-tag and
3-tag $\protect \raisebox{.3ex}{$\not$}E_{T}$+jets channels.  Systematic
uncertainties are listed by name; see the original references for a
detailed explanation of their meaning and on how they are derived.
Systematic uncertainties for $t\bar{t}H$ shown in this table are obtained
for $m_H=120$ GeV/$c^2$.  Uncertainties are relative, in percent, and are
symmetric unless otherwise indicated.}
\vskip 0.1cm
{\centerline{CDF: $t\bar{t}H$ $\protect \raisebox{.3ex}{$\not$}E_{T}$+jets 2-tag channel relative uncertainties (\%)}}
\vskip 0.099cm
\begin{ruledtabular}
\begin{tabular}{lccc}\\
Contribution              & non-$t\bar{t}$ & $t\bar{t}$ & $t\bar{t}H$   \\ \hline
Luminosity ($\sigma_{\mathrm{inel}}(p{\bar{p}})$)
                          & 0      & 3.8     & 3.8   \\
Luminosity Monitor        & 0      & 4.4     & 4.4   \\
Jet Energy Scale          & 0      & 2       & 11    \\
Trigger Efficiency        & 0      & 7       & 7     \\
$B$-Tag Efficiency        & 0      & 7       & 7     \\
ISR/FSR                   & 0      & 2       & 2     \\
PDF                       & 0      & 2       & 2     \\
$t{\bar{t}}$ Cross Section& 0      & 10      & 0     \\
$t{\bar{t}}b{\bar{b}}$ Cross Section  & 0    & 3       & 0    \\
Signal Cross Section      & 0      & 0       & 10    \\
Background Modeling       & 6      & 0       & 0     \\
Background $B$-tagging    & 5      & 0       & 0     \\
\end{tabular}
\end{ruledtabular}

\vskip 0.3cm
{\centerline{CDF: $t\bar{t}H$ $\protect \raisebox{.3ex}{$\not$}E_{T}$+jets 3-tag channel relative uncertainties (\%)}}
\vskip 0.099cm
\begin{ruledtabular}
\begin{tabular}{lccc}\\
Contribution              & non-$t\bar{t}$ & $t\bar{t}$ & $t\bar{t}H$   \\ \hline
Luminosity ($\sigma_{\mathrm{inel}}(p{\bar{p}})$)
                          & 0      & 3.8     & 3.8   \\
Luminosity Monitor        & 0      & 4.4     & 4.4   \\
Jet Energy Scale          & 0      & 3       & 13    \\
Trigger Efficiency        & 0      & 7       & 7     \\
$B$-Tag Efficiency        & 0      & 9       & 9     \\
ISR/FSR                   & 0      & 2       & 2     \\
PDF                       & 0      & 2       & 2     \\
$t{\bar{t}}$ Cross Section& 0      & 10      & 0     \\
$t{\bar{t}}b{\bar{b}}$ Cross Section  & 0    & 5       & 0    \\
Signal Cross Section      & 0      & 0       & 10    \\
Background Modeling       & 6      & 0       & 0     \\
Background $B$-tagging    & 10     & 0       & 0     \\
\end{tabular}
\end{ruledtabular}
\end{center}
\end{table}

\begin{table}[h]
\begin{center}
\caption{\label{tab:cdfsysttthalljets} Systematic uncertainties on the signal and
background contributions for CDF's $t\bar{t}H$ 2-tag and 3-tag all jets channels.
Systematic uncertainties are listed by name; see the original references for a
detailed explanation of their meaning and on how they are derived.  Systematic
uncertainties for $t\bar{t}H$ shown in this table are obtained for $m_H=120$
GeV/$c^2$.  Uncertainties are relative, in percent, and are symmetric unless
otherwise indicated.}
\vskip 0.1cm
{\centerline{CDF: $t\bar{t}H$ all jets 2-tag channel relative uncertainties (\%)}}
\vskip 0.099cm
\begin{ruledtabular}
\begin{tabular}{lccc}\\
Contribution              & non-$t\bar{t}$ & $t\bar{t}$ & $t\bar{t}H$   \\ \hline
Luminosity ($\sigma_{\mathrm{inel}}(p{\bar{p}})$)
                          & 0      & 3.8     & 3.8   \\
Luminosity Monitor        & 0      & 4.4     & 4.4   \\
Jet Energy Scale          & 0      & 11      & 20    \\
Trigger Efficiency        & 0      & 7       & 7     \\
$B$-Tag Efficiency        & 0      & 7       & 7     \\
ISR/FSR                   & 0      & 2       & 2     \\
PDF                       & 0      & 2       & 2     \\
$t{\bar{t}}$ Cross Section& 0      & 10      & 0     \\
$t{\bar{t}}b{\bar{b}}$ Cross Section  & 0    & 3       & 0    \\
Signal Cross Section      & 0      & 0       & 10    \\
Background Modeling       & 9      & 0       & 0     \\
Background $B$-tagging    & 5      & 0       & 0     \\
\end{tabular}
\end{ruledtabular}

\vskip 0.3cm
{\centerline{CDF: $t\bar{t}H$ all jets 3-tag channel relative uncertainties (\%)}}
\vskip 0.099cm
\begin{ruledtabular}
\begin{tabular}{lccc}\\
Contribution              & non-$t\bar{t}$ & $t\bar{t}$ & $t\bar{t}H$   \\ \hline
Luminosity ($\sigma_{\mathrm{inel}}(p{\bar{p}})$)
                          & 0      & 3.8     & 3.8   \\
Luminosity Monitor        & 0      & 4.4     & 4.4   \\
Jet Energy Scale          & 0      & 13      & 22    \\
Trigger Efficiency        & 0      & 7       & 7     \\
$B$-Tag Efficiency        & 0      & 9       & 9     \\
ISR/FSR                   & 0      & 2       & 2     \\
PDF                       & 0      & 2       & 2     \\
$t{\bar{t}}$ Cross Section& 0      & 10      & 0     \\
$t{\bar{t}}b{\bar{b}}$ Cross Section  & 0    & 6       & 0    \\
Signal Cross Section      & 0      & 0       & 10    \\
Background Modeling       & 9      & 0       & 0     \\
Background $B$-tagging    & 10     & 0       & 0     \\
\end{tabular}
\end{ruledtabular}
\end{center}
\end{table}


\begin{table}
\begin{center}
\caption{\label{tab:cdfsysttautau} Systematic uncertainties on the signal and background contributions for CDF's
$H\rightarrow \tau^+\tau^-$ channels.  Systematic uncertainties are listed by name; see the original references
for a detailed explanation of their meaning and on how they are derived. Systematic uncertainties for the Higgs
signal shown in these tables are obtained for $m_H=120$ GeV/$c^2$.  Uncertainties are relative, in percent, and
are symmetric unless otherwise indicated. Shape uncertainties are labeled with an "S".}
\vskip 0.1cm
{\centerline{CDF: $H \rightarrow \tau^+ \tau^-$ channel relative uncertainties (\%)}}
\vskip 0.099cm
\begin{ruledtabular}
\begin{tabular}{lccccccccccc}\\
Contribution & $Z/\gamma^* \rightarrow \tau\tau$ & $Z/\gamma^* \rightarrow ee$ & $Z/\gamma^* \rightarrow \mu\mu$  & $t\bar{t}$ & diboson  & fakes from SS
&W+jets & $WH$      & $ZH$  & VBF      & $gg\rightarrow H$ \\
\hline
PDF Uncertainty                                      &   1  & 1 & 1   &  1   &  1  &  -   &    -     &  1.2  &  0.9  &  2.2  &  4.9   \\
ISR 1 JET                                            &   -  & - & -   &  -   &  -  &  -   &    -     & -6.9  & -2.9  & -1.8  & 11.8 \\
ISR $\ge$ 2 JETS                                     &   -  & - & -   &  -   &  -  &  -   &    -     & -0.5  & 0.1   & -1.9  & 18.1 \\
FSR 1 JET                                            &   -  & - &  -  &  -   &  -  &  -   &    -     &  4.3  & 0.7   & 1.1   & -3.4 \\
FSR $\ge$ 2 JETS                                     &   -  & - &  -  &  -   &  -  &  -   &    -     & -0.9  & -0.5  & -1.0  & -5.0 \\
JES (S) 1 JET                                    &  7.9 & 7.6 & 3.9  &  -8.4   & 6.3  &  - &  -  &  -4.8 &  -5.3 &  -3.7 & 5.1   \\
JES (S) $\ge$ 2 JETS                             & 14.0 & 11.0 & 20.1 & 2.8   & 11.7 &  - &  -  &  5.4  &  4.8  &  -5.2 & 13.2   \\
Normalization 1 JET                                  &  2.2 &  2.2 &  2.2 & 10    &    6 & 10 &  25 &  5  &  5  & 10  & 23.5   \\
Normalization $\ge$2 JETS                            &  2.2 &  2.2 &  2.2 & 10    &    6 & 10 &  30 &  5  &  5  & 10  & 67.5   \\
MC Acceptance                                        &  2.3 &   2.3 &  2.3 &  -    &  -   &  - &  -  &  -  &  -  &  -  &  -   \\
$\varepsilon_{trig}$ (e/$\mu$ leg)                   &   -  &   0.3 & 1.0  &  -    &  -   &  - &  -  &  -  &  -  &  -  &  -   \\
$\varepsilon_{trig}$ ($\tau$ leg)                    &   -  &   3.0 & 3.0  &  -    &  -   &  - &  -  &  -  &  -  &  -  &  -   \\
$\varepsilon_{ID}lep$                                &   -  &   2.4 & 2.6  &  -    &  -   &  - &  -  &  -  &  -  &  -  &  -   \\
$\varepsilon_{vtx}$                                  &   -  &   0.5 & 0.5  &  -    &  -   &  - &  -  &  -  &  -  &  -  &  -   \\
e/$\mu \rightarrow \tau_h$ fake rate                 &   -  &   7.4 & 15.5 &  -    &  -   &  - &  -  &  -  &  -  &  -  &  -   \\
Luminosity                                           &   -  &   5.9 & 5.9  &  -    &  -   &  - &  -  &  -  &  -  &  -  &  -   \\
tau ID scale factor:                                 &      &       &      &       &      &    &     &     &     &     &      \\
N$_{obs}$                                            &  1.8 &    -  &   -  &  1.8  & 1.8  &  - &   - & 1.8 & 1.8 & 1.8 & 1.8 \\
N$_{SSdata}$                                         & -3.7 &     - &   -  &  -3.7 & -3.7 &  - &   - &-3.7 & -3.7 &-3.7 & -3.7 \\
N$_{W+jets}$                                         & -1.6 &     - &   -  & -1.6  & -1.6  & - &   - & -1.6& -1.6 &-1.6 & -1.6 \\
Cross section (DY)                                   & -2.1 &     - &   -  & -2.1  & -2.1  & - &   - & -2.1& -2.1 &-2.1 & -2.1 \\
MC Acceptance (DY)                                   & -2.2 &     - &   -  & -2.2  & -2.2  & - &  -  & -2.2& -2.2 &-2.2 & -2.2 \\
e/$\mu \rightarrow \tau_h$ fake rate                 & -0.1 &     - &   -  & -0.1  & -0.1  & - & -   & -0.1& -0.1 &-0.1 & -0.1 \\
\end{tabular}
\end{ruledtabular}
\end{center}
\end{table}


\begin{table}
\begin{center}
\caption{\label{tab:cdfsystVtautau} Systematic uncertainties on the signal and background contributions for CDF's
$WH \rightarrow \ell \nu \tau^+\tau^-$ and $ZH \rightarrow \ell^+ \ell^- \tau^+ \tau^-$ channels.  Systematic
uncertainties are listed by name; see the original references for a detailed explanation of their meaning and
on how they are derived. Systematic uncertainties for the Higgs signal shown in these tables are obtained for
$m_H=120$ GeV/$c^2$.  Uncertainties are relative, in percent, and are symmetric unless otherwise indicated.}
\vskip 0.1cm
{\centerline{CDF: $WH \rightarrow \ell \nu \tau^+\tau^-$ and $ZH \rightarrow \ell^+ \ell^- \tau^+ \tau^-$ $\ell\ell\tau_h+X$ channel relative uncertainties (\%)}}
\vskip 0.099cm
\begin{ruledtabular}
\begin{tabular}{lccccccccccccccc}\\
Contribution & $ZZ$ & $WZ$ & $WW$ & $DY(ee)$ & $DY(\mu\mu)$ & $DY(\tau\tau)$ & $Z\gamma$ & $t\bar{t}$ & $W\gamma$ & $W+jet$ & $WH$ & $ZH$ & $VBF$ & $gg \rightarrow H$\\
\hline
Luminosity                   & 5.9 & 5.9 & 5.9 & 5.9 & 5.9 & 5.9 & 5.9 & 5.9 & 5.9 & 5.9 & 5.9 & 5.9 & 5.9 & 5.9 \\
Cross Section                &11.7 &11.7 &11.7 & 5.0 & 5.0 & 5.0 &11.7 &14.1 &11.7 & 5.0 & 5.0 & 5.0 &10.0 &10.0 \\
Z-vertex Cut Efficiency      & 0.5 & 0.5 & 0.5 & 0.5 & 0.5 & 0.5 & 0.5 & 0.5 & 0.5 & 0.5 & 0.5 & 0.5 & 0.5 & 0.5 \\
Trigger Efficiency           & 1.1 & 1.1 & 1.0 & 1.0 & 1.0 & 1.1 & 1.1 & 1.0 & 0.8 & 1.0 & 1.2 & 1.2 & 1.2 & 1.1 \\
Lepton ID Efficiency         & 2.4 & 2.3 & 2.4 & 2.4 & 2.4 & 2.4 & 2.4 & 2.4 & 2.3 & 2.4 & 2.4 & 2.4 & 2.4 & 2.4 \\
Lepton Fake Rate             &10.7 & 8.0 &26.7 &26.0 &26.6 &15.1 &27.1 &22.4 &22.8 &28.7 & 2.9 & 2.3 &15.1 &13.6 \\
Jet Energy Scale             & 1.3 & 1.1 & 0.0 & 3.2 & 5.1 & 0.6 & 6.6 & 0.1 & 2.0 & 0.2 & 0.1 & 0.03& 0.6 & 0.4 \\
MC stat                      & 3.7 & 2.9 & 7.6 & 1.5 & 1.7 & 2.2 & 4.1 & 3.1 & 20.0& 3.1 & 1.5 & 1.4 & 3.8 & 9.4 \\
PDF Model                    &  -  &  -  &  -  &  -  &  -  &  -  &  -  &  -  &  -  &  -  & 1.2 & 0.9 & 2.2 & 4.9 \\
ISR/FSR Uncertainties        &  -  &  -  &  -  &  -  &  -  &  -  &  -  &  -  &  -  &  -  & 1.3 & 2.1 & 0.6 & 0.2 \\
\end{tabular}
\end{ruledtabular}

\vskip 0.3cm
{\centerline{CDF: $WH \rightarrow \ell \nu \tau^+\tau^-$ and $ZH \rightarrow \ell^+ \ell^- \tau^+ \tau^-$ $e\mu\tau_h+X$ channel relative uncertainties (\%)}}
\vskip 0.099cm
\begin{ruledtabular}
\begin{tabular}{lccccccccccccccc}\\
Contribution & $ZZ$ & $WZ$ & $WW$ & $DY(ee)$ & $DY(\mu\mu)$ & $DY(\tau\tau)$ & $Z\gamma$ & $t\bar{t}$ & $W\gamma$ & $W+jet$ & $WH$ & $ZH$ & $VBF$ & $gg \rightarrow H$\\
\hline
Luminosity                   & 5.9 & 5.9 & 5.9 & 5.9 & 5.9 & 5.9 & 5.9 & 5.9 & 5.9 & 5.9 & 5.9 & 5.9 & 5.9 & 5.9 \\
Cross Section                &11.7 &11.7 &11.7 & 5.0 & 5.0 & 5.0 &11.7 &14.1 &11.7 & 5.0 & 5.0 & 5.0 &10.0 &10.0 \\
Z-vertex Cut Efficiency      & 0.5 & 0.5 & 0.5 & 0.5 & 0.5 & 0.5 & 0.5 & 0.5 & 0.5 & 0.5 & 0.5 & 0.5 & 0.5 & 0.5 \\
Trigger Efficiency           & 1.4 & 1.4 & 1.1 & 1.1 & 1.3 & 1.1 & 1.4 & 1.1 & 1.0 & 0.7 & 1.3 & 1.3 & 1.2 & 1.2 \\
Lepton ID Efficiency         & 2.4 & 2.4 & 2.4 & 2.4 & 2.4 & 2.4 & 2.4 & 2.4 & 2.4 & 2.4 & 2.4 & 2.4 & 2.4 & 2.4 \\
Lepton Fake Rate             & 9.0 & 6.5 &26.6 &20.8 &31.4 &25.2 &39.4 &27.8 &19.3 &41.9 & 1.6 & 2.5 &28.5 &29.2 \\
Jet Energy Scale             & 0.0 & 0.3 & 2.2 & 0.0 & 0.8 & 1.5 & 0.5 & 0.8 & 0.0 & 0.0 & 0.2 & 0.1 & 1.7 & 0.0 \\
MC stat                      & 12.9& 7.2 & 20.9& 57.7& 12.6& 7.7 & 10.2& 12.4& 35.4& 25.8& 2.1 & 3.9 &13.0 &44.7 \\
PDF Model                    &  -  &  -  &  -  &  -  &  -  &  -  &  -  &  -  &  -  &  -  & 1.2 & 0.9 & 2.2 & 4.9 \\
ISR/FSR Uncertainties        &  -  &  -  &  -  &  -  &  -  &  -  &  -  &  -  &  -  &  -  & 0.6 & 0.2 & 0.1 & 0.0 \\
\end{tabular}
\end{ruledtabular}

\vskip 0.3cm
{\centerline{CDF: $WH \rightarrow \ell \nu \tau^+\tau^-$ and $ZH \rightarrow \ell^+ \ell^- \tau^+ \tau^-$ $\ell\tau_h\tau_h+X$ channel relative uncertainties (\%)}}
\vskip 0.099cm
\begin{ruledtabular}
\begin{tabular}{lccccccccccccccc}\\
Contribution & $ZZ$ & $WZ$ & $WW$ & $DY(ee)$ & $DY(\mu\mu)$ & $DY(\tau\tau)$ & $Z\gamma$ & $t\bar{t}$ & $W\gamma$ & $W+jet$ & $WH$ & $ZH$ & $VBF$ & $gg \rightarrow H$\\
\hline
Luminosity                   & 5.9 & 5.9 & 5.9 & 5.9 & 5.9 & 5.9 & 5.9 & 5.9 & 5.9 & 5.9 & 5.9 & 5.9 & 5.9 & 5.9 \\
Cross Section                &11.7 &11.7 &11.7 & 5.0 & 5.0 & 5.0 &11.7 &14.1 &11.7 & 5.0 & 5.0 & 5.0 &10.0 &10.0 \\
Z-vertex Cut Efficiency      & 0.5 & 0.5 & 0.5 & 0.5 & 0.5 & 0.5 & 0.5 & 0.5 & 0.5 & 0.5 & 0.5 & 0.5 & 0.5 & 0.5 \\
Trigger Efficiency           & 1.0 & 1.1 & 0.9 & 1.0 & 1.1 & 1.1 & 1.1 & 1.0 & 0.7 & 0.9 & 1.1 & 1.1 & 1.1 & 1.1 \\
Lepton ID Efficiency         & 3.3 & 3.3 & 3.3 & 3.3 & 3.3 & 3.3 & 3.3 & 3.3 & 3.3 & 3.3 & 3.3 & 3.3 & 3.3 & 3.3 \\
Lepton Fake Rate             &10.4 & 6.8 &38.1 &43.3 &39.9 &24.8 &32.8 &34.2 &28.8 &34.8 & 3.1 & 5.9 &28.1 &26.3 \\
Jet Energy Scale             & 5.5 & 0.0 & 0.0 & 3.3 & 1.6 & 1.2 & 1.6 & 0.0 & 0.0 & 1.1 & 0.1 & 0.6 & 1.8 & 1.7 \\
MC stat                      & 12.5& 8.1 & 16.9& 18.3& 12.5& 4.9 & 12.6& 14.7& 70.7& 8.7 & 2.0 & 3.3 & 9.4 &18.3 \\
PDF Model                    &  -  &  -  &  -  &  -  &  -  &  -  &  -  &  -  &  -  &  -  & 1.2 & 0.9 & 2.2 & 4.9 \\
ISR/FSR Uncertainties        &  -  &  -  &  -  &  -  &  -  &  -  &  -  &  -  &  -  &  -  & 1.2 & 0.5 & 0.4 & 0.04\\
\end{tabular}
\end{ruledtabular}
\end{center}
\end{table}


\begin{table}
\begin{center}
\caption{\label{tab:cdfallhadsyst} Systematic uncertainties on the signal and background contributions for CDF's
$WH+ZH \rightarrow jjbb$ and $VBF \rightarrow jjbb$ channels.  Systematic uncertainties are listed by name; see the original references for a
detailed explanation of their meaning and on how they are derived.  Uncertainties with provided shape systematics
are labeled with ``s''.  Systematic uncertainties for $H$ shown in this table are obtained for $m_H=115$ GeV/$c^2$.
Uncertainties are relative, in percent, and are symmetric unless otherwise indicated.  The cross section uncertainties
are uncorrelated with each other (except for single top and $t{\bar{t}}$, which are treated as correlated).  The QCD
uncertainty is also uncorrelated with other channels' QCD rate uncertainties.}
\vskip 0.1cm
{\centerline{CDF: $WH+ZH\rightarrow jjbb$ and $VBF \rightarrow jjbb$ channel relative uncertainties (\%)}}
\vskip 0.099cm
\begin{ruledtabular}
\begin{tabular}{lccccc} \\
Contribution              & $t\bar{t}$ & diboson & $W/Z$+Jets & VH  & VBF \\ \hline
Jet Energy Correction     &            &         &            & 7 s & 7 s \\
PDF Modeling              &            &         &            & 2   & 2   \\
SecVtx+SecVtx             & 7.6        & 7.6     & 7.6        & 7.6 & 7.6 \\
SecVtx+JetProb            & 9.7        & 9.7     & 9.7        & 9.7 & 9.7 \\
Luminosity                & 6          & 6       & 6          & 6   & 6   \\
ISR/FSR modeling          &            &         &            & 2 s & 3 s \\
Jet Moment                &            &         &            & s   & s   \\
Trigger                   & 4          & 4       & 4          & 4   & 4   \\
QCD Interpolation         &            &         &            & s   & s   \\
QCD MJJ Tuning            &            &         &            & s   & s   \\
QCD Jet Moment Tuning     &            &         &            & s   & s   \\
cross section             & 10         & 6       & 50         &     &     \\
\end{tabular}
\end{ruledtabular}
\end{center}
\end{table}


\begin{table}[h]
\begin{center}
\caption{\label{tab:cdfsystgg} Systematic uncertainties on the signal contributions for CDF's
$H\rightarrow \gamma \gamma$ channels.
Systematic uncertainties are listed by name; see the original references
for a detailed explanation of their meaning and on how they are derived.  Uncertainties are relative, in
percent, and are symmetric unless otherwise indicated.}
\vskip 0.1cm
{\centerline{CDF: $H \rightarrow \gamma \gamma$ channel relative uncertainties (\%)}}
\vskip 0.099cm
\begin{ruledtabular}
\begin{tabular}{lcccc} \\
Channel & CC & CP & CC Conv & PC Conv \\ \hline
\bf{Signal Uncertainties :} & & & & \\
Luminosity & 6 & 6 & 6 & 6 \\
$\sigma_{ggH}/\sigma_{VH}/\sigma_{VBF}$ &
    14/7/5 & 14/7/5 & 14/7/5 & 14/7/5 \\
PDF & 2 & 2 & 2 & 2 \\
ISR & 3 & 4 & 2 & 5 \\
FSR & 3 & 4 & 2 & 5 \\
Energy Scale & 0.2 & 0.8 & 0.1 & 0.8 \\
Trigger Efficiency & -- & -- & 0.1 & 0.4 \\
$z$ Vertex & 0.2 & 0.2 & 0.2 & 0.2 \\
Conversion ID & -- & -- & 7 & 7 \\
Detector Material & 0.4 & 3.0 & 0.2 & 3.0 \\
Photon/Electron ID & 1.0 & 2.8 & 1.0 & 2.6 \\
Run Dependence & 3.0 & 2.5 & 1.5 & 2.0 \\
Data/MC Fits & 0.4 & 0.8 & 1.5 & 2.0 \\ \hline
\bf{Background Uncertainties :} & & & & \\
Fit Function & 3.5 & 1.1 & 7.5 & 3.5 \\
\end{tabular}
\end{ruledtabular}
\end{center}
\end{table}


\begin{table}[h]
\begin{center}
\caption{\label{tab:d0systgg} Systematic uncertainties on the signal and background contributions for D0's
$H\rightarrow \gamma \gamma$ channel. Systematic uncertainties for the Higgs signal shown in this table are
obtained for $m_H=125$ GeV/$c^2$.  Systematic uncertainties are listed by name; see the original references
for a detailed explanation of their meaning and on how they are derived.  Uncertainties are relative, in
percent, and are symmetric unless otherwise indicated.}
\vskip 0.1cm
{\centerline{D0: $H \rightarrow \gamma \gamma$ channel relative uncertainties (\%)}}
\vskip 0.099cm
\begin{ruledtabular}
\begin{tabular}{lcc}\\
Contribution &  ~~~Background~~~  & ~~~Signal~~~    \\
\hline
Luminosity~~~~                            &  6     &  6    \\
Acceptance                                &  --    &  2    \\
electron ID efficiency                    &  2     &  --   \\
electron track-match inefficiency         & 10     &  --   \\
Photon ID efficiency                      &  3     &   3   \\
Photon energy scale                       &  2     &   1   \\
Cross Section                             &  4     &  10   \\
Background subtraction                    &  15 &  -       \\
\end{tabular}
\end{ruledtabular}
\end{center}
\end{table}


\begin{thebibliography}{000}
\bibitem{dzHWWtau} \DZero Collaboration,
`` Search for the Standard Model Higgs boson in $ \mu + \tau_{\rm{had}} \le 1$ jet final
state  with 7.3 fb$^{-1}$ of data'',
D0 Conference Note 6179.
\end{thebibliography}
\end{document}